\documentclass[11pt]{article}

\usepackage[letterpaper,margin=1in]{geometry}
\usepackage[T1]{fontenc}
\usepackage[utf8]{inputenc}
\usepackage{lmodern}
\usepackage{microtype}
\usepackage{graphicx}
\usepackage{amsmath,amssymb,bm}
\usepackage{booktabs,array,tabularx,longtable,multirow}
\usepackage{caption}
\usepackage{subcaption}
\usepackage{xcolor}
\usepackage{url}
\usepackage[numbers,sort&compress]{natbib}
\usepackage[hidelinks]{hyperref}
\usepackage{authblk}
\usepackage{enumitem}
\usepackage{float}
\usepackage{placeins}
\usepackage{setspace}

\title{\textbf{Bottom-up Systems at Scale: The Case of Reddit}}

\author[1,2]{Shambhobi Bhattacharya}
\author[2,3]{Jisung Yoon}
\author[5,7]{Andrew J. Stier}
\author[2,4,5,6]{Hyejin Youn\thanks{Corresponding author: \href{mailto:h.youn@snu.ac.kr}{h.youn@snu.ac.kr}}}

\affil[1]{McCormick School of Engineering and Applied Science, Northwestern University, Evanston, IL, USA}
\affil[2]{Northwestern Institute on Complex Systems, Evanston, IL, USA}
\affil[3]{KDI School of Public Policy and Management, Sejong, Republic of Korea}
\affil[4]{Seoul National University, Seoul, Republic of Korea}
\affil[5]{Santa Fe Institute, Santa Fe, NM, USA}
\affil[6]{Complexity Science Hub, Vienna, Austria}
\affil[7]{Booth School of Business, The University of Chicago, Chicago, IL, USA}

\date{}

\begin{document}
\maketitle

\begin{abstract}
How do human collectives navigate increasing regulatory challenges to maintain order and avoid dysfunction as they grow in size? Here, we quantify how measurable actions---from user-to-user interactions to top-down enforcement---scale with size in Reddit sub-communities, spanning five orders of magnitude from $10^2$ to $10^7$ users. We find regulatory actions scale systematically with community size across many different topics, with consistent scaling rates and empirically grounded governance modes that distinguish how communities regulate themselves. Observed scaling exponents align with well-known laws in urban systems: superlinear growth for peer interaction and enforcement ($\beta \approx 1.12$--$1.18$), and near-linear scaling for automated (bot) oversight ($\beta \approx 0.95$, with 95\% confidence intervals spanning 1.0)---regularities that invite cross-system comparison as a path toward identifying whether common generative processes underlie them. We identify three empirically grounded modes of regulatory functions: \textit{Intensity} (54\% of variance); \textit{One-way vs.\ Two-way Communication} (25\%); and \textit{Impersonal vs.\ Personal Moderation} (21\%). Our temporal analysis shows that increasing regulatory intensity is most likely absorbed by one-way coordination. These observations align with classic governance frameworks, including Ostrom's self-governance in commons and organizational theories of bureaucratic versus discretionary control, and quantify previous qualitative observations of online systems. Our findings provide an empirical starting point for understanding how different regulatory mechanisms interact in bottom-up systems as they scale.
\end{abstract}


\section*{Significance Statement}
Regulatory oversight is often viewed as institutional friction, yet its mathematically guided empirics remain largely uncharacterized. By analyzing a hundred thousand digital communities on Reddit, we identify mutual interactions, enforcement, and bot oversight and their scaling behaviors. We find that mutual interactions and enforcement scale superlinearly with community size while bot oversight scales near-linearly, suggesting it keeps pace with growth but does not yet offset rising human moderation burdens. These findings provide an empirical framework for characterizing how different regulatory mechanisms interact in online communities as they scale, and a basis for future comparative work across bottom-up collective systems.

\section{Introduction}

As human collectives grow, they become more innovative as larger groups share knowledge, divide tasks, and develop new ways of coordination \cite{malone2018superminds, hidalgo2015why, revolutionizing2016, malone2020collective, yang2026scaling, page2017diversity, bettencourt2007growth, Hong2020UniversalPathway, Hosseinioun2025SkillDependencies, Youn2025, youn2016scaling}. Scale, however, invites inherent instability. As social systems expand, the cognitive and social burdens of upholding shared expectations begin to outpace individual contributions, making cooperation and norm enforcement increasingly difficult. So, the same structural openness that allows communities to grow also undermines their capacity to remain coherent. The addition of even a few members, for example, can push the collective past a tipping point where consensus and cooperation begin to unravel  \cite{centola2018experimental}. This difficulty is especially pronounced when a large group consists of a broad range of participants with interdependent subgroups, adding layers of coordination complexity that fundamentally challenge the maintenance of order \cite{Yang2024Regulatory}. Consequently, problems including harassment, doxxing, and the spread of misinformation become more prevalent, threatening both collective stability and member safety \cite{Yasseri2012Dynamics, schrading2015analysis, buozis2019doxing, yang2020prevalence, BondGarrett2023, bozarth2023wisdom}

Regulation seems to be key to maintaining stability as communities scale\cite{daston2022rules, ostrom1990governing}, and understanding its costs motivates this study. Most communities, even decentralized and porous ones, eventually establish governance structures and norms and rules to manage coordination challenges. Such structures often involve designated roles such as moderators or leaders---who enforce norms, resolve conflicts, and anticipate emerging challenges \cite{gillespie2018custodians, fiesler2018reddit, king2009perspective, Kraut2012BuildingSO, simon1947administrative, kiene2025relational}.

These challenges are particularly salient in online platforms, which have opened up new ways to scale individual efforts, work as teams, and build communities, often at scales that would not be possible in traditional offline organizations such as firms or institutions \cite{malhotra2021socio, mcafee2007, lane2023teams, revolutionizing2016}. For example, Wikipedia's collaborative authorship, distributed software development on GitHub, and the

massive, multi-threaded conversations on Reddit demonstrate how large-scale coordination can arise even in the absence of centralized organization \cite{Medvedev2019, Matias2016GoingDark}. Through these online systems, interactions that initially begin as brief exchanges among strangers can accumulate through repeated engagement, gradually giving rise to meaningful connections with shared bottom-up norms, expectations, and commitments \cite{king2009perspective}. These ``communities at scale'' often span a wide range of social systems, from intimate circles to networks of millions. In some cases, the coordination they enable extends beyond digital space, influencing offline behavior, public discourse, and social movements \cite{Weninger2015, facebook_social, Burke2020, Haythornthwaite2018, mancini2022self}.

Reddit's governance evolution illustrates this trajectory. Founded in 2005 as a largely unregulated, community-run platform, Reddit initially relied on minimal interventions. As the platform expanded, however, this approach became harder to sustain: unmoderated expression frequently escalated into harassment, abuse, and harmful and policy-violating activity \cite{Matias2016GoingDark, chandrasekharan2022quarantined}. In response, Reddit introduced a series of governance reforms, including volunteer moderators, automated bots, formal site-wide policies, and expanded administrative authority for moderators to define and enforce community-specific rules at scale \cite{reddit_automoderator_2015, robertson2015reddit, chandrasekharan2017efficacy, jhaver2019humanmachine}. This transition---from a hands-off ethos of minimal moderation to a more structured governance model---has been experienced by many large online communities, raising a natural question about how regulatory effort itself grows as communities scale.

In this paper, we provide empirical evidence of how communities are regulated at scale by measuring regulatory actions. Using Reddit as a model system, we quantify measurable actions---from bottom-up mutual interactions to top-down enforcement---and measure how these actions
scale as a function with community size $N$. We categorize user actions into two-way interactions (e.g., user commenting) and one-way coordination actions (e.g., moderation or automated bot actions) (Fig.~\ref{fig1}). We aim to develop a quantitative framework that enables comparative analysis across different social systems~\cite{yoon2023individual, yang2026scaling}.

Our analysis shows that regulatory actions scale on Reddit (Fig.~\ref{fig2}). First, we find mutual interactions and enforcement scale super-linearly with community size. Larger communities generate disproportionately more bottom-up coordination through discussion, while directly facing increasing top-down enforcement demands. At the same time, we find bot oversight scales near-linearly, keeping pace with growth but not yet providing the kind of efficiency gains that would offset rising human moderation burdens. Together, these findings challenge the common assumption that larger communities naturally benefit from self-regulation or economies of scale \cite{gillespie2018custodians, grimmelmann2015virtues}. Instead, we find that growth imposes increased governance burdens, especially for tasks that require human judgment.

We next analyze variations in regulatory actions relative to the expected volume given size (Fig.~\ref{fig3}). While the scaling framework accounts for the average amount of actions as a function of community size, significant heterogeneity remains to be explained. We therefore examine the residuals, the variations in regulatory actions that deviate from the baseline predicted by size, to uncover the latent dimensions of online governance (Fig.~\ref{fig4}).

We find that these deviations are not random noise but are structured along three principal directions of covariation, which we interpret as regulatory modes, with labels grounded in both the loading
structure and existing governance theory:
\textit{Intensity} (54\% of variance) capturing shared regulatory demand where contentious communities require more of all regulatory modes;
\textit{One-way vs.\ Two-way Communication} (25\%), revealing a trade-off between top-down enforcement and peer conversation; and
\textit{Impersonal vs.\ Personal Moderation} (21\%), distinguishing automated uniform enforcement from human contextual judgment. Temporal trajectories in this space reveal consistent year-to-year changes in governance structure across communities, regardless of overall governance style (Fig.~\ref{fig5}). Together, these modes align with classic governance frameworks, ranging from self-governance in commons \cite{ostrom1990governing} to organizational theories that contrast bureaucratic standardization with discretionary judgment \cite{daston2022rules, simon1947administrative, chandrasekharan2017efficacy, bakcoleman2022case}

Our contributions are fourfold. First, we introduce a behavior-based typology of regulatory actions for cross-platform and cross-contextual comparison of how collective order is maintained \cite{yoon2023individual, Yang2024Regulatory}. Second, we demonstrate that regulatory behaviors in digital spaces are not arbitrary but follow consistent scaling laws, producing empirical regularities comparable to those documented in biological and urban systems \cite{bettencourt2007growth,bettencourt2013origins}. Third, we provide empirical evidence of structured governance modes across communities, including systematic dimensions of regulatory intensity, the choice of coordination between top-down enforcement and bottom-up mutual interaction, and the orientation toward automated (impersonal) versus human-centered (personal) modes of regulation\cite{ostrom1990governing, gillespie2018custodians, matias2019preventing, shaw2014laboratories, kraut2012building, yun2019early, jhaver2019humanmachine, chandrasekharan2018internet, geiger2016bot, chandrasekharan2019crossmod}. Fourth, our findings enable cross-system comparison of regulation across complex systems — from digital platforms to biological and socio-economic organizations — providing an empirical basis for future comparative work \cite{west2018scale, simon_sciences_1996, simon1947administrative, Yang2024Regulatory, yang2026scaling}.

\section{Background}
\subsection{Reddit as our Choice of Empirics}
Among online communities, we chose Reddit for our empirical studies because it offers properties uniquely well-suited to studying scaling in online social systems \cite{Medvedev2019}. First, the platform hosts thousands of interest-based communities, known as \emph{subreddits}, each organized around a shared topic and varying widely in size---from as few as 10 users to over 10 million. This wide scale provides the necessary statistical breadth to identify consistent scaling laws. Reddit is an open, bottom-up system between groups of random people and organizations. Anyone can join, create sub-communities (subreddits), and volunteer to moderate them. Nevertheless, they have their own rules, norms, and moderation practices through volunteer moderators, resulting in an open and porous ecosystem with a highly decentralized governance structure \cite{kiene2019technological, chandrasekharan2018internet, centivany2016popcorn}. Furthermore, Reddit is a system of over one million such communities whose topics range widely across sports, science, politics, humor, and more. These largely self-organized governance communities share common features of participatory norms while exhibiting distinct cultures. In this sense, they function like digital cities within a nation. Together, these features make Reddit a natural empirical setting for examining how regulatory actions emerge and scale across communities of vastly different sizes under a shared platform architecture.

Reddit’s regulatory evolution illustrates how bottom-up governance can gradually give rise to more formalized, top-down oversight. Founded in 2005, it was meant to be a largely unregulated self-organization. However, as the platform expanded, it faced mounting pressure to address harmful content and coordination breakdowns, including harassment and extremism. As later reflected by Reddit’s leadership, early concerns that stronger intervention might \textit{break} the platform gave way to the recognition that a hands-off approach could ultimately threaten its stability and survival \cite{crucible_reddit_huffman}.

In response to growing pressures, Reddit introduced incremental institutional mechanisms. Early interventions included the creation of the NSFW (Not Safe For Work) tag in 2006 to shield sensitive content from the front page and the introduction of volunteer moderators in 2008, allowing communities to self-organize moderation while largely preserving Reddit’s decentralized ethos. By 2015, however, governance demands had peaked, leading Reddit to formally integrate AutoModerator—previously a third-party bot—into the platform’s core infrastructure, enabling scalable and standardized rule enforcement across communities \cite{reddit_automoderator_2015,jhaver2019humanmachine}. Around this time, nearly a hundred thousand volunteer moderators were helping manage subreddits, and their negotiations and protests increasingly shaped platform decisions and policy changes \cite{chandrasekharan2017efficacy, Matias2016GoingDark}. Fig.~\ref{fig2}d shows Reddit's transition from its early ethos of minimal intervention to adopting a formal site-wide Content Policy and expanding automated moderation infrastructure \cite{robertson2015reddit, Xu2012MachineMod}.

This emergence of structured governance --- from a decentralized, largely unregulated self-organization to a system with increasingly formalized oversight --- makes Reddit a compelling case for our study \cite{Matias2016GoingDark}. Even more so with Reddit's recent public listing as a company, there is increased scrutiny around its regulatory mechanisms \cite{nishant2024reddit}, highlighting the importance of understanding how regulatory mechanisms function and scale in large online communities.
Lastly, Reddit provides a uniquely rich and publicly accessible dataset through platforms such as Pushshift \cite{baumgartner2020pushshift}, widely adopted by the research community \cite{proferes2021studying, Medvedev2019}. This archive includes detailed post- and comment-level metadata, enabling researchers to trace regulatory activity over time—from user interactions to moderator interventions \cite{chandrasekharan2022quarantined, fiesler2018reddit}. We detail this dataset in Section~S1 of the Supplementary Information and the Materials and Methods section.

In summary, the wide scale of community size and the emergence of structured governance are why we chose Reddit as a natural setting for examining how regulatory dynamics scale. In this study, we systematically measure and analyze regulatory actions across communities spanning several orders of magnitude using the scaling framework.

\subsection{Scaling Framework as Our Methodological Choice}
We use a scaling framework to quantify and compare regulatory actions in size. At its core, scaling analysis quantifies how a system's measurable properties, $Y$, respond to changes in its size, $N$ \cite{thompson1992growth}. This relationship is typically expressed through the power-law function:$$Y = Y_0 N^\beta$$ where $Y_0$ is a normalization constant ($Y_0 = Y(N=1)$) and $\beta$ is the scaling exponent. In biological and physical systems, this allometric scaling reveals the underlying constraints governing the system's architecture \cite{west2018scale, thompson1992growth}. When $\beta = 1$, the system scales linearly; however, the most profound insights arise from non-linearities: sub-linear scaling ($\beta < 1$), indicating economies of scale and increased efficiency, and super-linear scaling ($\beta > 1$), indicating increasing returns and accelerated activity \cite{bettencourt2007growth}.

The application of scaling laws to human collectives originates from the study of cities. On the one hand, urban socio-economic outputs (e.g., GDP, patent filings, crime rates) tend to scale super-linearly ($\beta \approx 1.15$), driven by the density of human interactions and increasingly productive urban life \cite{bettencourt2013origins, bettencourt2007growth, youn2016scaling, Bettencourt2014Professional, GomezLievano2012Statistics}.  On the other hand, physical infrastructure in urban areas (e.g., length of electrical cables, number of gas stations) scales sublinearly ($\beta \approx 0.85$), accounting for a shared use of resources as size increases \cite{bettencourt2013origins, bettencourt2007growth}. This framework suggests that as urban systems expand, they become more than just larger versions of their smaller selves.

As such, the scaling framework has been one of the most active analytic tools for understanding diverse complex systems, from biological metabolic systems~\cite{west1997general} and urban infrastructure~\cite{bettencourt2007growth, wadell2023ev}, to cognitive processes~\cite{holden2009fractal} and social phenomena. Recent work has extended these principles to mental health, demonstrating that depression rates scale sublinearly with city population~\cite{Stier2021}, and to legal systems, where municipal code complexity scales sublinearly with jurisdiction population size ~\cite{ash2024scaling, Fernandes2025USCaselaw}. Therefore, the value of these scaling laws lies not only in making disparate systems comparable but also in motivating formal models---for instance, models of how social interactions generate productivity \cite{bettencourt2013origins}. These findings suggest that scaling laws may represent common organizing principles that emerge across diverse complex systems \cite{yang2026scaling}. Recent work continues to apply this framework across a growing range of empirical systems~\cite{wadell2023ev, yang2024intercity}. Here, we extend this approach to the governance of online Reddit communities. %%We aim to contribute to this framework by analyzing online Reddit communities.

\begin{figure}[t]
\centering
\includegraphics[width=\textwidth]{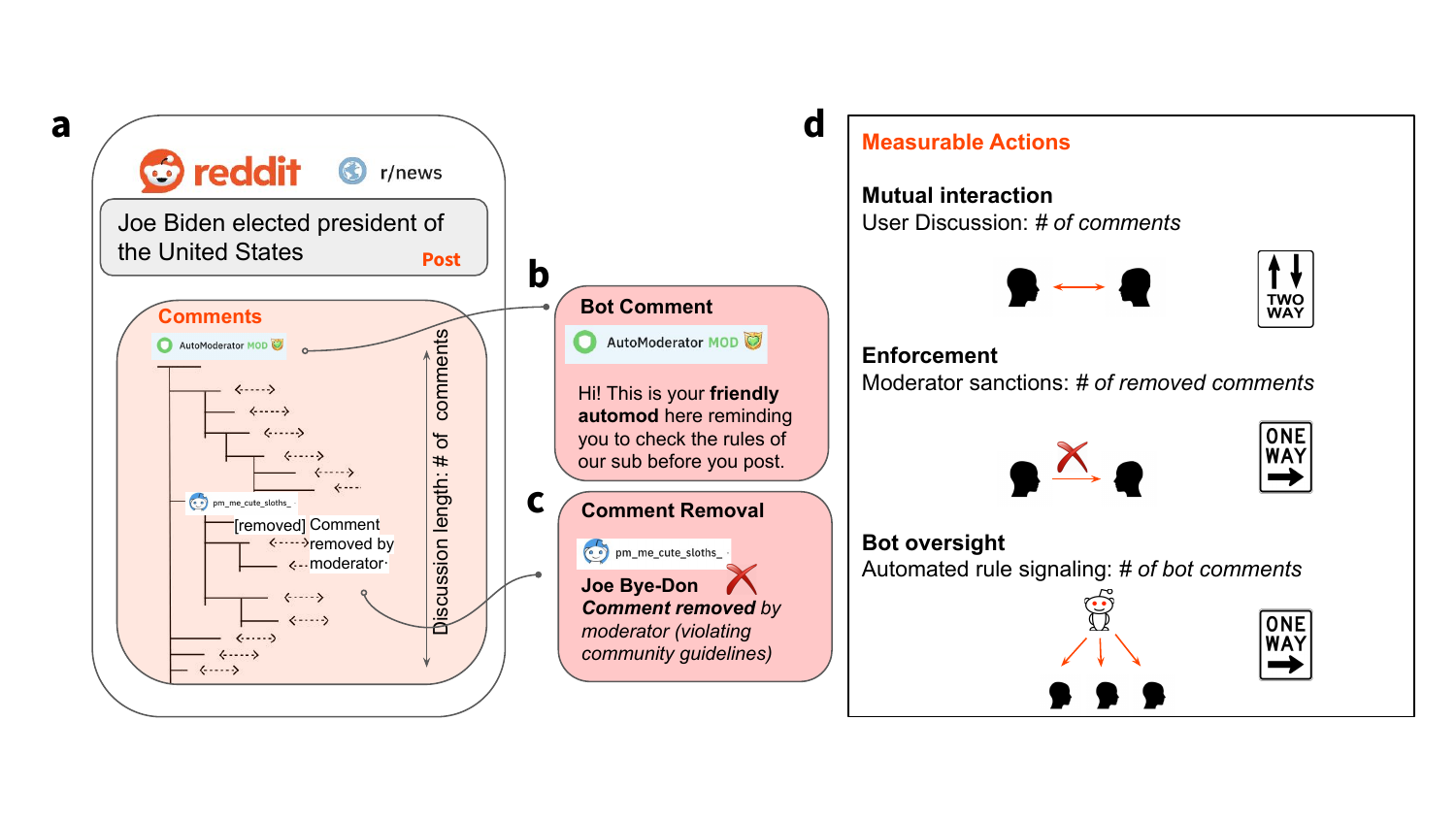}
\vspace{-30pt}
\caption{{\bf Measurable actions in Reddit.}
(\textbf{a}) Decentralized Units of Analysis: Subreddits serve as self-organized communities in which user activity and regulation occur without centralized authority. User discussion unfolds through comment threads attached to posts.
(\textbf{b}) Bot oversight: automated agents (e.g., AutoModerator) post standardized comments that signal rules, provide reminders, and apply codified moderation logic at scale.
(\textbf{c}) Enforcement: moderators apply discretionary sanctions by removing comments that violate community or platform guidelines.
(\textbf{d}) Analytical Framework of Measurable Actions. A typology of these actions categorized by their structural directionality, ranging from bottom-up mutual engagement (two-way coordination among users) to top-down enforcement and bot oversight (one-way coordination). Mutual interaction is operationalized as user discussion measured by the number of comments; enforcement as discretionary moderator sanctions measured by removed comments; and bot oversight as automated rule signaling measured by bot comments. Together, these metrics provide a quantitative basis for  characterizing how measurable actions scale with community size---from mutual interaction within which peer governance emerges, to direct top-down enforcement and automated oversight.}
\label{fig1}
\end{figure}

\section{Results}

\subsection{Regulatory functions and User Interactions in Reddit}
While urban scaling focuses on physical space, online platforms like Reddit provide a digital system where geography is replaced by network topology. Unlike cities, in these ``communities at scale", the absence of physical constraints allows for rapid expansion; yet coordination remains constrained by the social and cognitive capacities of their members. Recent work on Wikipedia has also found similar patterns of superlinear growth in contributions and sublinear growth in regulatory overhead \cite{yoon2023individual}. We therefore aim to contribute to this growing body of work by analyzing how regulatory actions scale across Reddit communities and comparing them with other complex systems.

Our dataset includes 107,006 subreddits spanning five orders of magnitude in size, from groups with 10 members to those with over 10 million members--- comprising 35 million comments and 700,000 moderator actions. Here, the subreddit serves as our unit of analysis, with each subreddit operating as a distinct community with its own norms, rules, and moderation structures, much like cities in an urban network. In our scaling framework, community size $N$ in the power-law function (introduced in the Background section) refers to the size of each subreddit, measured by its number of unique active users.

We first explain how these categories are defined, how the corresponding actions are operationalized and measured, and how the observed actions are situated within existing theory. First, the nature of the directionality of interactions distinguishes two-way versus one-way coordination. Two-way coordination is characterized by the absence of inherent asymmetry between two or more parties, such as in comment threads, where users engage reciprocally. In contrast, one-way coordination involves predominantly unilateral processes arising from inherent power differentials or asymmetries between individuals, such as enforcement and oversight by moderators or automated bots. We operationalize these coordination types into three measurable analytic constructs, as summarized in the empirical metrics shown in Fig.~\ref{fig1}:

\begin{enumerate}
  \item \textbf{Mutual Interaction (two-way coordination).} We measure the \emph{total volume of comments} within each subreddit as two-way coordination. In this study, we follow structuration theory as a theoretical foundation for our measures \cite{giddens1984constitution}. According to the theory, everyday conversations can serve as a channel for peer monitoring, the invocation of informal norms, and social approval or disapproval, not to mention users' engagement in the community.

  As such, mutual interactions, measured as the sequence of comments and replies, can function as more than simple self-expression and information exchange; they also serve as a channel for peer monitoring, the invocation of informal norms, and the expression of social approval or disapproval, reproducing the modalities that constitute governance in practice. These include: signification (the establishment and interpretation of shared systems of meaning within the community); legitimation (the evaluation of behavioral appropriateness through the application of norms and sanctions); and domination (the strategic use of resources) \cite{giddens1984constitution}.

  Empirically, subreddit threads are where users interpret content, explicitly cite community-specific rules (e.g., ``rule 3," off-topic warnings), provide reasons, exert informal pressure, and strategically use platform-specific resources such as karma/visibility. The literature on social norms characterizes these interactions as essential channels for decentralized, low-cost peer monitoring and sanctioning. Evidence shows that comments serve as a channel for peer monitoring, social approval/disapproval, and low-cost sanctions \cite{coleman1990foundations,bicchieri2006grammar,ostrom1990governing}.

  While not every comment is explicitly regulatory, evidence shows that more comments per post increase the \emph{opportunity and capacity} for norm enforcement by peers (more eyes and interventions per item), which is consistent with prior work on how discussion enables community self-governance \cite{Matias2016GoingDark,seering2020reconsidering} and that comments, more than posts, capture engaged participation \cite{horne2017social}. All in all, we treat these comments as mutual interactions that maintain collective order through reciprocal social feedback.

  \item \textbf{Enforcement (one-way removals):} We measure the \emph{number of comment removals} for one-way direct supervision. Reddit comments can be removed by either the original user, subreddit moderators, or Reddit administrators (site staff). However, Reddit provides a mark to distinguish between user-initiated deletions and administrative removals: content deleted by the original author is marked as ``[deleted]" and content removed by moderators for guideline violations is explicitly labeled as ``[removed]". We use these removal counts as an indicator of observable enforcement events through which volunteer moderators apply subreddit-specific rules and site-wide policies. This is a one-way regulatory action, central to the community's boundary-work, a mechanism used to maintain a  community identity and filter out non-compliant discourse \cite{fiesler2018reddit,gillespie2018custodians,chandrasekharan2018internet}. As such, in the context of structuration theory, the act of removal embodies the legitimation and authority dimensions of governance, where formalized norms are reinforced through active sanctions.

  \item \textbf{Bot Oversight (one-way automated moderations):}
  We measure the \emph{number of bot comments} (e.g., AutoModerator prompts, reminders, templated warnings) for one-way, automated enforcement. These bots operationalize codified rules at scale, performing tasks such as posting notices, rate-limiting, filtering, and providing structured guidance to users \cite{Tsvetkova2017Even,jhaver2019humanmachine, he2024platform}. By automating these processes, bots' infrastructure enforcement effectively reduces the routine cognitive and manual load on human moderators. Similar to human-led supervision, this is an asymmetric one-way coordination. Following structuration theory, bot interventions stabilize the legitimation and domination dimensions of governance through technical means, ensuring that community standards are consistently reinforced without direct human intervention in every instance. Therefore, bot comments provide a transparent and measurable trace of automated governance activity, allowing us to quantify the role of automation in the overall enforcement mix \cite{fiesler2018reddit,gillespie2018custodians}.

  Although both moderator removals and bot enforcement constitute one-way coordination, we purposefully distinguish between them because they rely on different governance logics. Bots primarily execute routine, codified tasks, whereas human moderators are required for contextual interpretation involving intent, civility, and relevance. Unlike automated systems that operate through predefined norms and rules, human supervision entails discretionary judgment under uncertainty. Understanding both bot comments and human-led removals thus allows us to differentiate how algorithmic and discretionary enforcement scale as communities grow in size and complexity.

\end{enumerate}

Figure~\ref{fig1} illustrates how we identify and quantify these three regulatory actions in a sample post from the \emph{r/news} subreddit. On Reddit, the prefix \emph{r/} denotes a subreddit, followed by its community name; we adopt this convention throughout the paper. \emph{r/news} is among the largest communities on the platform, with approximately one million unique commenting users during our sample period, and is ``the place to share news articles about current events in the United States and the rest of the world" \cite{subreddit-news}.

As shown in Fig.~\ref{fig1}a, comments are made to individual posts, generating multiple layers of engagement. The total volume of comments provides a measure of two-way coordination, as users respond to one another through reciprocal exchanges. Unlike posts, which primarily share and disseminate information, comment threads incorporate social feedback, debate, and the emergence of shared judgments that contribute to collective norm enforcement. In the early stages of the platform, user-driven interaction served as the primary governance mechanism, relying on distributed peer monitoring rather than centralized oversight. We quantify this mutual interaction as a bottom-up regulatory function by measuring discussion length, defined as the total number of comments associated with a post, which captures the intensity of community-wide engagement.

Some comments are generated by moderation bots. As illustrated in Fig.~\ref{fig1}b, the AutoModerator bot posts standardized reminders prompting users to review community rules before participating. As the platform expanded and challenges such as harassment and spam became more prevalent, moderation practices based solely on user interaction became insufficient. Moderation bots were therefore introduced to enable more scalable forms of governance. Now widely deployed across Reddit, these bots automate routine regulatory tasks, including posting rule reminders within threads, removing policy-violating content, limiting submission rates, and filtering spam. Moderators can configure bots to enforce community-specific rules at scale, allowing codified regulatory functions to be applied consistently as participation grows. As these are codified and relatively fixed, we differentiate them from moderators' supervision by counting the number of bot comments within each subreddit.

Finally, we measure the number of removed comments for direct supervision, indicated by [removed] as shown in Fig.~\ref{fig1}c. These are norm-violating comments. The example shows a removed comment with a notice explaining why it was removed, enforce norms by removing content that violates either subreddit-specific guidelines or Reddit's overarching content policy \cite{chandrasekharan2018internet}. As communities grew, this moderator function became increasingly important but also more challenging to scale effectively. We quantify this function by assessing the volume of removed comments in each subreddit.

Fig.~\ref{fig1}d summarizes our analytical framework. We categorize comments, removed comments and bot comments into mutual interaction, supervision and rule enforcement by their horizontal nature of actions. These are regulatory actions that adapt to ensure that subreddits remain functional and well-coordinated as they scale across orders of magnitude, and we examine the interplay between these three regulatory actions.

\subsection{Scaling Laws of Reddit Governance}

\begin{figure}[t]
\centering
\includegraphics[width=\textwidth]{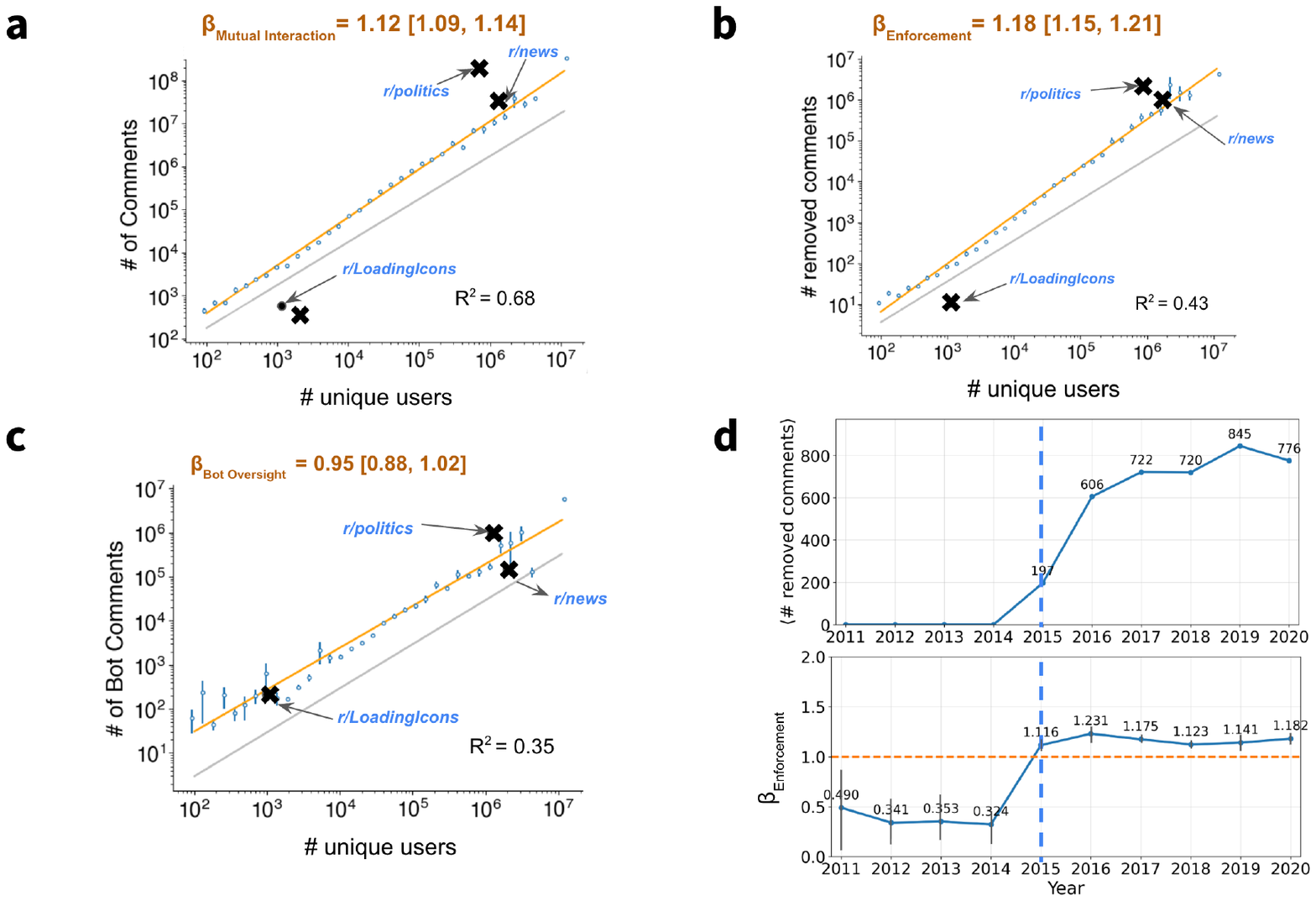}
\caption{\textbf{Scaling Analysis of the regulatory actions}
(\textbf{a}) Mutual interaction ($Y_{i,m}$, total comments),
(\textbf{b}) Enforcement ($Y_{i,e}$, \# removed comments), and
(\textbf{c}) Bot oversight ($Y_{i,b}$, \# bot comments) as functions of community size ($N_i$, unique active users).
For simplicity, individual subreddits are aggregated into logarithmic bins with error bars.
We provide a fitted scaling function ($Y \propto N^{\beta}$) in yellow with a gray linear line ($\beta = 1$) for reference.
Black crosses highlight three illustrative communities: \textit{r/news}, \textit{r/politics}, and \textit{r/LoadingIcons}, shown to indicate their relative positions in each scaling relationship.
(\textbf{d}) Temporal analysis of removed comments per subreddit (top) and enforcement exponents $\beta_e(t)$ (bottom).
The vertical dashed line marks 2015, when platform-wide policies and the widespread adoption of automated tools like AutoModerator were introduced.}
\label{fig2}
\end{figure}

We use scaling frameworks to identify and compare the underlying dynamics of diverse systems, addressing our research questions by quantifying how regulatory actions grow with a community size. More specifically, we examine the relationship between a quantity of interest ($Y$) and community size ($N$) to understand non-linear mechanisms that characterize complex systems \cite{west2018scale, thompson1992growth}.

For each subreddit $i$, we measure the volume of regulatory action $Y_{i,q}$ for each regulatory type $q$, such as mutual interaction (comments), enforcement (removed comments), and bot actions (bot comments). Community size $N_i$ is measured as the number of unique active users who have participated in $i$. These relationships are mathematically expressed as the power-law equation:

\begin{equation} \label{eq1}
Y_{i,q} = Y_{q}^{0} \cdot N_i^{\beta_q}
\end{equation}

Here, $Y_{q}^{0}$ is a normalization constant, which captures the baseline volume of regulatory actions independent of size, which could be thought of as an intrinsic quantity of action $q$. And the scaling exponent $\beta_q$ serves as another intrinsic quantity, determining the structural rate at which regulatory efforts evolve as the community expands.

We present the empirical scaling relationships between regulatory volume $Y_{i,q}$ and community size $N_i$ in Fig.~\ref{fig2}a--c for a curated subset of 2{,}828 labeled subreddits, and for the full universe of 107{,}006 subreddits in Fig.~S5 (SI Section~3). Each data point in each plot represents a log-binned aggregation, with specific subreddits identified by blue markers. Across both datasets, we find remarkably consistent power-law relationships spanning nearly five orders of magnitude. The absence of inflection points suggests that a unified set of scaling laws governs Reddit regulation, irrespective of community maturity or topic.  So, we reliably estimate the normalization constant $Y_{q}^0$ and the scaling exponent $\beta_q$ and we include these fitted lines (yellow) against gray linear reference lines ($\beta=1$) (See Methods).

The estimated scaling exponent $\beta_q$ quantifies whether a regulatory function benefits from economies of scale ($\beta < 1$) or burdens disproportionately ($\beta > 1$). For this comparative analysis, urban scaling serves as a good benchmark where socio-economic outputs scale superlinearly ($\beta \approx 1.15$) and infrastructure scales sublinearly ($\beta \approx 0.85$). We categorize Reddit's regulatory functions accordingly.

Fig. \ref{fig2} a and b show that \# comments ($q=$ mutual interactions) and  \# removed comments ($q = enforcement$)  scale superlinearly ($\beta_{\text{m}} = 1.12$ and $\beta_{\text{e}} = 1.18$). This superlinearity indicates these regulatory functions escalate faster than user growth, and thus dis-economies of scale. This dis-economies of scale in regulatory actions is not new. The legal sanctions, crime rates, and lawyer offices in urban areas show similar exponents of 1.15 \cite{youn2016scaling, GomezLievano2012, bettencourt2007growth}. While similar exponents across systems do not by themselves prove shared mechanisms, they raise the possibility of shared generative processes and motivate future cross-system investigation \cite{yang2026scaling}.

On the other hand, Fig.~\ref{fig2}c shows bot-generated comments (bot oversight) scale near-linearly ($\beta_{\text{bot}} = 0.95$, 95\% CI: [0.88, 1.02]; full-universe estimate $\beta = 0.984$, 95\% CI: [0.921, 1.048], see SI Section~S4). Because both confidence intervals include 1.0, we interpret this conservatively: bot oversight keeps pace with community growth but does not yet provide the kind of infrastructural efficiency that would offset the rising burden on human enforcement.

Ideally, automated oversight tools should supplement human moderators by scaling more effectively—providing increasing support as size grows to alleviate the regulatory burden. However, this observed near-linear pattern suggests that bots are merely keeping up with the rising user count rather than replacing or significantly reducing the need for human intervention. This dynamic likely persists because bots rely on codified knowledge; as communities expand, the variety of "abnormal" actions or nuanced violations may scale in ways that are not immediately codifiable, requiring continued reliance on human judgment for more complex governance tasks.

Finally, we show our findings are robust to different data sources and time windows (see SI Sec-1-3 for details). First, year-by-year analysis from 2011--2020 shows that post-2015 exponents are temporally stable, with removed comments consistently superlinear across all six years (Section~S2, Fig.~S4). Second, exponents estimated from our labeled sample of 2,828 subreddits closely match those from the full universe of 107,006 active subreddits ($\beta_{\text{comments}} = 1.12$, $\beta_{\text{mod}} = 1.16$, $\beta_{\text{bots}} = 0.98$), confirming that our findings are not artifacts of sample selection (Section~S3, Fig.~S5, Table~S3). Third, public moderator logs collected by Juneja et al (independent of Pushshift data) shows the similar exponent ($\beta = 1.17$, 95\% CI: [0.941, 1.406])  (Section~S4, Fig.~S7).

\subsection{Beyond Scale: Variation analysis}

\clearpage
\begin{figure}[p]
\centering
\includegraphics[width=\textwidth]{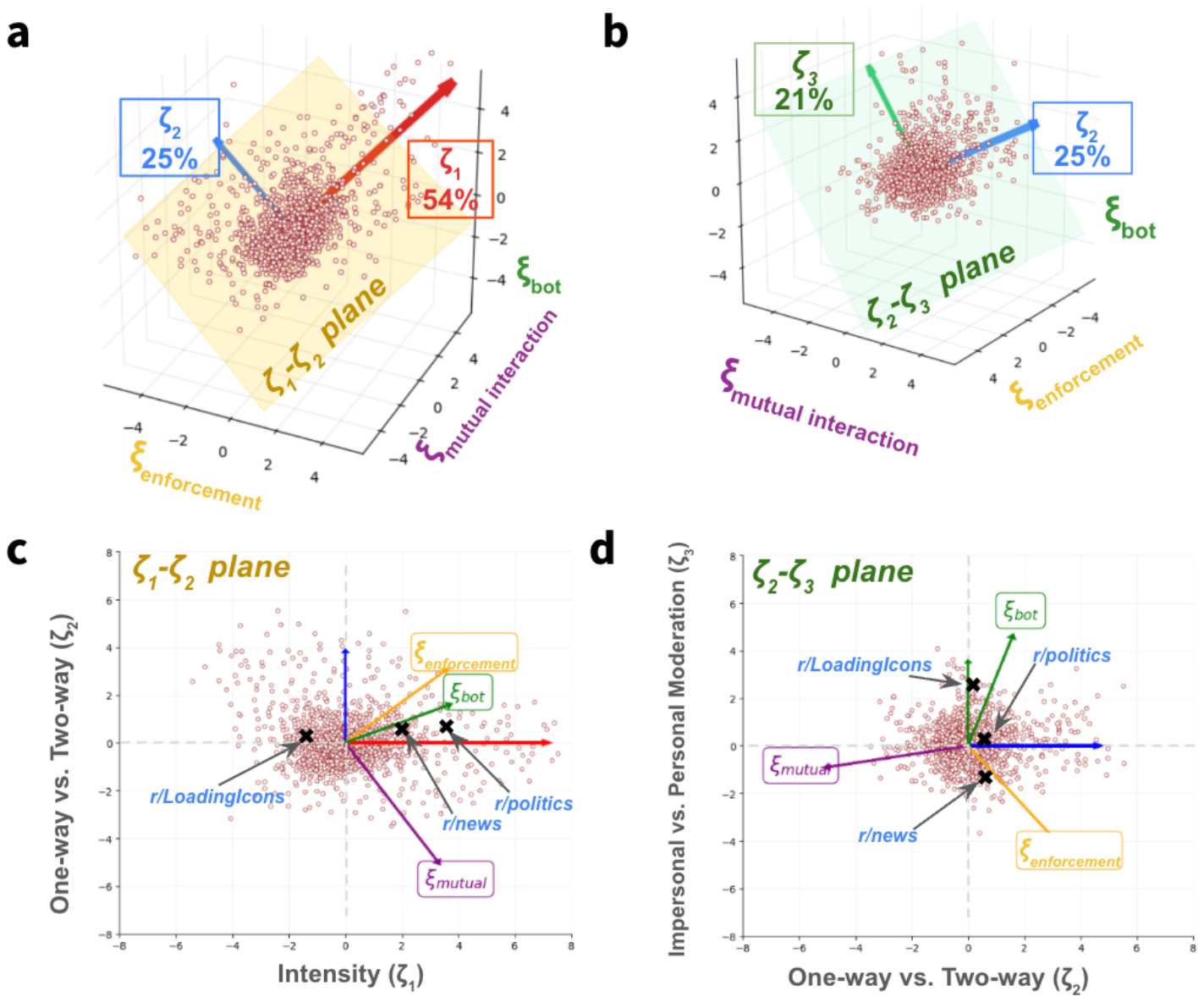}
\caption{\textbf{The regulation space: from regulatory actions to regulatory modes}
 Each subreddit $i$ (red marker) is embedded according to their regulatory action profile $\vec{\xi_i}$ that is $[\xi_{i,m},\xi_{i,e}, \xi_{i,b} ]$.   The red arrow represents the most two explaning vector, $\vec{\zeta_1}$ and  $\vec{\zeta_2}$,  for the variance distributions  $\vec{\xi_i}$.  The spread along  $\vec{\zeta_1}$, is roughly twice that of $\vec{\zeta_2}$, and their spanning plane is shared yellow.
\textbf{(b)} $\zeta_{1}$–$\zeta_{2}$ plane: a projection onto the yellow plane from (a), showing how subreddits distribute along these dominant dimensions. $\zeta_{1}$ indicates overall \textit{regulatory intensity}, where higher values indicate communities that employ all three mechanisms of mutual interaction, supervision, and bot enforcement more heavily. $\zeta_{2}$ seems to capture a trade-off or the \textit{one-way vs.\ two-way communication} axis, contrasting one-way enforcement (moderator and bot actions) with two-way conversational coordination (mutual interaction).
\textbf{(c)} The same space as (a) rotated to show the $\zeta_2$--$\zeta_3$ projection plane. Notice that the variance across both axes is comparable.
\textbf{(d)} $\zeta_2$--$\zeta_3$ plane: A projection that isolates a third interpretable structure of \textit{impersonal vs.\ personal moderation}. This axis distinguishes impersonal regulation (e.g., bots applying uniform rules automatically) from personal moderation (human moderators judging each case individually).
Together, these projections reveal a geometric decomposition of regulatory strategies that communities adopt - highlighting structured variation beyond what subreddit size alone can explain. The analysis shows how different modes of regulation interact, align, or trade off across communities with varying governance styles.}
\label{fig3}
\end{figure}
\clearpage

\clearpage
\begin{figure}[p]
\centering
\includegraphics[width=\textwidth]{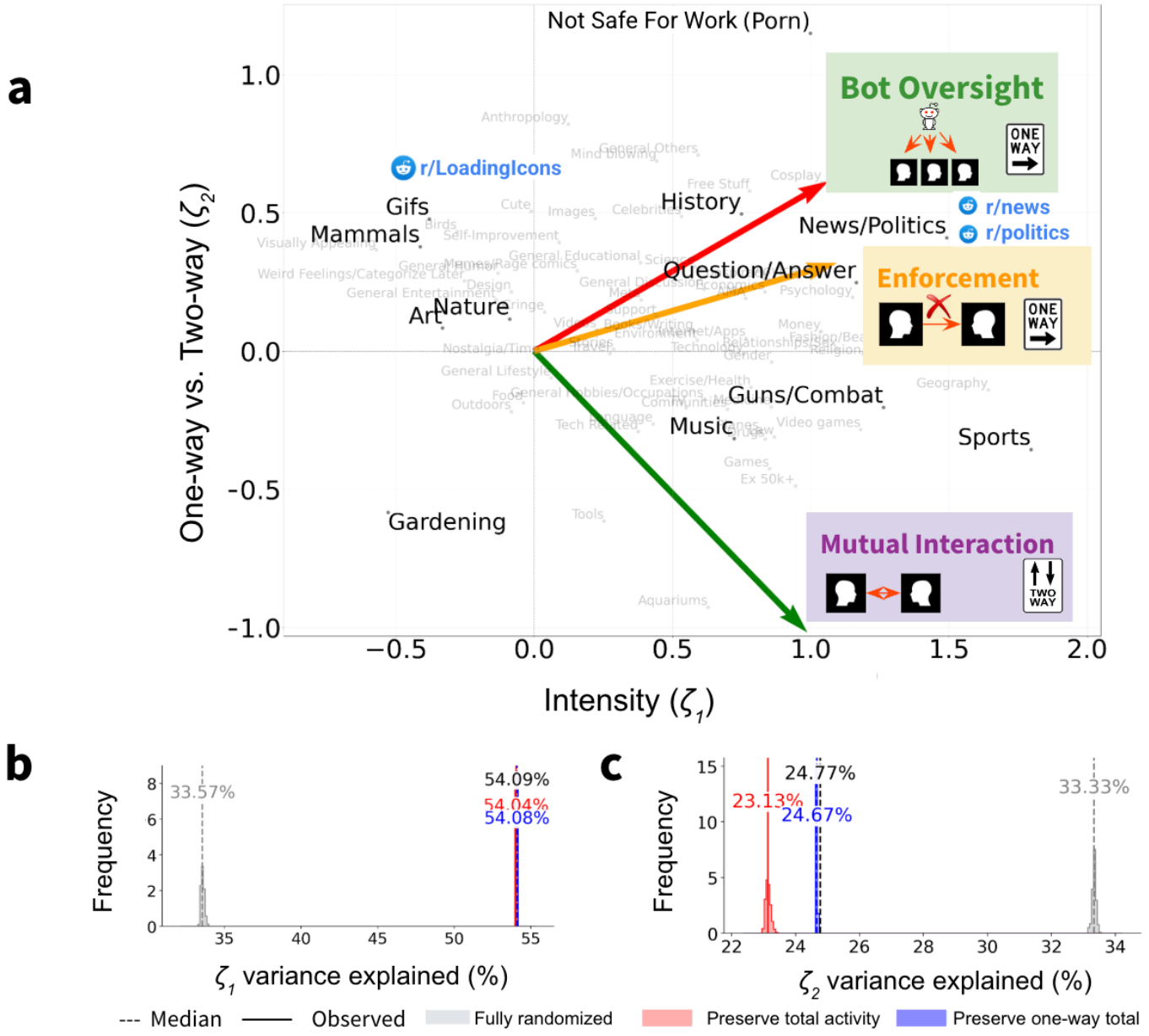}
\caption{\textbf{Regulatory mode space and robustness under constrained randomization.} \textbf{(a)} Projection of subreddits in the $(\zeta_{1},\zeta_{2})$ governance space based on scale-independent residual profiles $\vec{\xi}_i$ (Eq.~2–3). The horizontal axis $\zeta_1$ captures \textit{regulatory intensity}, indicating above- or below-baseline use of mutual interaction, enforcement, and bot oversight. The vertical axis $\zeta_2$ captures a \textit{one-way vs.\ two-way} coordination trade-off between one-way control (human enforcement and bot oversight) and conversational interaction. Colored loading vectors (Table~1) show contributions of each regulatory action. Contentious topics (e.g., \textit{News/Politics}, \textit{History}) cluster in the high-intensity, one-way region, whereas lower-conflict topics (e.g., \textit{Gardening}) occupy low-intensity, more conversational regions. Blue markers highlight illustrative communities (\textit{r/news}, \textit{r/politics}, \textit{r/LoadingIcons}). \textbf{(b)} Robustness of the intensity mode ($\zeta_1$): variance explained is compared to constrained randomizations preserving each community’s total regulatory activity while redistributing activity across action types. The observed variance ($\approx54\%$) aligns with this null, indicating balanced co-movement across regulatory functions. \textbf{(c)} Robustness of the one-way vs.\ two-way mode ($\zeta_2$): total one-way moderation is preserved while redistributing activity between human and bot actions. The observed variance ($\approx25\%$) matches this constrained null, confirming that the mode reflects the balance between one-way control and two-way interaction rather than technological substitution. Additional robustness tests are provided in SI Section~S5.}
\label{fig4}
\end{figure}
\clearpage

Our analysis reveals remarkably consistent scaling laws spanning nearly five orders of magnitude, but significant community-level heterogeneity still remains to be explained. As shown in Fig.~\ref{fig2}a-c, individual subreddits deviate from the fitted lines (yellow): some lie above (indicating excessive regulatory actions, such as r/news and r/politics) while others fall below (reduced regulatory activity, such as r/LoadingIcons). These deviations can be defined as the log-ratio of a subreddit $i$'s observed action volume $Y_{i,q}$ to the volume $\hat{Y}_q(N_i)$ predicted by the scaling model in Eq (1):

\begin{equation} \label{eq2}
\xi_{i,q} \equiv \log \frac{Y_{i,q}}{\hat{Y}_q(N_i)}
= \log \frac{Y_{i,q}}{\hat{Y}_{0,q} \, N_i^{\hat{\beta}_q}}.
\end{equation}

We then defined a scale-independent profile $\vec{\xi}$ by considering all three actions together.

\begin{equation} \label{eq3}
\vec{\xi}_i
= \big[\xi_{i,mutual},\xi_{i,supervision},\xi_{i,bot}\big].
\end{equation}

Fig.~\ref{fig3} maps each subreddit's standardized residuals $\vec{\xi}$ in a governance space defined by mutual interactions, enforcements, and bots' oversight. This representation allows systematic comparison across communities of different sizes, histories, and topics  (see Methods and SI Sec~5.2.2). For example, r/news ($\vec{\xi}_{news} = [0.1, 0.7, -0.6]$) and r/politics ($\vec{\xi}_{politics} = [0.6, 0.9, 0.7]$) both have positive residuals for mutual interaction and enforcement in their profiles (both above the baseline in Fig 2 a and b), but differ in automated oversight (one is above while the other is below the baseline in Fig 2 c). These cases illustrate r/politics relies heavily on bot-generated actions ($\xi_{politics, b} > 0$) while r/news does not ($\xi_{news, b} < 0$), indicating similar high-conflict communities can choose different technological paths. At the opposite end of the spectrum is r/LoadingIcons ($\vec{\xi}_{LoadingIcons} = [-1.3, -1.7, -0.2]$), characterized as a low-intensity governance where the inherent nature of the content requires minimal intervention. These individual cases suggest that the residuals are not random numbers; rather, they show empirically grounded modes of regulatory actions that communities adopt to manage their governance burden (see SI-Sec 5.4.2). This interpretation is consistent with latent mechanisms that shape collective behavior  \citep{stier2025bias}.

To understand these latent mechanisms of regulatory actions in our setting, we first analyze the covariance of the residuals across the entire platform. We find the most salient mode is the co-movement of all three regulatory actions, represented by the [1,1,1] direction in Fig. 3 (red arrows along the widest variance). Our Principal Component Analysis (PCA) confirms that the first principal component   $\vec{\zeta_1}$ aligns closely with this co-movement direction and explains the most variance (54\%).

All in all, these regulatory modes characterize how communities handle their governance burden beyond overall intensity and scale.  Given the strong covariance among regulatory actions (see SI-Sec S5, Table S3), we use these modes to infer organizational orientations, such as normative culture, topical dynamics, and emergent governance style, that manifest in the observed actions $q$ \cite{gelfand2018rulemakers, ostrom1990governing, weld2025perceptions}.

Table 1 summarizes the three modes by PCA results. The first mode, $\vec{\zeta_1}$, is indeed closely aligned with a systematic co-movement of regulatory actions; communities with high volume in one action tend to show high volume in the others. As a result, the primary differentiator between communities is the total intensity of their regulatory actions rather than the specific regulatory functions.  The second and third modes, detailed in Table 2, represent a community's organizational character once aggregate intensity is partitioned. With loadings [-0.8, 0.5, 0.3] over the action space  $\xi$ (represented by blue arrows in Fig. 3), the second mode $\vec{\zeta}_2$ exhibits opposing signs between mutual interaction and the remaining two functions. As a result, communities with positive $\zeta_2$ indicate a governance style of one-way control mechanisms, comprising both human enforcement and automated oversight, rather than two-way peer interactions. This axis identifies the structural trade-off between centralized coordination and horizontal discourse. On the other hand, the third mode ($\vec{\zeta}_3$) has loadings [-0.1, -0.6, 0.8], grouping mutual interaction and human enforcement against bot oversight. This mode therefore distinguishes between impersonal, rule-based algorithmic control and discretionary human enforcement. High values on this axis indicate a reliance on the deterministic efficiency of bots, whereas low values characterize prioritizing the nuanced, context-dependent judgment of human moderators (see SI Sec S5.4.1 and Fig. S16 for details).

\begin{table}[h!]
\begin{center}
 \small
\begin{tabular}{lccc}
\hline
& \textbf{Intensity $\vec{\zeta_1}$} & \textbf{One-way $\vec{\zeta_2}$} & \textbf{Impersonal $\vec{\zeta_3}$} \\
\hline
Mutual Int. & 0.543 & $-0.826$ & $-0.149$ \\
Enforcement & 0.585 & 0.500 & $-0.639$ \\
Bot Oversight & 0.602 & 0.260 & 0.755 \\
\hline
\label{tab:loadingmatrix}
\end{tabular}
\caption{Loading matrix for comparative governance profiles (between regulatory actions $\xi_q$ and regulatory modes $\zeta_q$): Each regulatory action  (mutual interaction, enforcement, and bot oversight) is differentially loaded to the regulatory modes  (intensity, one-way vs. two-way, and impersonal vs. personal). For example, the strong co-movement of all modes (Intensity) and the secondary trade-offs shown in Fig. 3 and Fig 4 a. The matrix is a transformation function between regulatory actions and modes. }
\end{center}
\end{table}

In Figure~4a, we map communities in regulatory mode space with regulatory intensity ($\zeta_1$) on the horizontal axis and the one-way versus two-way coordination mode ($\zeta_2$) on the vertical axis. Individual subreddits appear in gray, with selected examples marked in blue, and topic centroids shown in black (see Methods). This representation maps systematic differences across topical domains in how regulatory actions are combined within the shared governance structure.

Communities centered on contentious or norm-sensitive discussions—such as \textit{News/Politics} and \textit{History}—cluster in the upper-right quadrant ($\zeta_1>0$, $\zeta_2>0$), indicating high regulatory intensity combined with stronger reliance on one-way coordination through enforcement and automated oversight. Subreddits such as \textit{r/news} and \textit{r/politics} exhibit above-baseline levels across multiple regulatory channels, suggesting sustained governance demands in conflict-prone environments. In contrast, \textit{Sports} communities show high regulatory intensity ($\zeta_1>0$) but greater reliance on two-way conversational coordination relative to one-way control ($\zeta_2<0$). Finally, less contentious domains, such as \textit{Gardening}, appear at substantially lower values of $\zeta_1$, indicating lower overall regulatory intensity across all channels.

\clearpage
\begin{figure}[p]
\centering
\includegraphics[width=\textwidth]{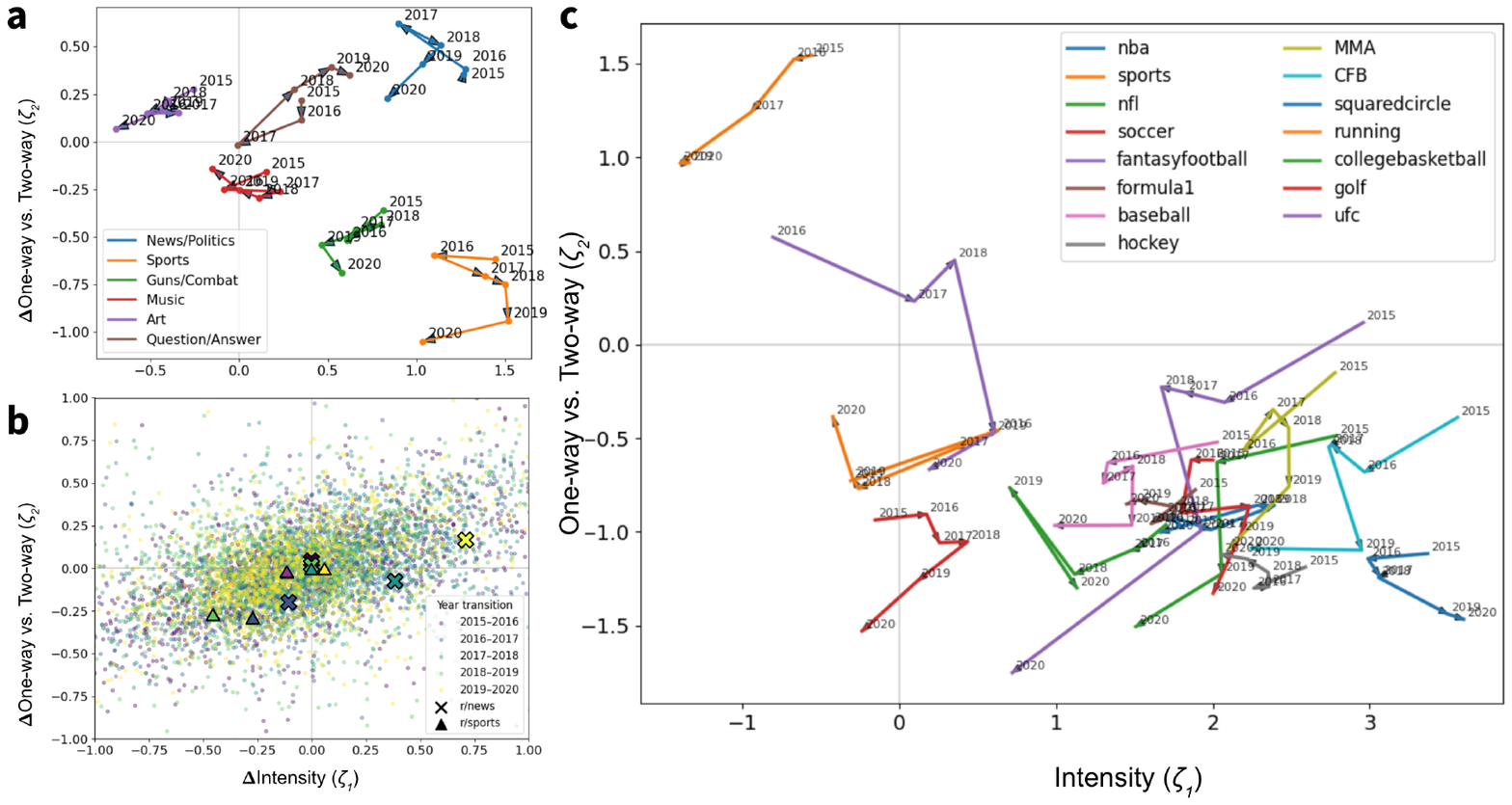}

\caption{\textbf{Temporal changes of individual communities' regulatory modes.}
\textbf{(a)} Yearly trajectories of topic category centroids in the $(\zeta_1,\zeta_2)$ regulatory mode space. Each point represents the mean position of subreddits within a topic category for a given year, and lines connect successive years. Categories occupy localized regions of the governance space and exhibit gradual within-domain movement over time. For example, sports and news/politics subreddits occupy quadrant 1 and quadrant 4, respectively, and remain there for five years, indicating their constant governance style.
\textbf{(b)} Distribution of year-to-year changes across all subreddits shown as $(\Delta \zeta_1,\Delta \zeta_2)$. Each point represents the change in regulatory intensity and coordination structure between two adjacent years for each subreddit. The overall positive covariance between $\Delta \zeta_1$ and $\Delta \zeta_2$ indicates that changes in intensity and one-way coordination generally move together. We indicate \textit{r/news} and \textit{r/sports} with markers.
\textbf{(c)} Individual subreddits' trajectories within-category of \textit{Sports} domain. Communities show heterogeneous paths, but movement remains local within the broader sports region rather than spanning the full regulatory mode space.}
\label{fig5}
\end{figure}
\clearpage

We test the structural robustness of the identified governance modes by comparing the variance explained by each mode with constrained randomized versions of the data (Fig.~4b–c). These tests evaluate whether the principal components reflect meaningful organizational structure rather than artifacts of the decomposition procedure itself.

For the first mode, regulatory intensity ($\vec{\zeta_1}$), all three regulatory actions co-move and load positively. To assess whether this dimension simply reflects uneven allocation across actions, we construct a null model that preserves each community’s total regulatory activity while randomly redistributing that activity across mutual interaction, human enforcement, and bot oversight. The observed variance explained by $\vec{\zeta_1}$ closely matches this constrained null distribution (Fig.~4b), indicating that the first mode captures balanced increases across all regulatory functions rather than patterns driven by any single channel.

For the second mode ($\vec{\zeta_2}$), which contrasts one-way control with two-way conversational coordination, we construct a second constrained null that preserves each community’s total one-way regulatory activity (human enforcement and automated oversight combined) while randomizing how that activity is distributed between human and bot actions. The observed variance explained again aligns with this constrained null (Fig.~4c), demonstrating that $\vec{\zeta_2}$ reflects the balance between total one-way control and two-way interaction rather than technological substitution between human and automated enforcement.

Together, these robustness tests show that the governance dimensions identified here are encoded in the empirical covariance structure of regulatory activity. They therefore represent meaningful organizational orientations adopted by communities rather than statistical artifacts of the principal component analysis method.

Methodological details are provided in Methods and Section~S5 of the Supplementary Information, including the standardization procedures used to construct comparable residual profiles across regulatory actions (Section~S5.2.2), geometric validation tests confirming that the principal components align with theoretically motivated reference directions (Section~S5.4.1, Fig.~S13), and additional constrained randomization tests demonstrating that the observed variance structure is robust to alternative redistributions of regulatory activity across action types (Section~S5.4.2, Fig.~S14).

Together with the scaling laws and governance modes identified above, these results provide a unified view of how online communities regulate as integrated systems. Rather than treating individual regulatory actions in isolation, the mode decomposition reveals a structured governance space in which different regulatory functions co-vary and combine into distinct organizational orientations. In this representation, communities are not defined solely by their scale or activity level, but by how they position themselves within a multidimensional system of regulatory intensity and coordination structure.

The analysis presented thus far is based on the pooled structure of communities observed across the platform over the 2015–2020 period, providing a cross-sectional view of how regulatory actions co-vary and organize into shared modes. This representation captures the dominant geometry of governance across communities, but does not yet address how individual communities evolve within this space over time. A natural next step is therefore to examine how communities move within this governance space as they grow and adapt.

\subsection{Temporal trajectories in governance space}

We follow communities year by year and trace their trajectories in the $(\zeta_1,\zeta_2)$ plane. This temporal perspective allows us to observe whether communities change primarily by shifting overall intensity, by adjusting coordination structure, or by moving along both dimensions together. Figure~5 shows these trajectories at multiple levels of aggregation. Two broad patterns emerge.

First, communities tend to occupy relatively stable regions of the governance space and exhibit gradual, local movement rather than large structural shifts. Category centroids such as \textit{News/Politics} and \textit{Sports} remain well separated across the entire period (Fig.~5a--b), and individual subreddits largely move within bounded regions associated with their topical domain (Fig.~5c). While trajectories show visible year-to-year fluctuation and heterogeneity across communities, most movement occurs as local circulation within a region rather than displacement across fundamentally different governance orientations. Some communities, such as \textit{r/fantasyfootball}, display more pronounced directional drift over time, whereas others, such as \textit{r/golf}, remain tightly clustered within a narrow range.

Overall, however, trajectories tend to fluctuate within a characteristic neighborhood, rarely crossing from a strongly one-way to a strongly two-way regime, or vice versa. Movement, therefore, appears locally structured rather than globally random: communities adjust their regulatory configurations within a limited neighborhood rather than traversing the full governance landscape. Second, Fig.~5d shows year-to-year adjustments in regulatory intensity and coordination structure are positively correlated, though they need not have been as shown no correlation (by construct) in Fig 3 C. This pattern of results holds across communities with different governance styles (Spearman's correlation $r_{\zeta_{1,>0},\zeta_{2,>0}} =0.44;\ p<0.001$, $r_{\zeta_{1,>0},\zeta_{2,<0}} =0.31;\ p<0.001$,
$r_{\zeta_{1,<0},\zeta_{2,>0}} =0.59;\ p<0.001$, $r_{\zeta_{1,<0},\zeta_{2,<0}} =0.54;\ p<0.001$).

Taken together, these temporal patterns complement the cross-sectional structure identified above. The governance modes inferred from covariance not only summarize differences across communities at a given moment, but also organize how communities adjust over time. Communities appear to move through this space through incremental adjustments rather than abrupt transitions between fundamentally different governance regimes. In particular, communities with different governance styles show similar patterns of incremental adjustments, regardless of their overall governance style. Additional analyses of temporal coupling, persistence, and trajectory structure are reported in SI Section~S6.

\section{Discussion and Conclusions}

Large collective systems need regulatory mechanisms that coordinate behavior, enforce norms, and sustain interactions. Yet regulation is often studied as isolated actions rather than as a system due to methodological and empirical convenience, such as event-level analysis of court rulings or banning a single user. Here, we examine regulation as a system-level property of online communities. Using scaling analysis, residual analysis, covariance analysis, and temporal analysis, we identify regulatory structures: scaling laws with size, regulatory modes, governance styles of communities, and their temporal evolution.

First, we find that interactions and enforcement scale superlinearly with community size. Larger communities generate disproportionately more bottom-up coordination through
discussion, while directly facing increasing top-down enforcement demands that outpace an individual moderator's capability. Automated moderation, by contrast, scales near-linearly with size, keeping pace with community growth without yet providing efficiency gains that would offset rising human enforcement demands. Furthermore, we show these relationships remain stable across years and samples, indicating that the findings are manifestations of a persistent organizational structure rather than transient fluctuations. These scaling exponents can be placed alongside similar patterns found in Wikipedia, GitHub, legal caselaw, and urban systems \cite{yoon2023individual, ash2024scaling, Fernandes2025USCaselaw, yang2026scaling}. While similar exponents across systems do not by themselves prove shared mechanisms, they give us a concrete basis for comparison.

Second, our residual analysis shows covariances among regulatory actions, resulting in regulatory modes. These modes include overall regulatory intensity, coordination structure, and the orientation toward automated (impersonal) versus human-centered (personal) forms of regulation, distinguishing communities that rely more heavily on bot oversight from those that rely more on direct human supervision. Together, these dimensions define a governance space in which communities occupy distinct regions corresponding to characteristic regulatory configurations. Communities differ not only in how much regulation they deploy but in how regulatory actions are combined. In other words, these systems operate not in terms of individual actions but in terms of modes. These modes are consistent with classic governance frameworks like Ostrom’s theory of self-governance in commons \cite{ostrom1990governing} and organizational theories of bureaucratic versus discretionary control, and provide quantitative evidence that online communities adopt distinct coordination and enforcement orientations \cite{chandrasekharan2017efficacy, bakcoleman2022case}.

Finally, we find that the temporal evolution of communities' regulatory modes, $\zeta(t)$, maintains their governance style, $\vec{\zeta(t)}$. These include category-level positions as well as individual communities within $\zeta$ space. Additionally, we find that changes in how much regulation communities deploy are systematically linked to how that regulation is structured. For example, we find that regulatory intensity is highly coupled with one-way coordination in its temporal evolution.

In summary, regulatory actions scale predictably with collective activity, organize into dominant modes given scale, and evolve through structured adjustments. The observed covariance among these residual actions suggests that regulation is not a series of isolated interventions but an integrated system given scale. In the end, governance manifests as a high-level epistemological orientation---such as top-down enforcement and bottom-up coordination---rather than a collection of individual, isolated tasks. As such, our work complements the existing literature by providing details on events and corresponding actions with system-level structure. Furthermore, our system-level framework allows the comparison of regulatory structures across disparate platforms and organizations through their scaling exponents, residuals, and modes \cite{yang2026scaling}. This is not a claim that Reddit, Wikipedia, and cities are the same kind of system — rather, it is an invitation: by measuring regulatory actions in each, extracting their scaling exponents, and decomposing their residuals into modes, we can ask whether comparable patterns — intensity, one-way versus two-way coordination, and impersonal versus personal moderation — emerge across them, and whether common generative processes underlie them.

We acknowledge that this analysis of regulatory actions does not account for many important aspects of regulatory systems: the success of communities' governance measures and formal rules. First, our framework, methods, and empirics do not assess whether members of these communities are happy about their governance. This is important but very difficult to handle, as notions of success are multifaceted. And yet, one can combine a survey of perceptions of moderators with large language models to evaluate success measures \cite{weld2025perceptions}.  Second, the formal rules and codified norms that significantly shape governance styles \cite{reddy2023evolution}. There are site-wide policies such as hate speech and doxing that all subreddits are expected to follow, but there are many specific rules and guidelines that differ by subreddit regarding submissions, comments and user behaviors \cite{fiesler2018reddit, reddy2023evolution}. For instance, while both r/news and r/politics mandate fresh, English-language articles with matching titles, they differ in paywalls, content scope, and user eligibility. r/news requires users to have a certain level of karma and age, whereas r/politics relies on an approved domains list to manage source reliability. In our study, we find r/politics relies more on bots' actions than r/news, that is, $\zeta_{bot}$(r/politics) $> \zeta_{bot}$ (r/news). It is unclear how this $\zeta$ space interacts with the difference in rule. Future research, therefore, should account for these formal rules, voting mechanisms, and moderator decision-making processes to examine whether comparable governance structures emerge in other bottom-up collective systems.

\section{Materials and Methods}

\subsection{Datasets: Pushshift, Modlog, and r/ListofSubreddits}

Our study uses three datasets: Pushshift, Mod-log, and  r/ListofSubreddits. First dataset is Pushshift, a public archive covering approximately 10,700 subreddits with more than 35 million comments and 700,000 moderator actions from 2015 to 2020 \cite{baumgartner2020pushshift}. The second dataset is r/ListofSubreddits that provides subreddit-level topical categories on a smaller subset of communities. Finally, we use the Mod-log dataset \cite{Juneja2020}, which includes moderators' actions in detail, for robustness checks on a smaller subset of communities. Our main findings remain consistent across all three sources.

First, Pushshift provides the most coverage of data records, where two entries are the most relevant to our study: submissions and comments. Submissions refer to original content posted to a subreddit, including links, text posts, images, or videos, serving as focal points for subsequent interactions. Comments are individual responses to submissions. Both are stored as monthly newline-delimited JSON files containing unique identifiers, author information, timestamps, subreddit metadata, textual content, and user feedback indicators (upvotes and downvotes).

For our analysis, we construct three measures of regulatory actions (Fig 1). First, mutual interaction (two-way coordination) is measured as the \# of comments associated with each submission, aggregated to the subreddit level $i$. This quantity indicates the volume of community engagement and peer-to-peer interaction. Second, we measure the \# of removed comments as records of enforcing the norms and regulations, identified by the text marker of \texttt{[removed]} in the selftext column of the dataset. This marker denotes content removed by moderators or administrators for rule violations \cite{chandrasekharan2018internet}. Third, we measure bot-generated comments as automated oversight. Following Jhaver et al.\ \cite{jhaver2019transparency}, we identify bot accounts using standardized naming conventions (e.g., usernames ending in “Bot,” “\_bot,” “--bot,” or “Modbot”), supplemented with curated lists of known Reddit bots, including AutoModerator \cite{redditbotlist}. Aggregating these bot comments at the subreddit level provides a proxy measure of automated rule enforcement.

To ensure that the \texttt{[removed]} flag provides a valid proxy for moderator intervention, we validate our supervision measure against an independent data source, Modelog. In Section~S5, we compare our estimates with public moderator logs collected by Juneja et al., which directly record moderator actions for 187 subreddits that opted into transparent logging.

Despite substantial differences in measurement approach (inferring removals rather than using direct action logs), temporal scope (a six-year aggregation versus a rolling three-month window in 2018), and coverage differences (50,690 subreddits versus 187 subreddits in Mod-log during the overlapped time window), the two measures are in good agreement. For example, on overlapping subreddits, removal counts are strongly correlated (\(r \approx 0.80\)), and the estimated scaling exponent from Mod-log (\(\beta = 1.174\), 95\% CI: [0.941, 1.406]) is consistent with our Pushshift-based estimate (\(\beta = 1.18\)). This consistency confirms that the observed superlinear scaling of supervision is not a measurement artifact (Fig.~S7).

In addition, we use a smaller sample of 2,828 communities, subreddit listed and curated in \textit{r/ListofSubreddits} for metadata \cite{redditlistofsubreddits}. The \textit{r/ListofSubreddits} community creates and maintains a public directory intended to help new users discover and navigate Reddit's ecosystem by listing and categorizing subreddits. The user-generated curation provides structured metadata for each community on the list, including topical classification across 88 categories (e.g., Science, Politics, Images; see Section~S6 and Table~S4 for the complete list). This metadata allows us to examine heterogeneity across communities and to interpret scaling residuals meaningfully.

To ensure our results are not driven by sample selection, we conduct extensive robustness checks using the full universe of 107,006 active subreddits meeting minimal activity thresholds in Pushshift (at least 10 unique users and 100 comments over 2015–2020).  Section~S4 reports these results in detail: estimated exponents of both the full universe and the curated sample ($\beta_{\text{comments}} = 1.117$ [95\% CI: 1.092, 1.142], $\beta_{\text{removals}} = 1.163$ [95\% CI: 1.117, 1.209], $\beta_{\text{bots}} = 0.984$ [95\% CI: 0.921, 1.048]), all of which is in a good agreement with the main results within confidence intervals overlapping substantially (Fig.~S5, Table~S2).   Together, these results confirm that the curated dataset does not introduce systematic bias into the estimated scaling relationships.

Note that we intentionally restrict our analysis to the post-2015 period because 2015 marks a major shift in Reddit's governance regime. Prior to 2015, moderation relied predominantly on manual effort by volunteer moderators with minimal interventions, consistent with Reddit's early laissez-faire philosophy. As the platform expanded, however, this approach proved unsustainable: unmoderated expression frequently escalated into harassment, abuse, and illegal activity, while the limited number of moderators had insufficient enforcement capacity.

In response, Reddit implemented major governance reforms in 2015. First, the platform adopted a formal Content Policy explicitly prohibiting ``attacks and harassment of individuals” \cite{robertson2015reddit} and second, officially introduced AutoModerator (Automod), a tool that enabled moderators to implement programmable, rule-based enforcement at scale \cite{chandrasekharan2017efficacy}. At the same time, moderators were granted expanded administrative authority and means to define and enforce community-specific rules.

Our empirical finding that confirms a pronounced discontinuity at this transition (see SI-Section~S1), and year-by-year scaling estimates verify the structural break in 2015 (See SI-Section~S3). For example, Fig. 2d shows a huge discontinuity in the average number of removed comments per subreddit that jumped from 0.026 (2014) to 197 (2015) within one year, while total comments showed no corresponding discontinuity (see SI Section~S1, Table~S1, and Fig.~S3 for more details). Fig 2 d and Fig.~S4 in SI also further confirm this regime change through year-by-year scaling analysis: exponents for removed comments jumped sharply at 2015 (from $\beta \approx 0.35$ to $\beta \approx 1.15$) and remained stable thereafter (see Section~S3 for details).

Thus, these changes marked Reddit's transition from largely manual moderation with little intervention, which could not scale sustainably, to a structured regulatory system capable of operating at platform scale. The latter state is what we are interested in for this study. Therefore, it is our deliberate choice to focus on the post-2015 period, where removals offer a more interpretable measure of realized enforcement under a more structured moderation regime --- one no longer bottlenecked by capacity constraints --- allowing us to ask how
enforcement, and regulatory actions more broadly, scale with community size.

\subsection{Scaling Analysis}

Scaling analysis explores how various quantities change as a system grows. In Reddit’s subreddits, where community sizes and moderation efforts vary greatly, this approach helps us understand how regulatory functions scale with expanding communities. By applying scaling analysis, we identified patterns and dynamics similar to those seen in urban systems and biological processes~\cite{bettencourt2007growth, koonin2006power}. In our analysis, we conceptualized each subreddit as a distinct organizational unit within the Reddit platform. This framework enabled us to examine how regulatory needs evolve with the number of users in each community.

To measure the size of each subreddit, we used the number of active unique contributors ($N$). For a deeper examination of the scaling dynamics, we applied the following power-law scaling model, $Y_{i,q} = Y_q^{0} N_i^{\beta_q}$. In this model, $Y$ represents the regulatory costs (such as moderation actions or bot interventions), and $Y_0$ is a normalization constant. The exponent $\beta$ quantifies how rapidly $Y$ increases relative to increases in $N$, allowing us to assess whether regulatory costs grow faster ($\beta > 1$) or slower ($\beta < 1$) than expected in a linear relationship.

Our data shows a distinct characteristic: most subreddits have a relatively small number of contributors, typically around 100 users, while a few subreddits attract significantly larger contributors. If we use ordinary least squares (OLS) regression for model fitting, the model would naturally focus on explaining the majority of the data. However, our research objective is to understand the growth patterns of subreddits, not just to model the predominant data.

To address this, we employed a specific approach for fitting the data to the scaling law, tailored to our dataset. Before model fitting, we applied logarithmic binning to the data, ensuring that every scale—from small to large subreddits—was given equal consideration. This approach allows us to capture growth patterns across the full range of subreddit sizes, providing a comprehensive understanding of how coordination costs scale within online communities.

\subsection{Residual Analysis}

Residuals capture deviations between observed and predicted regulatory activity. For each community $i$, we compute the scale-adjusted residual, $\xi_i =\log {Y_i}/{\hat{Y}_0 N_i^{\hat{\beta}}}$, where $Y_i$ is observed regulatory activity and $\hat{Y}_0 N_i^{\hat{\beta}}$ is the predicted value from the fitted scaling model. Residuals reveal systematic deviations from average scaling behavior. Section~S5 provides a detailed analysis of the residual structure. Correlations among standardized residuals are positive across regulatory actions, indicating shared regulatory demand while leaving room for coordination trade-offs revealed through PCA.

\subsection{Comparative Governance Style: Regulatory Modes}
To characterize variation in governance beyond scale, we analyze the joint structure of scale-adjusted residuals across regulatory actions. For each community $i$, we construct a three-dimensional residual profile $\vec{\xi}_i = [\xi_{i,\text{mutual}},\, \xi_{i,\text{enforcement}},\, \xi_{i,\text{bot}}]$, where each component measures the log-deviation of observed activity from the value predicted by the scaling baseline. These residual profiles capture how communities allocate regulatory effort across mutual interaction, human enforcement, and automated oversight relative to platform-wide expectations.

We then apply Principal Component Analysis (PCA)
\cite{pearson1901lines, hotelling1933analysis} to the correlation matrix of standardized size-adjusted residuals to identify dominant directions of variation in this three-dimensional governance space. Residuals are standardized to zero mean and unit variance prior to PCA, ensuring each regulatory action contributes equally regardless of scale (see SI Sec~S5.2.2 for details).

Rather than treating regulatory actions independently, this approach allows us to identify empirically grounded modes that summarize how regulatory functions co-vary across communities. The decomposition yields three orthogonal governance modes that together capture the full variance in the residual structure. The first mode ($\zeta_1$) explains 54.1\% of the variance and corresponds to \textit{regulatory intensity}: a balanced co-movement of all three regulatory actions. Communities with high $\zeta_1$ values exhibit above-baseline levels across mutual interaction, enforcement, and bot activity simultaneously, whereas those with low values show below-baseline levels across all channels. The second mode ($\zeta_2$), explaining 24.8\% of the variance, captures the primary coordination contrast between one-way control and two-way interaction. Positive values indicate greater reliance on one-way regulatory mechanisms (human enforcement and automated oversight) relative to conversational coordination, while negative values indicate greater reliance on mutual interaction. This mode, therefore, reflects how communities structure coordination once overall intensity is accounted for. The third mode ($\zeta_3$), explaining 21.1\% of the variance, distinguishes between impersonal and personal forms of regulation by contrasting automated oversight with discretionary human enforcement. Together, these three modes define a governance mode space in which each community occupies a position determined by its residual regulatory profile, that is, how communities combine mutual interaction, human enforcement, and automated oversight to maintain collective order. This interpretation is consistent with prior work showing that online communities vary systematically in their governance values and organizational orientations \cite{weld2022what, weld2025perceptions, chandrasekharan2018internet}.

We interpret these principal components as axes along which governance choices systematically co-vary. The labels we assign to each axis---regulatory intensity, one-way versus two-way coordination, and impersonal versus personal moderation---are theoretically
informed interpretations of the loading structure, grounded in existing governance frameworks. To assess whether the identified modes reflect meaningful organizational structure, we conduct geometric validation and constrained randomization tests (Section~S5). Geometric alignment tests confirm that the first mode closely follows the balanced co-movement direction across all regulatory actions, while the second and third modes align with theoretically interpretable contrasts between coordination mechanisms. We further evaluate robustness using constrained randomization analyses that preserve key structural properties of the data while disrupting cross-action covariance. A null model that destroys all correlations produces approximately equal variance across components, whereas preserving each community’s total regulatory intensity recovers the first mode. Additional constrained redistributions that preserve total one-way control while randomizing its composition recover the second mode. These tests demonstrate that the governance modes arise from structured covariance in the empirical data rather than from methodological artifacts of the decomposition method. Full methodological details and validation results are provided in Section~S5 of the Supplementary Information.

\section*{Data Availability}
All data and code underlying this study are publicly available via a GitHub repository archived on Zenodo at \url{https://doi.org/10.5281/zenodo.18651781}. The repository contains processed datasets, analysis scripts, and detailed instructions for reproducing all figures and results reported in this study.

\section*{Author Contributions}
H.Y. and J.Y. conceived the research. S.B. and J.Y. collected the data. S.B. conducted the analysis. S.B. and H.Y. wrote the manuscript. All authors edited and reviewed the manuscript.

\section*{Competing Interests}
The authors declare no competing interests.

\section*{Acknowledgments}
The authors would like to acknowledge the support of the National Science Foundation Grant Award Number 2133863 and the Emergent Political Economies Grant from the Omidyar Network. The authors thank Prerna Juneja for sharing the Modlog Dataset \cite{Juneja2020} and sharing valuable insights into moderator actions. The authors also extend their appreciation to Bruce Ankenman and Chris Kempes for their detailed suggestions and feedback on the manuscript. The authors thank the PhD committee consisting of Joshua Becker, Noshir Contractor, and Aaron Shaw for their suggestions. H.Y. and J.Y. acknowledge the Global Humanities and Social Sciences Convergence Research Program through the National Research Foundation of Korea (NRF), funded by the Ministry of Education (2024S1A5C3A02042671). H.Y.~acknowledges the support from the Institute of Management Research at Seoul National University.

\bibliographystyle{unsrtnat}
\bibliography{references}

\end{document}

\maketitle

\vspace{-0.5em}
\textbf{This PDF file includes:}
\begin{itemize}
  \item Supplementary Text
  \item Supplementary Figures S\textit{1}--S\textit{17}
  \item Supplementary Tables S\textit{1}--S\textit{4}
\end{itemize}

\newpage

\section*{Supplementary Text}

\section*{Overview}
\addcontentsline{toc}{section}{Overview}

This Supplementary Information provides detailed documentation supporting the main text's analysis of regulatory scaling in Reddit communities. Together, these sections establish that the scaling relationships reported in the main text are: (1) grounded in valid data from a stable regulatory regime, (2) temporally consistent across years, (3) robust to sample selection, (4) validated against independent measurement, and (5) accompanied by interpretable residual structure that reveals governance patterns beyond size effects.

\paragraph{Section~\ref{sec:S1}: Reddit's Pushshift Dataset and the 2015 Regulatory Transition.}
We describe our data source (the Pushshift Reddit archive), document how we operationalize the three measurable actions analyzed in the main text, and explain the sample construction from archive to analysis. We then present evidence of a structural break in 2015—when Reddit introduced scalable moderation tools that fundamentally changed what removal data measures. This section establishes why our scaling analysis focuses on the 2015--2020 period: pre-2015 removal counts reflect enforcement \emph{capacity}, while post-2015 counts offer a more interpretable measure of \emph{realized enforcement} under a more structured moderation regime.

\paragraph{Section~\ref{sec:S3}: Scaling Analysis for 2011--2020.}
We examine how scaling exponents evolved year by year across the full decade. This analysis provides independent confirmation of the 2015 regime change (exponents for removals jump sharply at 2015) and demonstrates that post-2015 exponents are temporally stable. The relationships reported in the main text therefore reflect durable organizational properties rather than transient fluctuations.

\paragraph{Section~\ref{sec:S4}: Scaling Analysis 2015--2020 (Combined).}
We present scaling results for the full universe of 107,006 active subreddits, demonstrating that exponents estimated from our labeled sample of 2,828 subreddits generalize to the broader platform. This establishes robustness to sample selection.

\paragraph{Section~\ref{sec:robustness}: Public Moderator Logs.}
We validate our Pushshift-based removal measure against an independent data source: public moderator logs that directly record moderation actions. This validation confirms that our proxy tracks actual moderation activity and that the superlinear scaling of human enforcement ($\beta > 1$) is not an artifact of our measurement approach.

\paragraph{Section~\ref{sec:pca}: PCA on Size-Adjusted Residuals.}
We conduct Principal Component Analysis on the residuals from our scaling regressions to examine how regulatory actions co-vary across communities \emph{beyond} size-predicted levels. This analysis reveals three interpretable governance dimensions—regulatory intensity, one-way vs.\ two-way coordination, and impersonal vs.\ personal moderation. We validate these dimensions through geometric alignment tests and constrained randomization analyses.

\paragraph{Section~\ref{sec:temporal_dynamics}: Temporal Dynamics in Governance Space.}
We extend the cross-sectional analysis (2015-2020) of the residuals by examining how communities move within the governance mode space over time. Tracking year-to-year changes in regulatory intensity and coordination structure reveals that communities exhibit gradual, locally bounded movements rather than large structural transitions, and that changes in intensity and coordination are positively coupled. These temporal patterns support the interpretation of governance modes as meaningful organizational orientations rather than purely descriptive statistical constructs.

\paragraph{Section~\ref{sec:S7}: Labeled Subreddit List.}
We provide the list of 2,828 subreddits across 88 content categories used in the main analysis, enabling replication and extension of our work.

\newpage

\section{Reddit's Data Landscape and the 2015 Regulatory Transition}
\label{sec:S1}

\subsection{Overview: Connecting Data to Claims}

The main text reports scaling exponents for Reddit's regulatory actions: Mutual Interaction ($\beta = 1.12$), Enforcement ($\beta = 1.18$), and Bot Oversight ($\beta = 0.95$). These estimates were derived from analyzing 2,828 labeled subreddits across 88 content categories during 2015–2020. This section provides the empirical foundation for those claims by:

\begin{enumerate}
    \item Describing the source data (Pushshift Reddit archive) and how we constructed our analysis samples.
    \item Demonstrating Reddit's growth patterns using our archived data to contextualize the scale of our analysis.
    \item Documenting the 2015 moderation regime change that justifies our focus on the post-2015 period.
    \item Showing that the 2015 discontinuity appears in removal \emph{rates}, not in underlying user behavior - establishing that post-2015 data offers a more interpretable measure of \emph{realized enforcement} under a more structured moderation regime, no longer bottlenecked by enforcement \emph{capacity} constraints.
\end{enumerate}

Throughout, we present figures generated from our Pushshift archive to verify that the patterns we analyze in the main text are grounded in the actual data we collected.

\subsection{Data Source: The Pushshift Reddit Archive}

As described in the main text, we use the Pushshift Reddit corpus \cite{baumgartner2020pushshift}, a publicly available archive of Reddit submissions and comments. Our local copy spans January 2011 through December 2020 and includes submissions and comments. \textit{Submissions} are original posts to subreddits (links, text posts, images, videos), with metadata including author, timestamp, subreddit, title, body text, and vote scores. \textit{Comments} are responses within discussion threads, with the same metadata plus parent identifiers preserving reply structure. A point to note is that comments removed by moderators display \texttt{[removed]} in the body field.

\paragraph{Operationalizing measurable actions.} From these raw records, we construct the three measures that we use throughout the main text to understand regulatory actions:

\begin{itemize}
    \item \textbf{Mutual Interaction} = Total comments per subreddit. As stated in the main text: ``In Reddit threads, replies are where users interpret posts, invoke norms (e.g., `rule 3,' off-topic warnings), provide reasons, and exert informal pressure.'' We aggregate comment counts to the subreddit level to capture the volume of peer-to-peer dialogue.

    \item \textbf{Enforcement} = Count of comments with body text equal to \texttt{[removed]}. The main text notes: ``Removals are observable enforcement events through which volunteer moderators apply subreddit rules and site policy.'' This provides a direct trace of human moderation activity.

    \item \textbf{Bot Oversight} = Count of comments from bot accounts. Following \cite{jhaver2019humanmachine}, we identify bots by usernames ending in ``Bot,'' ``\_bot,'' ``--bot,'' or ``Modbot,'' plus a curated list including AutoModerator. The main text describes these as ``moderation bots [that] operationalize codified rules at scale-posting notices, rate-limiting, filtering, and giving structured guidance.''
\end{itemize}

\paragraph{Community size.} We measure subreddit size ($N$) as the number of unique authors contributing at least one comment or submission during the observation period. This captures \emph{active participation} rather than passive subscription counts, ensuring a like-to-like mapping: we count the people whose actions generate the regulatory activity we measure.

\subsection{Sample Construction: From Archive to Analysis}

The main text analyzes ``2,828 labeled subreddits across 88 categories'' (\emph{Fig. 2 in the main text}). Here we clarify how this sample relates to the broader Reddit ecosystem and why it is appropriate for scaling analysis.

\paragraph{The full archive.} Table~\ref{tab:pushshift_stats} presents basic statistics for our archived Pushshift dataset, illustrating the growth in the number of subreddits and unique users over time. These figures represent \emph{all subreddits with recorded activity} in each year-not a sample, but the full observable Reddit ecosystem as captured by Pushshift. The table shows, for instance, that our archive contains 62,053 subreddits with activity in 2020, with an average of 3,777 unique active users per subreddit.

\paragraph{Active subreddit universe.} For scaling analysis, we require subreddits with sufficient activity to reliably estimate regulatory actions. Applying minimal activity thresholds (at least 10 unique authors and 100 comments over 2015–2020), we obtain a working universe of \textbf{107,006 subreddits}. Section~\ref{sec:S4} reports scaling results for this full universe, demonstrating that our main findings are not artifacts of sample selection.

\paragraph{Labeled analysis sample.} For interpretable residual analysis (identifying which \emph{types} of communities deviate from scaling predictions), we use a curated list of 2,828 subreddits organized into 88 topic categories, obtained from the r/ListOfSubreddits community wiki (see Section~\ref{sec:S7} for the complete list). This labeled sample:
\begin{itemize}
    \item Spans the same five orders of magnitude in size as the full universe (from $\sim$10$^2$  to $\sim$10$^7$ active users)
    \item Covers diverse content domains (news, science, gaming, support communities, etc.)
    \item Enables substantive interpretation of residual patterns (e.g., ``news/politics communities show elevated enforcement'')
\end{itemize}

Summary statistics for the labeled sample are reported in Table~\ref{tab:labeled_stats}.

\begin{table}[h!]
\centering
\begin{tabular}{lc}
\toprule
\textbf{Metric (Labeled sample: 2,828 subreddits)} & \textbf{2015--2020 average per subreddit} \\
\midrule
Number of subreddits & 2{,}828 \\
Average unique active users per subreddit & 1.67 million \\
Average no. of comments per subreddit & 1.64 million \\
Average no. of removed comments per subreddit & 52{,}657 \\
Average no. of posts per subreddit & 34{,}847 \\
Average no. of removed posts per subreddit & 551 \\
\bottomrule
\end{tabular}
\caption{\textbf{Summary statistics for the labeled subreddit sample (2015--2020).}
Values represent per-subreddit averages across the 2,828 categorized communities used in the main analysis .}
\label{tab:labeled_stats}
\end{table}

\paragraph{Why the labeled sample is sufficient.} Scaling analysis requires variation across the size range, not exhaustive coverage. Our labeled sample preserves the full dynamic range of Reddit community sizes while enabling content-based interpretation. Section~\ref{sec:S4} confirms that scaling exponents estimated from the labeled sample ($\beta_{\text{comments}} = 1.12$, $\beta_{\text{removals}} = 1.18$, $\beta_{\text{bots}} = 0.95$) match those from the full 107,006-subreddit universe ($\beta_{\text{comments}} = 1.117$, $\beta_{\text{removals}} = 1.163$, $\beta_{\text{bots}} = 0.984$), establishing robustness.

\begin{table}[tph]
    \centering
    \includegraphics[width=\textwidth]{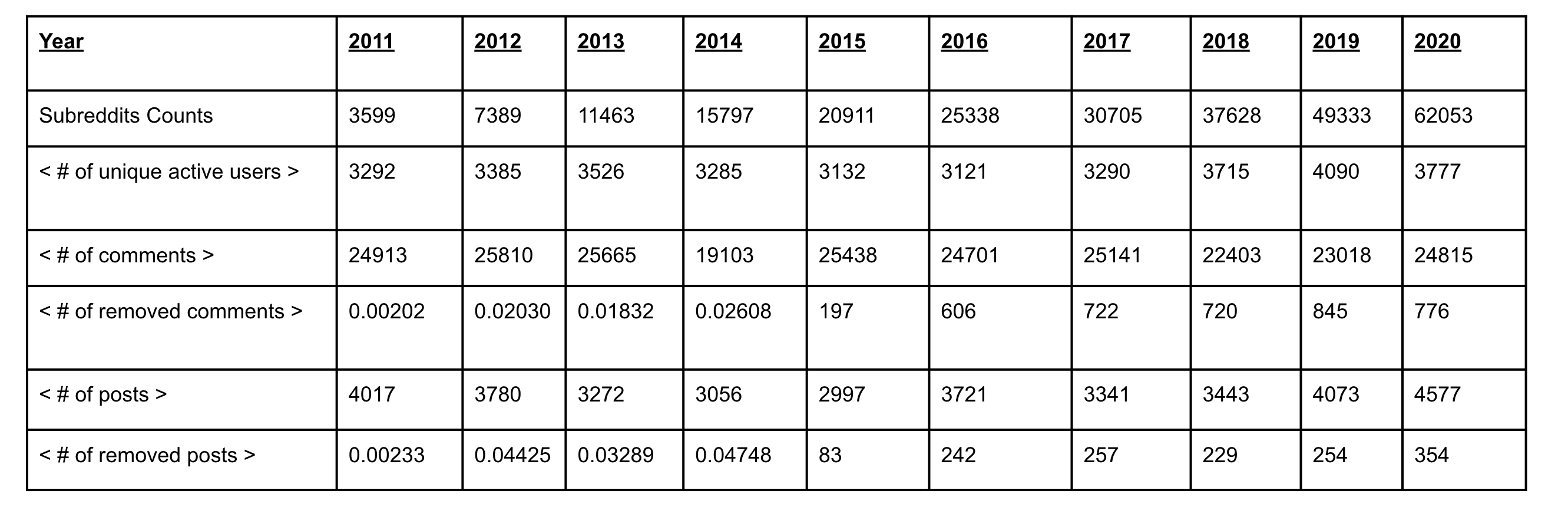}
    \caption{\textbf{Pushshift Dataset Statistics.} Each row reports the average value of the corresponding metric across all active subreddits in our archive for that year. The table illustrates the growth in the number of subreddits and unique users over time. Note the dramatic discontinuity in removed content beginning in 2015: average removed comments per subreddit jumped from 0.026 (2014) to 197 (2015)---a 7,500-fold increase; while total comments showed no corresponding change. This discontinuity, analyzed in Section~\ref{sec:2015_shift}, motivates our focus on post-2015 data.}
    \label{tab:pushshift_stats}
\end{table}

\subsection{Reddit's Growth Trajectory in Our Data}
\label{sec:growth_trajectory}

Before examining regulatory dynamics, we document how Reddit evolved during our observation period. The main text notes that Reddit hosts ``thousands of interest-based communities---each organized around a shared topic and varying widely in size.'' We show the growth trends from 2011 to 2020 in Figures \ref{fig:growth_subs_users} and \ref{fig:growth_posts_comments} across all of Reddit to provide a sense of the platform's scale and expansion.

\subsubsection{Platform Expansion: Supply vs.\ Density}

\begin{figure}[h!]
    \centering
    \includegraphics[width=\textwidth]{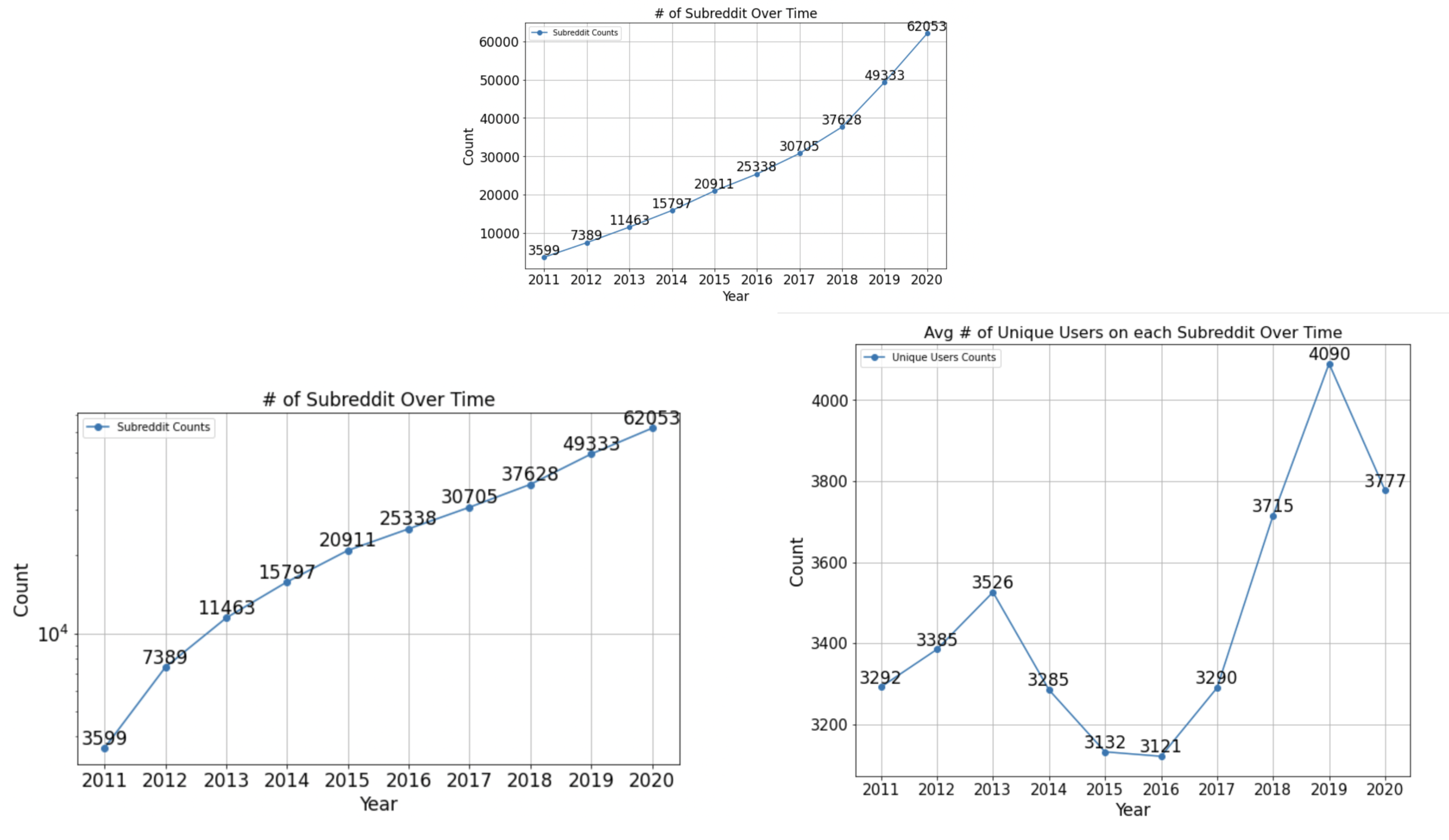}
    \caption{\textbf{Growth of total Subreddits and Users per Subreddit over time.} Data computed from our Pushshift archive. \emph{Left:} Active subreddits grew 17-fold---from 3,599 (2011) to 62,053 (2020). \emph{Right:} Despite this expansion, average unique users per subreddit remained stable at $\sim$3,100–4,100. Reddit grew by adding communities, not by inflating existing ones. This stability ensures that our cross-sectional scaling analysis is not confounded by temporal trends in what constitutes a ``typical'' subreddit.}
    \label{fig:growth_subs_users}
\end{figure}

Figure \ref{fig:growth_subs_users} reveals a striking pattern in our data: \emph{the supply of communities exploded while participation density remained stable.}

\paragraph{Subreddit proliferation.} The left panel (in Figure \ref{fig:growth_subs_users})shows that the number of active subreddits in our archive grew from 3,599 in 2011 to 62,053 in 2020---a 17-fold increase. This growth reflects Reddit's low barriers to community creation: any user can establish a new subreddit at no cost. The main text describes this as an ``open system: there are no barriers to joining, and anyone with a Reddit account may post to any subreddit.''

\paragraph{Stable community density.} The right panel (in Figure \ref{fig:growth_subs_users}) shows that \emph{average} unique active users per subreddit fluctuated narrowly between approximately 3,100 and 4,100 across the entire decade. Despite massive platform growth, the typical subreddit did not systematically grow larger. New users dispersed across an expanding ecosystem rather than concentrating in existing communities.

\paragraph{Interpretation.} Reddit's growth occurred through \emph{proliferation} (more communities serving more niches) rather than \emph{densification} (existing communities becoming larger). This has methodological implications: our cross-sectional scaling analysis compares communities of different sizes at similar points in Reddit's evolution, not communities observed at fundamentally different platform stages.

\subsubsection{Activity Levels: Comments and Posts}

\begin{figure}[h!]
    \centering
    \includegraphics[width=\textwidth]{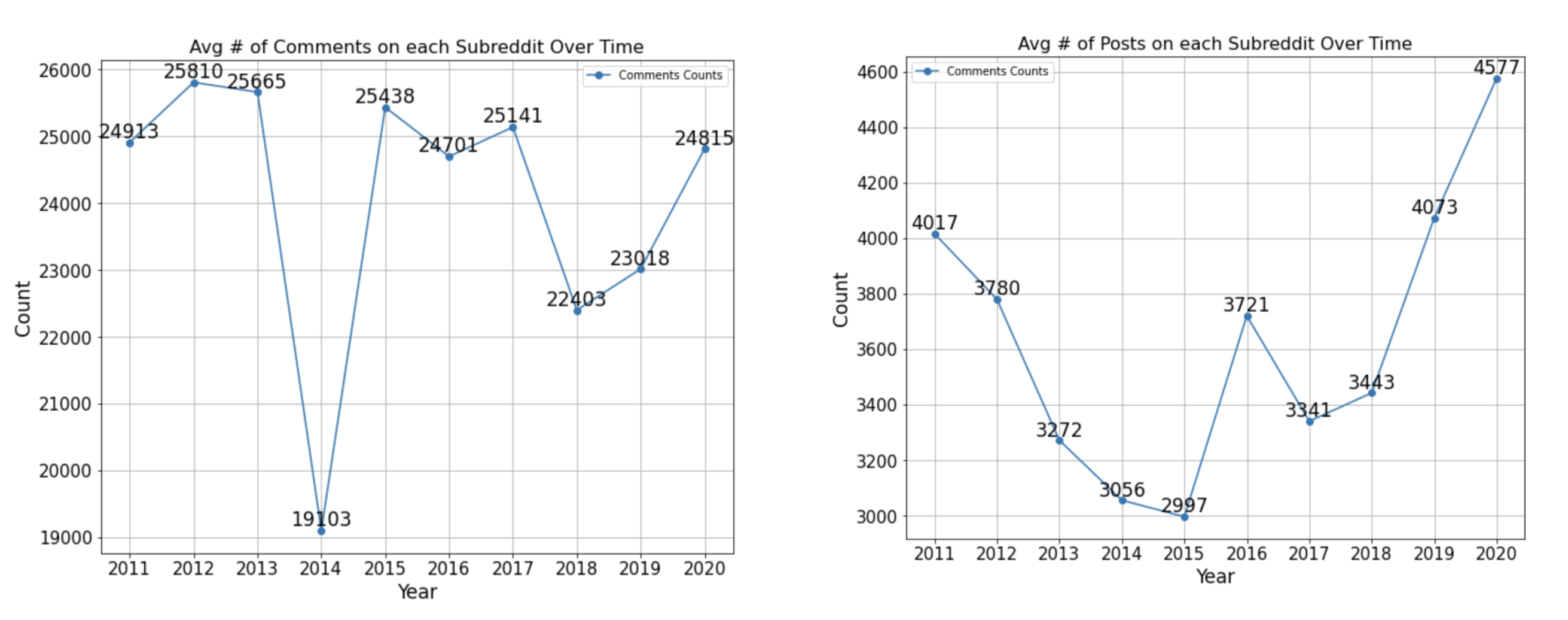}
    \caption{\textbf{Growth of average number of posts and comments over time.} Data computed from our Pushshift archive. \emph{Left:} Average comments per subreddit ($\sim$19,000–26,000) showed no systematic trend. \emph{Right:} Average posts per subreddit ($\sim$3,000–4,600) were similarly stable. Combined with Figure~\ref{fig:growth_subs_users}, this establishes that the ``typical subreddit'' maintained consistent characteristics throughout our observation period, supporting the validity of cross-sectional comparisons.}
    \label{fig:growth_posts_comments}
\end{figure}

Figure~\ref{fig:growth_posts_comments} extends this analysis to content production, showing that activity levels in our data were similarly stable.

\paragraph{Comments per subreddit.} The left panel (Figure~\ref{fig:growth_posts_comments} shows average comments per subreddit fluctuated between approximately 19,000 and 26,000 across the decade with no systematic trend. The dip in 2015–2016 and subsequent recovery does not represent a platform-wide behavioral shift but rather compositional effects as many small, low-activity subreddits entered our archive.

\paragraph{Posts per subreddit.} The right panel (Figure~\ref{fig:growth_posts_comments}) shows average posts per subreddit ranged from approximately 3,000 to 4,600, again with no clear trend. Content production rates were as stable as participation density.

\paragraph{Why stability matters.} If activity levels had systematically increased over time, our scaling analysis would face a confound: large subreddits (disproportionately observed later) might show different regulatory patterns simply because they existed during a different era of Reddit culture. The observed stability in our data indicates that size variation across communities reflects genuine organizational differences rather than temporal artifacts.

\subsection{The 2015 Moderation Regime Change}
\label{sec:S2}
\label{sec:2015_shift}

Table~\ref{tab:pushshift_stats} reveals two variables that \emph{did not} remain stable: removed comments and removed posts. In 2015, there was a sudden increase in the number of removed comments and posts, as shown in Figure~\ref{fig:2015_shift}. This coincided with a shift in moderation policy and the introduction of new moderation tools, such as the Automoderator \cite{reddit_automoderator_2015, jhaver2019humanmachine, chandrasekharan2017efficacy, Matias2016GoingDark, robertson2015reddit}. Interestingly, the total number of comments and posts did not show significant changes (Figure~\ref{fig:growth_posts_comments}), indicating that the policy shift primarily influenced removal actions. Prior to 2015, moderation processes struggled to scale effectively, but with the introduction of Automoderator, a stable moderation process was established. Therefore, we conduct our scaling analysis using data from post-2015. This section expands on why this discontinuity occurred and what it implies for our scaling analysis.

\subsubsection{Institutional Context: What Changed in 2015}

The main text notes: ``We concentrated on the period after 2015 for our final scaling analysis to better understand Reddit's more stable regulation system as post-2015 the platform adopted bot support and sitewide regulatory policies.'' Here we expand on this institutional context.

\paragraph{AutoModerator integration.} In 2015, Reddit officially integrated AutoModerator; previously a third-party bot-into the platform's core infrastructure \cite{reddit_automoderator_2015, jhaver2019humanmachine}. AutoModerator enables moderators to define rule-based triggers for automatic content removal, warnings and flagging. This transformed moderation from a purely manual, reactive process to a partially automated, proactive system. The main text describes this shift as ``Moderators can program these bots to scale individual regulatory actions, enabling them to effectively maintain rules specific to their communities.''

\paragraph{Site-wide Content Policy.} Reddit introduced its first formal Content Policy in 2015 \cite{robertson2015reddit}, replacing the platform's earlier hands-off approach with explicit prohibitions on harassment, violence, and illegal content. This gave moderators clearer enforcement grounds and created expectations for action.

\paragraph{Enhanced moderator tools.} Beyond AutoModerator, Reddit expanded moderator capabilities: improved queues, better logging, streamlined workflows, and subreddit-specific rule configuration. These tools reduced the per-action cost of moderation.

\subsubsection{Evidence in Our Data: Removals jumped, Activity did not}

\begin{figure}[h!]
    \centering
    \includegraphics[width=\textwidth]{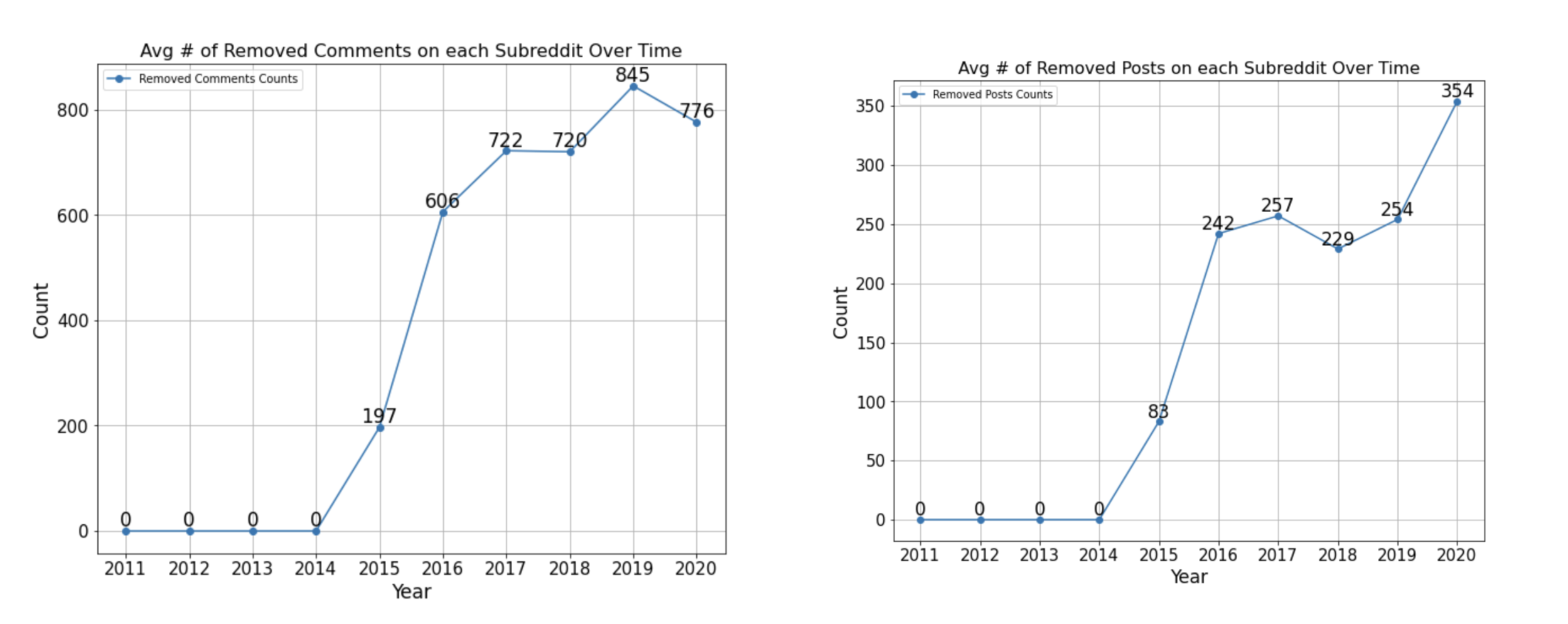}
    \caption{\textbf{2015 increase in removed posts and comments.} Data computed from our Pushshift archive. \emph{Left:} Average removed comments per subreddit were effectively zero before 2015 ($<0.03$), then jumped to 197 and continued rising to 606–845 in subsequent years. \emph{Right:} Removed posts show an identical pattern. This discontinuity does \emph{not} appear in total comments or posts (Figure~\ref{fig:growth_posts_comments}) meaning that users did not suddenly misbehave more. The discontinuity reflects lifted capacity constraints or how moderators gained tools allowing them to address violations that previously went undetected or unactioned.}
    \label{fig:2015_shift}
\end{figure}

Figure~\ref{fig:2015_shift} visualizes the discontinuity in our archive. The pattern is unambiguous:

\begin{itemize}
    \item \textbf{Removed comments}: Averaged 0.002–0.026 per subreddit from 2011–2014, then jumped to 197 in 2015 and continued rising from 606 to peaking at 845 for 2016–2020 (Figure~\ref{fig:2015_shift} left panel).

    \item \textbf{Removed posts}: Similarly near-zero number of posts pre-2015 (0.002–0.047), then jumped to 83 in 2015 and stabilized at around 229–354 for 2016–2020 (Figure~\ref{fig:2015_shift} right panel).

    \item \textbf{Total comments and posts}: No corresponding discontinuity (Figure~\ref{fig:growth_posts_comments}). Content production continued its stable trajectory through 2015.
\end{itemize}

\paragraph{Interpretation: Capacity constraints, not behavioral change.} If the 2015 jump in removals reflected a sudden increase in rule-violating behavior, we would expect a corresponding increase in total activity (more users posting more content, including problematic content). We observe no such increase in our data. Instead, the discontinuity has the signature of a \emph{capacity constraint being lifted}: rule violations were always occurring, but moderators lacked the tools to address them at scale. AutoModerator and enhanced tooling broke this bottleneck, and removals rose to more interpretably reflect realized enforcement under a more structured moderation regime.

This interpretation aligns with the main text's description of Reddit's evolution: ``Prior to 2015, moderation processes struggled to scale effectively, but with the introduction of Automoderator, a stable moderation process was established.''

\subsubsection{Implications for Scaling Analysis}

The 2015 discontinuity has direct consequences for interpreting scaling exponents:

\paragraph{Pre-2015: Removals measure capacity, not demand.} Before scalable tools existed, removal counts reflected what moderators \emph{could} remove given limited bandwidth---not what communities \emph{needed} removed. A small subreddit with one dedicated moderator might achieve high removal rates; a large subreddit with the same moderator team would fall behind, leaving violations unaddressed. Any scaling exponent estimated from pre-2015 data would conflate regulatory demand with moderator exhaustion.

\paragraph{Post-2015: Removals as a more interpretable measure of realized enforcement.} With AutoModerator handling routine violations and enhanced tools streamlining human review, post-2015 removal counts more interpretably reflect realized enforcement under a more structured moderation regime. The constraint shifted from ``what can moderators physically do'' to a setting where removal counts are no longer bottlenecked by the capacity constraints of the earlier, transitive phase.

\paragraph{Justification for 2015--2020 focus.} This is not arbitrary sample selection but a principled methodological choice. The scaling exponent $\beta = 1.18$ for enforcement reported in the main text  measures how realized enforcement under a more structured moderation regime grows with community size---a more interpretable quantity than what pre-2015 data could provide. An exponent estimated from pre-2015 data would measure something different and less interpretable: the interaction between latent enforcement activity and capacity constraints of the earlier, transitive phase.

Section~\ref{sec:S3} provides further evidence by showing how scaling exponents \emph{themselves} changed across the 2015 boundary---the exponent for removed comments jumped from $\sim$0.35 pre-2015 to $\sim$1.15 post-2015, confirming that the pre-2015 and post-2015 data generating processes differ fundamentally.

\subsection{Summary of section: Establishing the foundation for scaling analysis}

This section has established four foundational points that support the main text's scaling analysis:

\begin{enumerate}
    \item \textbf{Data source and operationalization.} We use the Pushshift Reddit archive (2011–2020) \cite{baumgartner2020pushshift} operationalizing Mutual  Interaction as comment counts, Enforcement as \texttt{[removed]} flags, and Bot Oversight as comments from identified bot accounts as described in the Materials and Methods section in the main text.

    \item \textbf{Sample construction.} Our labeled sample of 2,828 subreddits across 88 categories preserves the full five-order-of-magnitude size range present in Reddit, enabling interpretable analysis. Section~\ref{sec:S4} confirms that scaling exponents match those from the full 107,006 subreddit universe.

    \item \textbf{Temporal stability.} Platform growth occurred through community proliferation, not densification. Average community size and activity levels remained stable across the decade in our data, ensuring that cross-sectional size comparisons are not confounded by temporal trends.

    \item \textbf{The 2015 structural break.} A discontinuity in removal data was caused by the introduction of scalable moderation tools, not by changes in user behavior which means pre-2015 removals reflect enforcement capacity constraints while post-2015 removals offer a more interpretable measure of realized enforcement under a more structured moderation regime. Our focus on 2015–2020 ensures that scaling exponents capture meaningful organizational dynamics.
\end{enumerate}

With these foundations established, Section~\ref{sec:S3} examines whether scaling exponents are stable across individual years, testing whether the patterns in the main text represent durable properties of community governance rather than transient fluctuations.
\newpage

\section{Scaling Analysis for 2011--2020: All Subreddits}
\label{sec:S3}

\subsection{Overview: Testing Temporal Stability}

The main text reports scaling exponents estimated from 2015–2020 data: Mutual Interaction ($\beta = 1.12$), Enforcement ($\beta = 1.18$), and Bot oversight ($\beta = 0.95$). If these exponents reflect genuine organizational dynamics, that is, fundamental relationships between community size and regulatory demand, they should be stable properties, not transient fluctuations. This section tests that claim by examining how scaling exponents evolved year-by-year from 2011 to 2020.

We conducted a scaling analysis for all subreddits (entire active subreddits of Reddit) from 2011 to 2020. The jump in the exponent post-2015, as indicated by the arrows in the right panel of Figure~\ref{fig:scaling_yearly}, demonstrates the impact of the new moderation tools introduced that year. These plots illustrate how a more regulated Reddit emerged post-2015, due to the ability to scale moderation actions effectively. Additionally, we observe that the exponents post-2015 are consistently superlinear or sublinear (across all panels in Figure~\ref{fig:scaling_yearly}), highlighting the remarkable consistency in our main results. This consistency helps validate our findings.

The year-by-year analysis serves three purposes:
\begin{enumerate}
    \item \textbf{Validating the 2015 discontinuity}: Section~\ref{sec:S1} argued that pre-2015 removal data reflects capacity constraints rather than regulatory demand. If correct, scaling exponents for removals should show a discontinuity at 2015---and they do.
    \item \textbf{Confirming post-2015 stability}: The exponents reported in the main text should not be artifacts of a particular year. If the scaling relationships are stable, exponents should remain consistent from 2015 through 2020.
    \item \textbf{Establishing that size-scaling is a durable property}: Year-to-year consistency suggests that the relationships we observe reflect structural features of online community governance, not idiosyncratic patterns in our data.
\end{enumerate}

\subsection{Methods: Year-by-Year Scaling Estimation}

For each year from 2011 to 2020, we estimate the scaling relationship:
\[
Y = Y_0 \cdot N^{\beta}
\]
where $Y$ is the regulatory action (comments, removed comments or bot comments), $N$ is the number of unique active authors, and $\beta$ is the scaling exponent. We apply the same logarithmic binning procedure described in the Methods section of the main text to ensure that communities across all size ranges contribute equally to the estimate, preventing large communities from dominating the fit. This yields a time series of exponents $\beta_t$ for each regulatory action, allowing us to track how the size-scaling relationship evolved as Reddit grew and its governance infrastructure matured.

\subsection{Results: Temporal Evolution of Scaling Exponents}

\begin{figure}[h!]
    \centering
    \includegraphics[width=\textwidth]{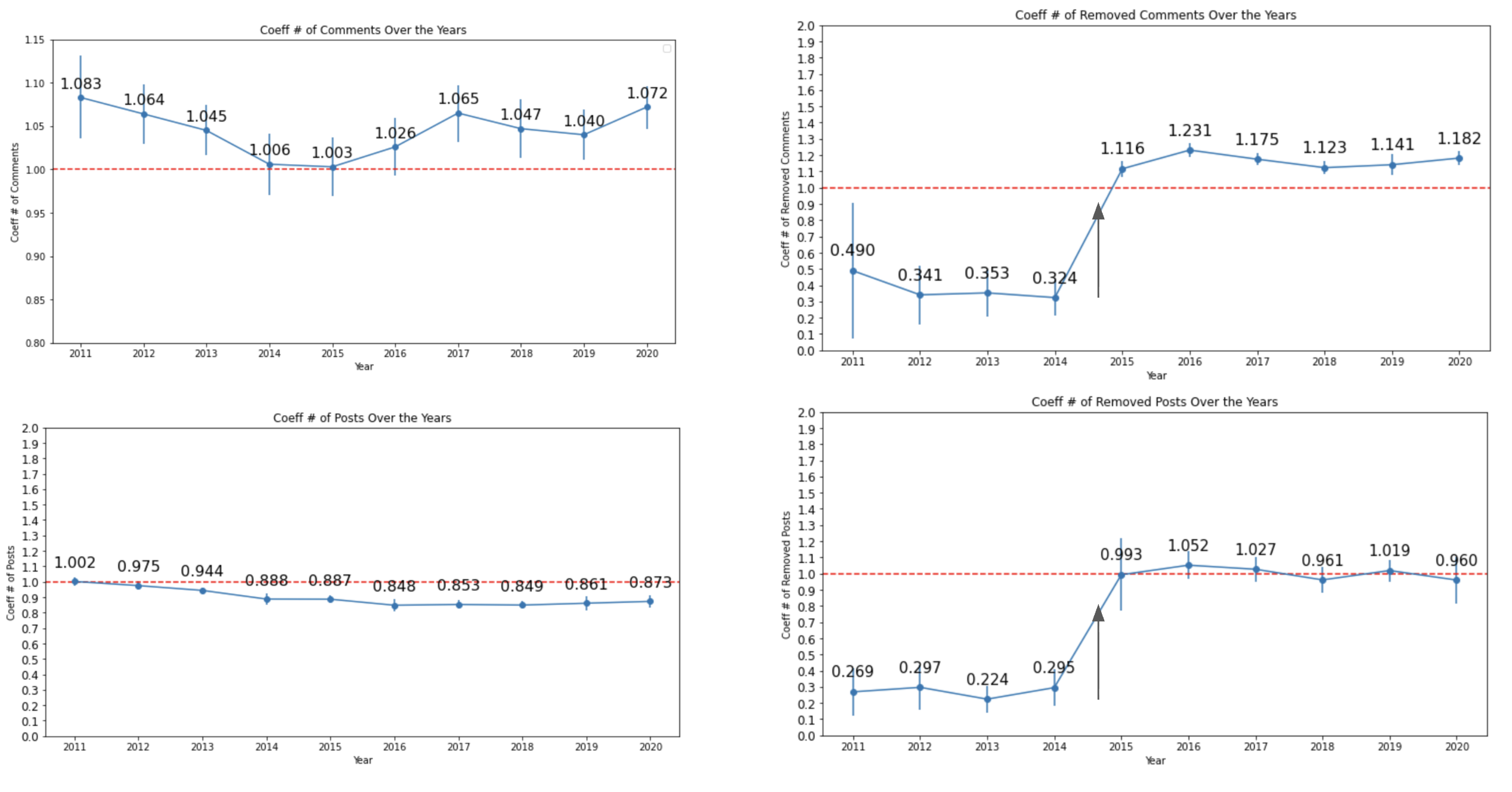}
    \caption{\textbf{Scaling coefficients 2011 to 2020.} Each panel shows the estimated scaling exponent ($\beta$) for each of our regulatory actions across all subreddits in our Pushshift archive. \emph{Top left:} Comments (Mutual Interaction) show stable superlinear scaling throughout ($\beta \approx 1.00$--$1.07$). \emph{Top right:} Removed comments (enforcement) show a dramatic jump at 2015 (arrows), from $\beta \approx 0.34$--$0.49$ to $\beta \approx 1.12$--$1.23$, confirming the regime change documented in Section~\ref{sec:S1}. \emph{Bottom left:} Posts show stable sublinear scaling ($\beta \approx 0.85$--$1.00$). \emph{Bottom right:} Removed posts mirror the pattern of removed comments, jumping from $\beta \approx 0.22$--$0.30$ to $\beta \approx 0.96$--$1.05$ at 2015. The post-2015 consistency validates the exponents reported in the main text.}
    \label{fig:scaling_yearly}
\end{figure}

Figure~\ref{fig:scaling_yearly} presents the complete results. Each panel shows the estimated scaling exponent for each of our regulatory actions across the ten-year period. We now interpret each panel in detail.

\subsubsection{Mutual Interaction (Comments): Stable Superlinear Scaling}

The top-left panel of Figure~\ref{fig:scaling_yearly} shows the scaling exponent for total comments over time.

\paragraph{Pre-2015 pattern.} From 2011 to 2014, the exponent fluctuated between approximately 1.00 and 1.06. Comments already showed (weakly) superlinear scaling: larger communities generated disproportionately more discussion than smaller ones, even before the moderation infrastructure matured.

\paragraph{Post-2015 pattern.} From 2015 to 2020, the exponent stabilized in the range 1.03--1.07, with remarkably little year-to-year variation. The main text reports $\beta = 1.12$ for the labeled sample; the slight difference reflects the broader universe analyzed here and the aggregation across years.

\paragraph{Interpretation.} Unlike removal data, comment counts were never capacity-constrained---users could always post comments regardless of moderator bandwidth. The stability of the comments exponent across the entire decade confirms that superlinear scaling of mutual interaction is an intrinsic property of Reddit communities, not an artifact of the 2015 policy changes. Larger communities generate disproportionately more peer-to-peer dialogue, consistent with the urban scaling literature where socioeconomic outputs scale superlinearly with city population \cite{bettencourt2007growth}.

\subsubsection{enforcement (Removed Comments): The 2015 Discontinuity}

The top-right panel shows the most striking pattern: a dramatic discontinuity in the scaling exponent at 2015.

\paragraph{Pre-2015 pattern.} From 2011 to 2014, the exponent for removed comments was severely depressed: $\beta \approx 0.34$ (2011), 0.35 (2012), 0.32 (2013), and 0.49 (2014). These values are far below linear scaling, suggesting that larger communities had \emph{fewer} removals per capita than smaller communities.

\paragraph{Why pre-2015 exponents were artificially low.} This counterintuitive pattern makes sense in light of Section~\ref{sec:S1}'s argument about capacity constraints. Before scalable moderation tools existed, removal counts reflected moderator bandwidth, not regulatory need. Small communities with dedicated moderators could maintain high removal rates; large communities with the same moderator-to-user ratio fell behind. The low exponents do not mean large communities needed less moderation, they mean large communities \emph{couldn't achieve} the moderation they needed.

\paragraph{Post-2015 pattern.} The exponent jumped to 1.12 in 2015 and remained stable thereafter: 1.18 (2016), 1.23 (2017), 1.18 (2018), 1.14 (2019), 1.18 (2020). The arrows in the right panels of Figure~\ref{fig:scaling_yearly} highlight this discontinuity. Post-2015 exponents are consistently superlinear, indicating that moderation demand grows faster than community size meaning larger communities require disproportionately more enforcement.

\paragraph{Interpretation.} The 2015 jump confirms that the introduction of AutoModerator and enhanced tools lifted a binding capacity constraint. Once moderators could scale their efforts, the observed removal counts began more interpretably reflecting realized enforcement under a more structured moderation regime. The post-2015 exponent of $\beta \approx 1.15$--$1.23$ represents the ``true'' scaling of enforcement with community size. This is the regime captured by the main text's estimate of $\beta = 1.18$ (Fig. 2).

\subsubsection{Content Production (Posts): Stable Sublinear Scaling}

The bottom-left panel shows scaling exponents for total posts (submissions).

\paragraph{Pattern across all years.} Post counts show stable sublinear scaling throughout the decade, with exponents ranging from approximately 0.85 to 1.00. Unlike comments, posts do not show superlinear scaling---larger communities do not generate disproportionately more submissions per capita.

\paragraph{Interpretation.} The distinction between comments (superlinear) and posts (sublinear to linear) reflects their different roles in Reddit's structure. Posts initiate discussions; comments sustain them. The superlinear scaling of comments suggests that discussions become proportionally more elaborate in larger communities---more replies, deeper threads, more back-and-forth conversations. Posts, as the raw material for discussion rather than the discussion itself, scale more modestly with size.

\subsubsection{Removed Posts: Mirrors the enforcement Pattern}

The bottom-right panel shows that removed posts exhibit the same discontinuity as removed comments.

\paragraph{Pre-2015 pattern.} Exponents were severely depressed: $\beta \approx 0.27$ (2011), 0.30 (2012), 0.22 (2013), 0.30 (2014). As with removed comments, this reflects capacity constraints rather than low regulatory demand.

\paragraph{Post-2015 pattern.} The exponent jumped to 0.99 in 2015, then fluctuated between 0.96 and 1.05 through 2020----approximately leading to linear scaling. This is notably lower than the superlinear scaling of removed comments ($\beta \approx 1.18$), suggesting that comment moderation demands scale more steeply with community size than post moderation.

\paragraph{Interpretation.} The difference between removed comments ($\beta \approx 1.18$) and removed posts ($\beta \approx 1.00$) is substantively meaningful. Comments, nested within threads, may escalate more readily into problematic territory (arguments, personal attacks) as communities grow and discussions become more contentious. Posts, as standalone submissions, may face more uniform moderation demands across community sizes. This justifies the main text's focus on comment-level regulatory actions.

\subsection{The 2015 Discontinuity in Exponents: Direct Evidence for the Regime Change}

The year-by-year analysis provides independent confirmation of the regime change documented in Section~\ref{sec:S1}. There, we observed that \emph{removal counts} jumped at 2015 while \emph{total activity} remained stable. Here, we observe that \emph{scaling exponents} for removals jumped at 2015 while exponents for total activity remained stable.

These two patterns tell the same story from different angles:
\begin{itemize}
    \item \textbf{Level change}: More content was removed post-2015 (Figure~\ref{fig:2015_shift})
    \item \textbf{Slope change}: Removals scaled more steeply with size post-2015 (Figure~\ref{fig:scaling_yearly})
\end{itemize}

Both are signatures of a capacity constraint being lifted. Pre-2015, moderators in large communities were overwhelmed meaning they removed less content (lower levels) and could not keep pace with growth (lower slopes). Post-2015, scalable tools lifted the moderator bandwidth constraint, allowing removal counts to more interpretably reflect realized
enforcement and revealing the underlying organizational relationship between community size and regulatory actions.

\subsection{Post-2015 Consistency: Validating Main Text Estimates}

The main text reports scaling exponents based on 2015–2020 data from 2,828 labeled subreddits. Figure~\ref{fig:scaling_yearly} shows that these estimates are robust in two senses:

\paragraph{Temporal stability.} The post-2015 exponents are remarkably consistent year-over-year:
\begin{itemize}
    \item Comments: 1.03--1.07 (6 years)
    \item Removed comments: 1.12--1.23 (6 years)
    \item Posts: 0.85--0.87 (6 years)
    \item Removed posts: 0.96--1.05 (6 years)
\end{itemize}
The narrow ranges indicate that scaling relationships are stable properties of Reddit's community structure, not fluctuations driven by particular events or platform changes.

\paragraph{Full-universe consistency.} The exponents in Figure~\ref{fig:scaling_yearly} are computed from \emph{all} subreddits in our archive, not just the 2,828 labeled communities. The consistency between full-universe and labeled-sample estimates (detailed in Section~\ref{sec:S4}) confirms that our main findings are not artifacts of sample selection.

\subsection{Summary: What temporal analysis establishes}

\begin{enumerate}
    \item \textbf{The 2015 discontinuity is real.} Scaling exponents for removals jumped sharply at 2015, confirming that pre-2015 and post-2015 data reflect different data-generating processes.

    \item \textbf{Pre-2015 removal exponents measure capacity, not demand.} The artificially low exponents ($\beta \approx 0.3$--$0.5$) reflect moderator bandwidth constraints, not the true relationship between size and regulatory need.

    \item \textbf{Post-2015 exponents are stable.} Year-to-year consistency from 2015--2020 indicates that the scaling relationships reported in the main text reflect durable organizational properties.

    \item \textbf{Activity exponents were always stable.} Comments and posts, which were never capacity-constrained, show consistent scaling throughout the decade, validating our measurement approach.
\end{enumerate}

Section~\ref{sec:S4} complements this temporal analysis by examining the \emph{combined} 2015--2020 estimates for the full subreddit universe, establishing robustness to sample definition.

\newpage
\section{Scaling Analysis 2015--2020 (Combined): All Subreddits}
\label{sec:S4}

\subsection{Overview: Robustness to Sample Selection}

The main text analyzes 2,828 labeled subreddits to enable interpretable residual analysis, helping in identifying which \emph{types or topics} of communities deviate from scaling predictions. But this raises a natural question: are the scaling exponents artifacts of this particular sample of 2,828 subreddits? This section addresses that concern by presenting results for the full universe of active subreddits in the period of 2015-2020.

These results below represent the combined exponents for all of Reddit, encompassing all subreddits. In the main paper, we provide detailed results for a subset of labeled subreddits to allow for more in-depth interpretation. As demonstrated in Figure~\ref{fig:scaling_combined}, the results for all of Reddit are consistent with those for our labeled subset, reinforcing the robustness of our findings.

\subsection{Sample Definition: The Full Active Universe}

For this analysis, we include all subreddits meeting minimal activity thresholds over the 2015--2020 period:
\begin{itemize}
    \item At least 10 unique active authors (contributors of at least one comment or submission)
    \item At least 100 total comments
\end{itemize}

These thresholds exclude dormant or trivially small communities where regulatory actions cannot be reliably measured. The resulting universe comprises \textbf{107,006 subreddits}-approximately 38 times larger than the labeled sample used in the main text.

The summary statistics for this full universe are presented in the table within Figure~\ref{fig:scaling_combined}:
\begin{itemize}
    \item 107,006 subreddits
    \item Average of 5,802 unique active users per subreddit
    \item Average of 50,954 comments per subreddit
    \item Average of 1,486 removed comments per subreddit
    \item Average of 8,521 posts per subreddit
    \item Average of 551 removed posts per subreddit
\end{itemize}

\subsection{Methods: Aggregated Scaling Estimation}

Rather than estimating year-by-year exponents as in Section~\ref{sec:S3}, here we aggregate all 2015--2020 activity for each subreddit and estimate a single scaling relationship. This approach:
\begin{itemize}
    \item Maximizes statistical power by pooling across years
    \item Provides estimates directly comparable to the main text's 2015--2020 analysis
    \item Smooths over year-to-year fluctuations to reveal the underlying relationship
\end{itemize}

We apply the same logarithmic binning and OLS regression procedure described in the main text Methods, fitting:
\[
\log_{10} Y = \log_{10} Y_0 + \beta \log_{10} N
\]
for each regulatory action $Y$ against community size $N$ (unique authors).

\subsection{Results: Full-Universe Scaling Exponents}

\begin{figure}[h!]
    \centering
    \includegraphics[width=\textwidth]{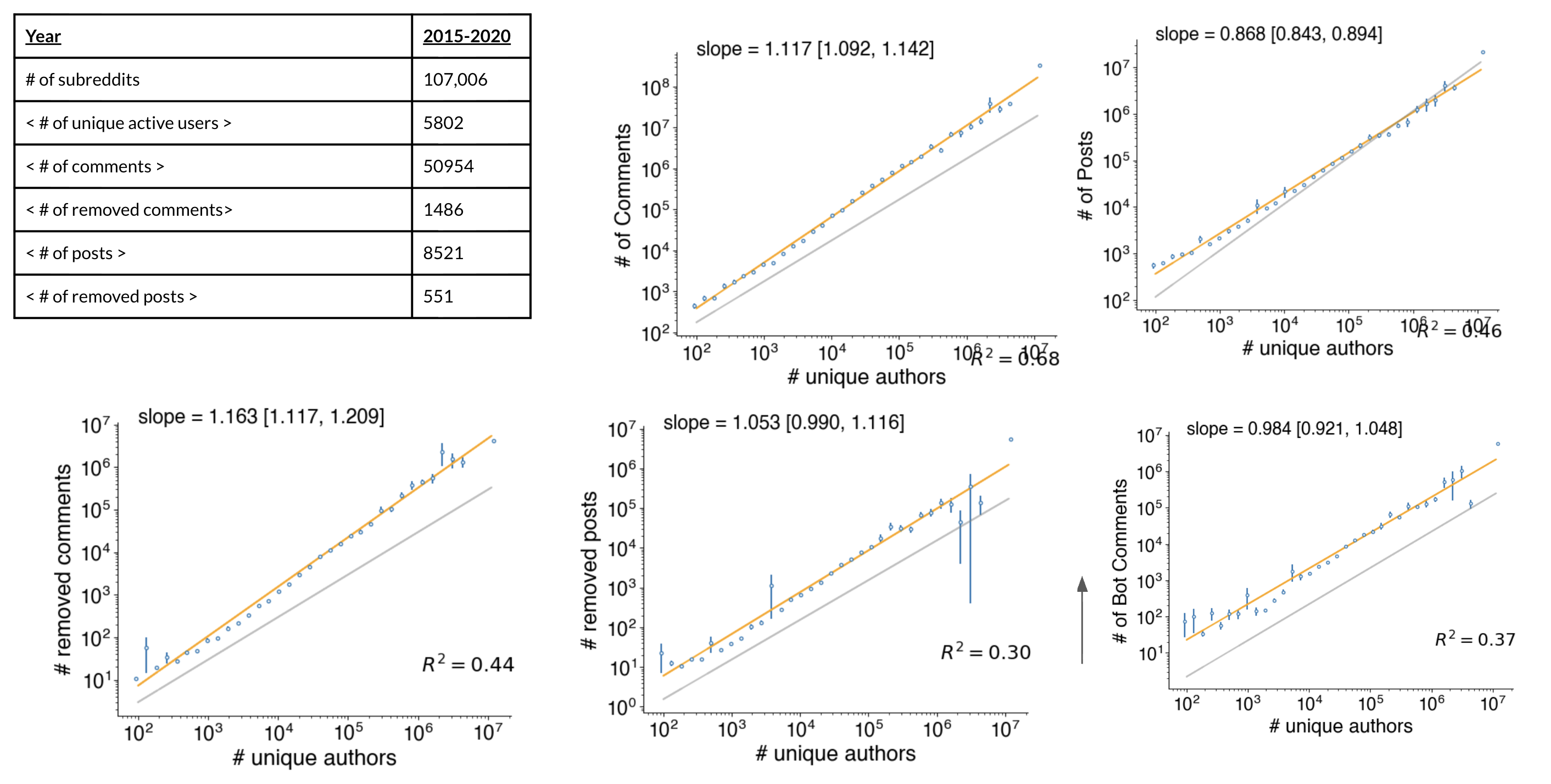}
    \caption{\textbf{Scaling coefficients 2015--2020 (combined).} Results for the full universe of 107,006 active subreddits in our Pushshift archive. Each panel shows the log-log relationship between a regulatory action and community size (unique authors). \emph{Top left:} Comments scale superlinearly with $\beta = 1.117$ [95\% CI: 1.092, 1.142]. \emph{Top right:} Posts scale sublinearly with $\beta = 0.868$ [95\% CI: 0.843, 0.894]. \emph{Bottom left:} Removed comments scale superlinearly with $\beta = 1.163$ [95\% CI: 1.117, 1.209], $R^2 = 0.44$. \emph{Bottom right:} Removed posts scale approximately linearly with $\beta = 1.053$ [95\% CI: 0.990, 1.116], $R^2 = 0.30$. \emph{Bottom center:} Bot comments scale approximately linearly with $\beta = 0.984$ [95\% CI: 0.921, 1.048], $R^2 = 0.37$. These estimates closely match those from the labeled sample in our main analysis, confirming robustness of our sampling.}
    \label{fig:scaling_combined}
\end{figure}

Figure~\ref{fig:scaling_combined} presents the scaling relationships for all 107,006 subreddits in the period 2015-2020 (cross-sectional). Each panel shows the log-log relationship between a regulatory action and community size, with the estimated exponent and 95\% confidence interval.

\subsubsection{Comments (Mutual Interaction)}

The top-left first panel in Figure~\ref{fig:scaling_combined} shows that comments scale superlinearly with community size:
\[
\beta_{\text{comments}} = 1.117 \quad [95\% \text{ CI: } 1.092, 1.142]
\]

This closely matches the main text estimate of $\beta = 1.12$ from the labeled sample. The tight confidence interval indicates a robust relationship: across five orders of magnitude in community size, from $\sim$10$^2$ to $\sim$10$^7$ users, larger communities generate disproportionately more peer-to-peer discussion.

\paragraph{Interpretation.} Superlinear scaling of Mutual Interaction suggests that larger communities do not simply have more conversations---they have more \emph{per capita} conversations. This could reflect network effects (more potential interaction partners), discussion dynamics (longer threads as debates escalate), or topic properties (larger communities tackling more contentious subjects). The consistency between labeled ($\beta = 1.12$) and full-universe ($\beta = 1.117$) estimates indicates that this is a universal property of Reddit communities, not specific to particular content domains.

\subsubsection{Posts}

The top-right panel in Figure~\ref{fig:scaling_combined} shows that posts scale sublinearly:
\[
\beta_{\text{posts}} = 0.868 \quad [95\% \text{ CI: } 0.843, 0.894]
\]

Larger communities generate fewer posts per capita than smaller ones. A community with 10$\times$ the users does not produce 10$\times$ the submissions-it produces approximately $10^{0.87} \approx 7.4\times$ the submissions.

\paragraph{Interpretation.} The contrast between superlinear comments and sublinear posts suggests different dynamics for initiating versus sustaining discussion. Posts require generating novel content; comments respond to existing content. The sublinear scaling of posts may reflect diminishing returns to content production as communities grow-not every user needs to post for the community to thrive, but many users engage in discussion.

\subsubsection{Removed Comments (Enforcement)}

The bottom-left panel in Figure~\ref{fig:scaling_combined} shows that removed comments scale superlinearly:
\[
\beta_{\text{removed comments}} = 1.163 \quad [95\% \text{ CI: } 1.117, 1.209], \quad R^2 = 0.44
\]

This closely matches the main text estimate of $\beta = 1.18$ from the labeled sample. The exponent significantly exceeds 1.0, indicating that moderation demand grows faster than community size---a dis-economy of scale for governance.

\paragraph{Interpretation.} Superlinear scaling of Enforcement means that moderator workload per user \emph{increases} as communities grow. A community with 10$\times$ the users requires approximately $10^{1.16} \approx 14.5\times$ the comment removals or that moderators must work proportionally harder in larger communities. This aligns with the urban scaling literature, where regulatory actions like legal sanctions scale superlinearly with city population \cite{bettencourt2007growth, youn2016scaling}.

\paragraph{Note on $R^2$.} The $R^2 = 0.44$ for removed comments is lower than for total comments, reflecting greater heterogeneity in moderation practices across communities. Some communities moderate heavily; others take a hands-off approach. This variation, analyzed via residuals in the main text and Section~\ref{sec:pca}, is substantively meaningful rather than problematic.

\subsubsection{Removed Posts}

The bottom-center panel in Figure~\ref{fig:scaling_combined} shows that removed posts scale approximately linearly:
\[
\beta_{\text{removed posts}} = 1.053 \quad [95\% \text{ CI: } 0.990, 1.116], \quad R^2 = 0.30
\]

The confidence interval includes 1.0, so we cannot reject linear scaling. This contrasts with removed comments ($\beta = 1.163$), which are clearly superlinear.

\paragraph{Interpretation.} The difference between post removal ($\beta \approx 1.0$) and comment removal ($\beta \approx 1.16$) suggests that comment moderation faces steeper scaling challenges. Comments, embedded in threaded discussions, may be more prone to escalation-arguments, attacks, tangents that require moderator intervention. Posts, as standalone submissions, may generate more uniform moderation demands across community sizes.

\subsubsection{Bot Comments (Bot Oversight)}

The bottom-right panel in Figure~\ref{fig:scaling_combined} shows that bot comments scale approximately linearly:
\[
\beta_{\text{bot comments}} = 0.984 \quad [95\% \text{ CI: } 0.921, 1.048], \quad R^2 = 0.37
\]

This matches the main text estimate of $\beta = 0.95$. The confidence interval spans 1.0, consistent with linear scaling.

\paragraph{Interpretation.} Near-linear scaling of bot oversight suggests that bots ``keep pace'' with community growth but do not disproportionately increase their activity. This contrasts with human moderation (superlinear), suggesting that bots and humans play different roles in the governance ecosystem. Bots handle imopersonal, routine, scalable tasks (posting rule reminders, flagging keywords); humans handle personal, nuanced, context-dependent judgments (evaluating borderline content, mediating disputes). The different scaling patterns quantify this division of labor.

\subsection{Comparison: Full Universe vs.\ Labeled Sample}

Table~\ref{tab:exponent_comparison} compares scaling exponents from the full universe (107,006 subreddits) to those from the labeled sample (2,828 subreddits) reported in the main text.

\begin{table}[h!]
\centering
\begin{tabular}{lccc}
\toprule
\textbf{Regulatory Action} & \textbf{Full Universe} & \textbf{Labeled Sample} & \textbf{Difference} \\
 & (107,006 subs) & (2,828 subs) & \\
\midrule
Comments (Mutual Interaction) & 1.117 [1.092, 1.142] & 1.12 [1.09, 1.14] & $<0.01$ \\
Removed Comments (Enforcement) & 1.163 [1.117, 1.209] & 1.18 [1.15, 1.21] & 0.02 \\
Bot Comments (Bot Oversight) & 0.984 [0.921, 1.048] & 0.95 [0.88, 1.02] & 0.03 \\
Posts & 0.868 [0.843, 0.894] & -- & -- \\
Removed Posts & 1.053 [0.990, 1.116] & -- & -- \\
\bottomrule
\end{tabular}
\caption{\textbf{Scaling exponent comparison: Full universe vs.\ labeled sample.} Estimates from the full universe of 107,006 subreddits closely match those from the 2,828 labeled subreddits analyzed in the main text. All confidence intervals overlap substantially. The main text's focus on labeled subreddits enables interpretable residual analysis without sacrificing generalizability; the scaling relationships are universal properties of Reddit communities.}
\label{tab:exponent_comparison}
\end{table}

\paragraph{Key finding: Estimates are nearly identical.} The largest difference is 0.03 (for bot comments), well within confidence intervals. The labeled sample, despite being 38$\times$ smaller than the full universe, accurately captures Reddit's scaling relationships.

To summarise this section, the main text's use of labeled subreddits is methodologically sound. The labeled sample enables content-based interpretation of residuals (e.g., ``news/politics communities show elevated enforcement'') without introducing bias into the scaling estimates. The patterns we observe are not artifacts of sample selection; they characterize Reddit as a whole.

\subsection{Variance Explained: What Scaling Laws Capture and What They Miss}

The $R^2$ values in Figure~\ref{fig:scaling_combined} indicate how much of the variation in regulatory actions is explained by community size alone:

\begin{itemize}
    \item Comments: High $R^2$ (typically $>0.6$), size strongly predicts discussion volume
    \item Removed comments: $R^2 = 0.44$, size explains about half the variation
    \item Removed posts: $R^2 = 0.30$, size explains less than a third
    \item Bot comments: $R^2 = 0.37$, size explains about a third
\end{itemize}

\paragraph{Interpretation.} Size is a powerful predictor of regulatory activity, but it is not the whole story. The unexplained variance that is captured by residuals reflects community-specific factors: topic contentiousness, moderator philosophies, rule complexity, user culture. Section~\ref{sec:pca} analyzes this residual variation systematically, revealing that it has interpretable structure rather than being random noise.

\paragraph{Connection to main text.} The main text's residual analysis (Figure 3, Figure 4) examines why some communities lie above or below the scaling line. The $R^2$ values here quantify how much variation those residuals represent: approximately half the variation in enforcement, and approximately two-thirds of the variation in bot oversight, reflects factors \emph{other} than community size.

\subsection{Summary: Robustness Established}

This section has established three key points:

\begin{enumerate}
    \item \textbf{Full-universe estimates match labeled-sample estimates.} Scaling exponents computed from 107,006 subreddits are nearly identical to those from 2,828 labeled subreddits. The main text's findings generalize to all of Reddit.

    \item \textbf{Scaling relationships are statistically robust.} Confidence intervals are tight, and the relationships span five orders of magnitude in community size. These are not fragile correlations but strong, consistent patterns.

    \item \textbf{Size explains substantial but not all variation.} Residuals from the scaling relationships ($R^2 \approx 0.3$--$0.5$ for moderation variables) represent meaningful community-specific variation that the main text and Section~\ref{sec:pca} analyze further.
\end{enumerate}

Together with Section~\ref{sec:S3}'s temporal analysis, these results establish that the scaling laws reported in the main text are both temporally stable (consistent across 2015--2020) and sample-robust (consistent across labeled and full-universe samples). The relationships reflect genuine organizational properties of Reddit communities, not methodological artifacts.

\newpage
\newpage

\section{Public Moderator Logs}
\label{sec:robustness}

\subsection{Overview: Why External Validation Matters}

The scaling analysis in the main text and Sections~\ref{sec:S3}--\ref{sec:S4} relies on a specific operationalization of enforcement: counting comments whose body text equals \texttt{[removed]} in the Pushshift archive. This proxy is intuitive as removed comments are a direct trace of moderator enforcement, but it is derived from a single data source using a single detection method. How confident can we be that this measure captures actual moderation activity?

This section addresses that concern by validating our removal proxy against an independent data source: public moderator logs collected by Juneja et al.~\cite{Juneja2020}. These logs provide a direct record of moderator actions, recorded by a bot with read access to moderation queues. If our Pushshift-based proxy accurately measures enforcement intensity, it should:
\begin{enumerate}
    \item \textbf{Align in levels}: Subreddits with high removal counts in our data should also show high removal counts in the mod-log.
    \item \textbf{Align in scaling}: The relationship between removals and community size should exhibit the same superlinear exponent ($\beta > 1$) in both datasets.
\end{enumerate}

We find strong support for both predictions, establishing that our removal measure tracks genuine moderation activity and that the superlinear scaling of enforcement is not an artifact of our measurement approach.

The validation proceeds in four steps. First, we describe how the two datasets are collected (Section~\ref{sec:modlog_collection}). Second, we examine what the mod-log contains and confirm that comment removal is the dominant moderation action (Section~\ref{sec:modlog_contents}). Third, we test whether our removal counts align with mod-log counts on overlapping subreddits (Section~\ref{sec:modlog_alignment}). Fourth, we estimate scaling exponents from the mod-log and compare them to our Pushshift-based estimates (Section~\ref{sec:modlog_scaling}).

\subsection{How the Two Datasets Are Collected}
\label{sec:modlog_collection}

We validate our \textbf{removed comments} proxy against \textbf{public moderator logs} from Juneja et al.~\cite{Juneja2020}. We ask: (i) what does the mod-log actually contain (is \texttt{removecomment} dominant and stable?), (ii) do our cross-sectional ``removals'' \emph{levels/rankings} align on the overlap set, and (iii) is the \textbf{scaling exponent} $\beta$ (removals vs.\ size) consistent across sources.

\paragraph{Public moderator logs.}
Juneja et al.~\cite{Juneja2020} collected moderation logs from subreddits that \emph{opt in} by inviting a read-only bot (\texttt{u/publicmodlogs}). The bot exposes a rolling 3-month feed of moderator actions (e.g., \texttt{removecomment}, \texttt{removelink}) with timestamps and target IDs. We use their dataset covering Mar--Sep 2018 for 187 subreddits (overlap with our sample: 143). This provides a direct record of actions, but only for subreddits that opted in and only for the most recent months.

\paragraph{Our dataset (Pushshift comments).}
We use the Pushshift Reddit corpus and treat a comment as removed when its stored body equals \texttt{[removed]} (a comment-level signal). We aggregate by subreddit over 2015--2020. For this validation analysis, we use a subset of 50{,}690 subreddits that have non-zero removed comments, and measure size by unique authors (at least one contribution). This yields broad coverage and a consistent \emph{comment-removal} measure that we use as enforcement intensity.

\paragraph{Key differences between sources.} The two datasets differ in coverage, timing, and measurement:
\begin{itemize}
    \item \textbf{Coverage}: Our Pushshift data covers 50,690 subreddits; the Juneja et al.\ ~\cite{Juneja2020} dataset covers 187 subreddits that opted into public logging.
    \item \textbf{Timing}: Our data aggregates 2015--2020; the mod-log provides a rolling 3-month window in 2018.
    \item \textbf{Measurement}: We infer removal from the \texttt{[removed]} flag in comment bodies; the mod-log directly records moderator actions.
\end{itemize}
These differences make the validation conservative: if our proxy aligns with mod-log data despite different coverage, timing, and measurement approaches, we can be confident it captures the underlying phenomenon.

\subsection{What Does the Mod-Log Measure? Is \texttt{removecomment} the Main enforcement Unit?}
\label{sec:modlog_contents}

\paragraph{What the public mod-log contains.}
The public mod-log is a stream of moderator \emph{actions} with time, moderator, and the moderated target. In the Juneja et al.\ dataset we observe three action labels: \texttt{removecomment}, \texttt{removelink}, and \texttt{unignorereports}.

\paragraph{Like-to-like with our proxy.}
Our measure comes from Pushshift comment records and flags a removal when the stored body equals \texttt{[removed]}. This is \emph{comment-level enforcement}, so the appropriate mod-log comparator is \texttt{removecomment} (not \texttt{removelink}, which is post-level majorly but also comment level if the comment is just a link). We still report link removals for completeness, but apples-to-apples comparisons use \texttt{removecomment}.

\medskip
\noindent\begin{minipage}{\linewidth}
\centering
\small
\begin{tabular}{lrr}
\toprule
Action & Count & Share of all \\
\midrule
\texttt{removecomment}   & 376{,}055 & 0.787 \\
\texttt{removelink}      & 101{,}528 & 0.212 \\
\texttt{unignorereports} & 498       & 0.001 \\
\bottomrule
\end{tabular}
\end{minipage}

\medskip
\noindent
\textbf{Dominance of removecomment as a enforcement metric.} We examine whether \texttt{removecomment} is the dominant action both over time and across community sizes.

\begin{figure}[t]
  \centering
  \includegraphics[width=.95\linewidth]{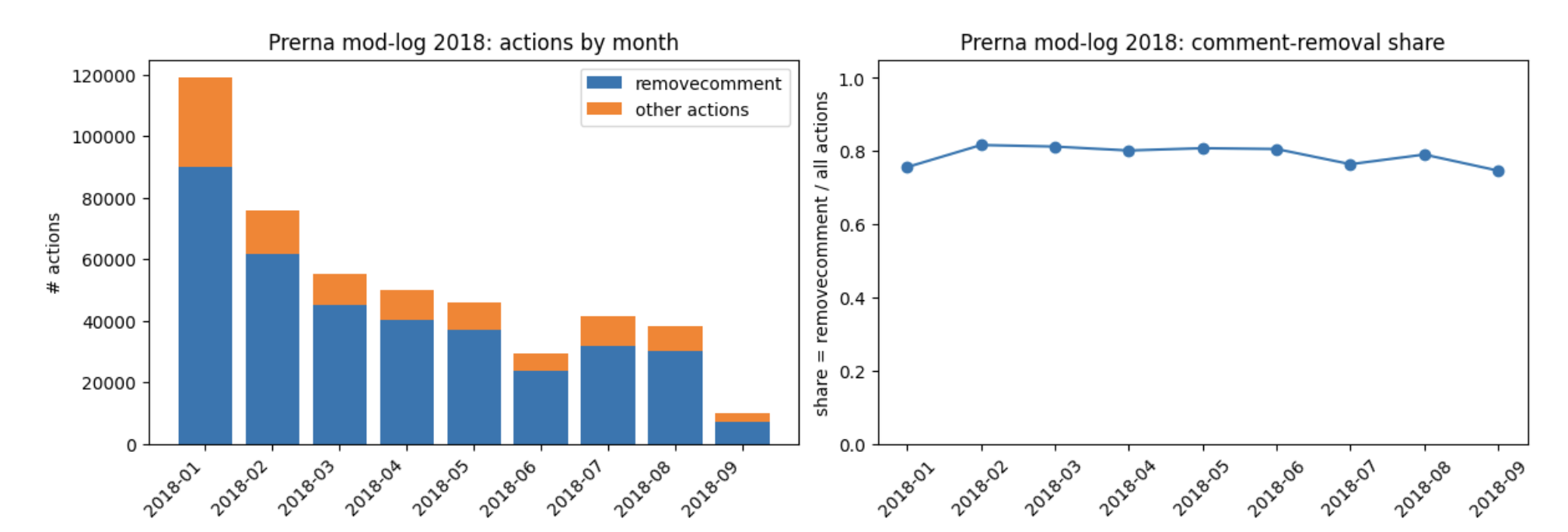}
  \caption{\textbf{Public mod-log (2018): comment removal is the dominant action and stable over time.}
  \emph{Left:} monthly counts, stacked as \texttt{removecomment} vs.\ all other actions.
  \emph{Right:} monthly share $\texttt{removecomment}/\text{all actions}$, which stays in the $\sim$0.75--0.82 band despite the rolling 3-month window. Data from Juneja et al.~\cite{Juneja2020}.}
  \label{fig:si_modlog_monthly}
\end{figure}

\emph{Stability across time.} By month in 2018, the \texttt{removecomment} share is stable at \textbf{$\sim$0.75--0.82} (Fig.~\ref{fig:si_modlog_monthly} right panel), showing no systematic drift despite the rolling 3-month observation window. This consistency is notable: the absolute volume of moderation activity varies substantially month-to-month (from $\sim$20,000 to $\sim$120,000 actions), yet the \emph{composition} of that activity remains constant. Whether moderation demand is high or low, approximately four out of five enforcement actions (80\%) are comment removals.

\emph{Stability across size.} Across community size bins, \texttt{removecomment} remains the majority action throughout (57--87\%), though the median share is lower in small communities ($\sim$0.48) than in larger ones ($\sim$0.71). This difference reflects sample composition: small subreddits in the opt-in Juneja et al.~\cite{Juneja2020} dataset are disproportionately cryptocurrency and link-sharing communities (e.g., \texttt{pythoncoding}, \texttt{CardanoCoin}) where spam-filtering posts is the primary moderation task, and such communities have particular incentive to opt into public logging to demonstrate transparent moderation.

\emph{Implications for measurement.} The presence of \texttt{removelink}-heavy communities highlights that comment removal, while dominant overall, does not capture the full spectrum of moderation activity. Ideally, a comprehensive measure of regulatory intensity would include post removals, bans, and other enforcement actions. Nevertheless, \texttt{removecomment} accounts for $\sim$80\% of logged actions in this sample, and the Pushshift archive provides reliable detection only for comment-level removals. We therefore treat removed comments as a \emph{lower bound} on total enforcement---one that is consistently measurable across 50,690 communities and, as Section~\ref{sec:modlog_scaling} shows, produces scaling estimates consistent with direct mod-log measurement.

\emph{Conclusion:} Comment removals are the dominant unit of enforcement in the mod-log, stable over time, and the apparent size-dependence reflects community-type sampling in the opt-in data rather than a fundamental difference in how small vs.\ large communities moderate.

\paragraph{Interpretation.} The mod-log data confirm that comment removal is not just one of many moderation tools; it is the primary enforcement mechanism. This validates our focus on comment removals as the measure of enforcement in the main text. The stability of this share across months, and the explainable variation across community types, suggests that comment removal serves a consistent function across the Reddit ecosystem.

\subsection{Do Our Removal Data Align with the Mod-Log on the Overlap Set?}
\label{sec:modlog_alignment}

Despite different coverage windows (3 months vs.\ all years), the \textbf{cross-sectional ranking} is similar:
\begin{itemize}
  \item Log--log correlation between our removed-comments and mod-log \texttt{removecomment}: \textbf{$r\approx0.80$} (N=129 subs).
  \item \textbf{Opt-in coverage caveat.} The public mod-log sample omits many of the largest subreddits that did not invite the bot (opt-in). Notable examples missing from the Juneja et al.\ dataset \cite{Juneja2020} include: \texttt{AskReddit}, \texttt{funny}, \texttt{pics}, \texttt{gaming}, \texttt{aww}, \texttt{worldnews}, \texttt{news}, \texttt{politics}, \texttt{memes}, \texttt{videos}, \texttt{todayilearned}, \texttt{mildlyinteresting}, \texttt{gifs}, \texttt{movies}, \texttt{WTF}, \texttt{pcmasterrace}, \texttt{IAmA}, \texttt{interestingasfuck}, \texttt{RoastMe}, \texttt{tifu}. This absence reflects \emph{opt-in} exposure, not a measurement disagreement.
  \item Non-overlap is spread across size bins, but many giants are missing; therefore level comparisons are restricted to the intersection (ours: 50{,}690 subs; Juneja et al. \cite{Juneja2020}: 187; overlap: 143).
\end{itemize}

\paragraph{Interpretation.} The correlation of $r \approx 0.80$ is notably strong given the differences between data sources. Our Pushshift-based proxy, despite measuring removal through a different mechanism (detecting \texttt{[removed]} flags rather than recording actions directly), ranks subreddits similarly to the closer to ground-truth mod-log. Subreddits that our data identifies as high-removal are also high-removal in the mod-log, and vice versa.

\emph{Conclusion:} Our proxy tracks the same enforcement signal in the overlap, with level differences explained by time-window and opt-in coverage.

\subsection{Is the Scaling Exponent \texorpdfstring{$\beta$}{beta} Consistent Across Sources?}
\label{sec:modlog_scaling}

The most important validation for our main text claims is whether the \emph{scaling relationship}---not just the levels is consistent across data sources. Section~\ref{sec:S4} reported $\beta \approx 1.18$ for removed comments in our Pushshift data. Does the mod-log show the same superlinear pattern?

\begin{figure}[t]
  \centering
  \includegraphics[width=.85\linewidth]{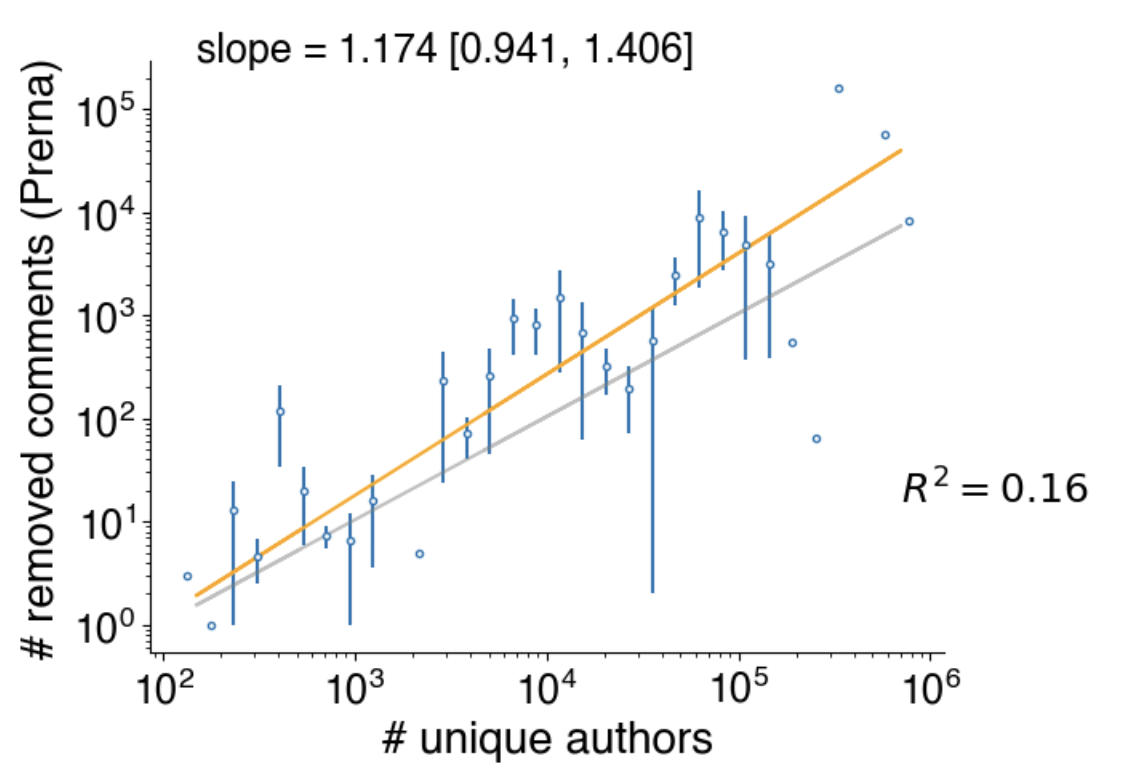}
  \caption{\textbf{Public mod-log (\texttt{removecomment}) shows \emph{superlinear} scaling.}
  Each point is a subreddit; $x$ is unique authors and $y$ is the number of \texttt{removecomment} actions from the public mod-log (2018, rolling 3-month coverage). The OLS line in log--log has slope $\beta=1.174$ (95\% CI $[0.941,\,1.406]$), $R^2=0.16$. Despite higher noise from the short, opt-in window, the slope is $>1$ and aligns with our Pushshift-based estimate ($\beta\approx1.18$). Mod-log data from Juneja et al.~\cite{Juneja2020}.}
  \label{fig:si_modlog_scaling}
\end{figure}

We fit $\log R = \alpha + \beta \log S$ with $S=$ unique authors and $R=$ removals:
\begin{itemize}
  \item \textbf{Juneja et al.\ mod-log, \texttt{removecomment} (2018 monthly)}: $\beta=1.174$ \,[95\% CI: $0.941,\,1.406$], $R^2=0.16$ (Figure ~\ref{fig:si_modlog_scaling}).
  \item \textbf{Ours (overlap), removed-comments (aggregate)}: $\beta\approx1.18$, $R^2\approx0.75$.
\end{itemize}

\paragraph{Interpretation.} The mod-log estimate ($\beta = 1.174$) in Figure ~\ref{fig:si_modlog_scaling} is pretty close to our Pushshift-based estimate ($\beta \approx 1.18$), the difference being less than 0.01. The confidence intervals overlap substantially. The lower $R^2$ in the mod-log analysis (0.16 vs.\ 0.75) reflects the smaller sample size (143 vs.\ 50,690 subreddits) and shorter time window (3 months vs.\ 6 years), not a fundamental difference in the relationship.

Critically, to note here is that both estimates are \textbf{superlinear} ($\beta > 1$). The mod-log independently confirms the main text's central finding that moderation demand grows faster than community size. This is not an artifact of how we detect removals in Pushshift---it is a property of how Reddit communities actually moderate.

\subsection{Summary}

\begin{itemize}
  \item In public mod-logs, \texttt{removecomment} is the main and stable moderator action ($\sim$80\% of actions). It is a natural enforcement unit.
  \item Our Pushshift-based removed-comments proxy aligns with mod-log \texttt{removecomment} across subreddits in the overlap ($r \approx 0.80$).
  \item The scaling exponent is consistent ($>1$) across sources and sample definitions, supporting the robustness of using \textbf{removed comments} as the moderation-intensity measure in size-scaling analyses.
\end{itemize}

\paragraph{Connection to subsequent analysis.} Having established that our removal measure is valid in this section, temporally stable (in Section~\ref{sec:S3}), and robust to sample selection (in Section~\ref{sec:S4}), we now turn to analyzing the \emph{residuals} from our scaling regressions. Section~\ref{sec:pca} examines how regulatory actions co-vary across communities beyond what size predicts, revealing interpretable governance dimensions that characterize different community regulatory styles.

\section{PCA on Size-Adjusted Residuals: What PC1, PC2, and PC3 Tell Us About Subreddit Governance}
\label{sec:pca}

\subsection{Overview: Principal dimensions of residuals}
\label{sec:pca_overview}

\begin{figure}[!ht]
\centering
\includegraphics[width=\linewidth]{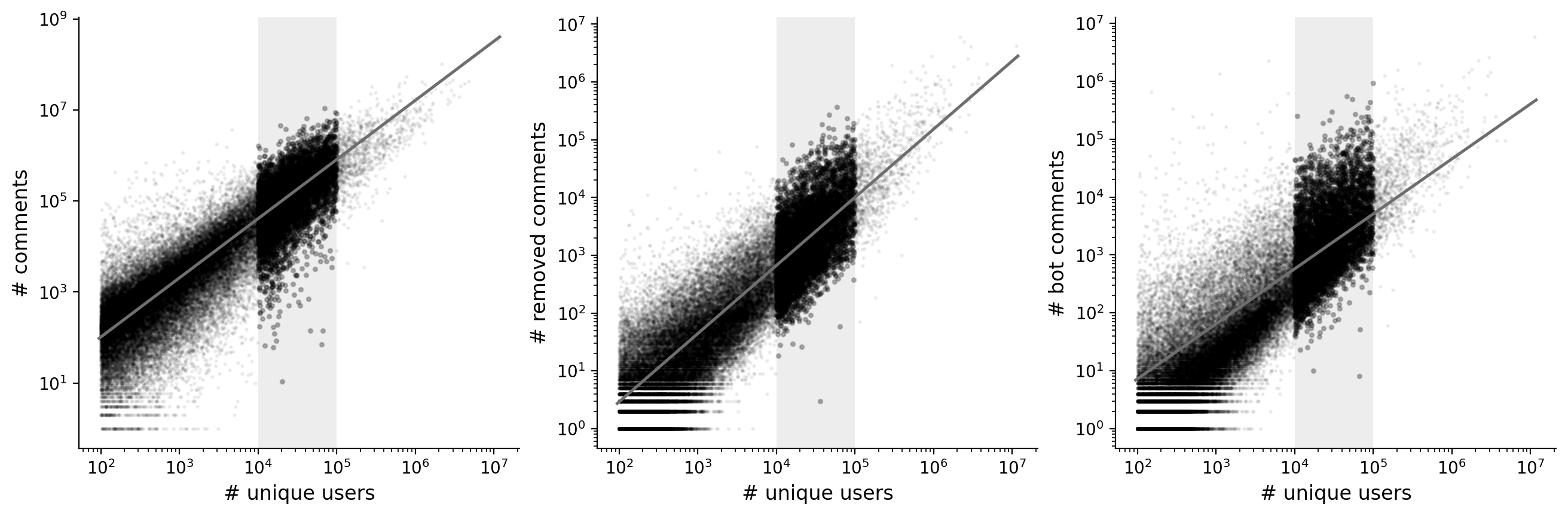}
\caption{\textbf{Same-size spread in regulatory activity (raw data).} Each panel shows raw counts versus unique users on log--log axes (left: comments, middle: removals, right: bot comments). The grey line is the fitted scaling relation from Figure~2 (slopes $\beta=1.12$, $1.18$, and $0.95$, respectively). The shaded vertical band marks a representative size cohort (e.g., $10^4$--$10^5$ users). Within a single cohort, activity differs by orders of magnitude across communities, revealing large same-size heterogeneity that averages in Figure~2 (log-binned means) cannot show. This motivates adjusting for size and analyzing residuals.}
\label{fig:same_size_spread}
\end{figure}

Fig. 2 in the main text shows that regulatory actions scale with community size (measured as active users): Mutual interaction with $\beta = 1.12$, enforcement with $\beta = 1.18$, and bot oversight with $\beta = 0.95$. For the residual and PCA analyses that follow, we focus on 44,471 subreddits that have non-zero values for all three regulatory actions (comments, removed comments, and bot comments), ensuring that log-residuals can be computed for each dimension. Although the scaling relationship holds consistently for average behavior across almost five orders of magnitude in community size, the average trend masks notable differences among communities. As Figure~\ref{fig:same_size_spread} shows, many subreddits of similar size may operate with very different levels of interaction, moderation, or bot oversight, pointing to a broader heterogeneity in how communities regulate themselves. For example, in \textbf{Fig.\ 2a--c} in the main text, the subreddit r/news ($\sim$10$^6$ users) lies above the scaling line in all three panels. Its mutual interaction, enforcement, and bot oversight all exceed size-based expectations. This makes sense: news discussions are contentious, users comment frequently to debate current events, moderators must remove rule-violating content, and bots post reminders to keep discussions civil. The topic generates regulatory demand across all three actions simultaneously.

In contrast, r/LoadingIcons ($\sim$10$^3$ users) shows a different pattern. This community shares looping GIFs and videos with minimal discussion. In \textbf{Fig.\ 2a}, r/LoadingIcons lies below the Mutual Interaction scaling line: users come to view content, not to debate. But in \textbf{Fig.\ 2c}, r/LoadingIcons lies slightly below (practically on) the bot oversight scaling line: the community uses bots to remind users of specific formatting rules. Here, one regulatory action is suppressed while another is elevated. These examples show that residuals do not vary randomly. Some communities run hot on all three actions; others show mixed patterns. What is the guiding force behind this variance? How do the three residuals co-vary across 44,471 subreddits?

We want to know beyond size, what are the other governance factors? Governance is not a single action but a system of actions. How do mutual interaction, enforcement, and bot oversight combine together? What guides their variation? To see beyond size, we examine residuals. A residual measures how much a subreddit deviates from the scaling line. Positive residuals mean the subreddit lies above the line which means more regulatory activity than expected for its size. Negative residuals indicate that it lies below the expected level, implying less activity than anticipated.

To answer this, we extract size-adjusted residuals. For each subreddit $i$ and action $j$, the residual $\xi$ measures how far the observed value lies from the scaling prediction:
\[
\xi_{i,j} = \log_{10} Y_{i,j} - \left(\hat{\alpha}_j + \hat{\beta}_j \log_{10} N_i\right)
\]

These residuals formalize the contrasts already visible in Figure~2 (e.g., \textit{r/news} high on all three vs.\ \textit{r/LoadingIcons} low on conversation but average on bot oversight). A residual of $\xi = +1$ means the observed value is 10$\times$ the size-based expectation; $\xi = -1$ means 0.1$\times$ the expectation. We have two contrasting cases: r/news, where all three residuals are positive, and r/LoadingIcons, where Mutual Interaction is negative but bot oversight residual is nearly zero. Which pattern is more common? The answer matters because governance is not a single action but a system of actions. If residuals vary independently, each action operates in isolation. But if residuals co-vary systematically, the three actions form coordinated governance profiles: communities would have distinct regulatory styles or mechanims, not just ``more'' or ``less'' moderation. As a first step, we compute pairwise correlations among the three residuals across all 44,471 subreddits.

The correlation matrix shows that all three pairwise correlations are positive (Table~\ref{tab:correlation_matrix}). Communities that run hot on one action tend to run hot on the others. The r/news pattern is not an anomaly: across the dataset, positive residuals cluster together. In governance terms, this makes sense. A community facing high regulatory demand in one area likely faces demand in others. Contentious topics generate more user comments, which require more moderation, which in turn benefits from more bot support.

\begin{table}[h]
\centering
\begin{tabular}{lccc}
\toprule
 & $\xi_{\text{Mutual}}$ & $\xi_{\text{enforcement}}$ & $\xi_{\text{Bot}}$ \\
\midrule
$\xi_{\text{Mutual}}$ & 1.00 & 0.26 & 0.29 \\
$\xi_{\text{Enforcement}}$ & 0.26 & 1.00 & 0.36 \\
$\xi_{\text{Bot}}$ & 0.29 & 0.36 & 1.00 \\
\bottomrule
\end{tabular}
\caption{Pairwise correlations among size-adjusted residuals for Mutual interaction, enforcement, and bot oversight across 44,471 subreddits. Positive, moderate correlations indicate a shared demand component (high on one action tends to be high on others) while leaving room for structured differences that PCA will separate.}
\label{tab:correlation_matrix}
\end{table}

But positive correlations do not mean all communities behave like r/news. The r/LoadingIcons pattern, where one action is elevated while another is suppressed, also exists in the data. The correlations are positive but moderate (0.26 to 0.36), leaving room for substantial variation. How do we reconcile these two observations? We need a method that can identify the dominant pattern (the positive correlations that produce r/news like communities) while also revealing secondary patterns (the trade-offs that produce r/LoadingIcons like communities).

Principal Component Analysis (PCA) provides this decomposition. Rather than treating the three correlations separately, PCA asks: what is the largest source of shared variation across all three actions? And after removing that, does the remaining variation have structure? PCA reorients the coordinate system to reveal orthogonal axes of variation, ordered by how much variance each explains. We apply PCA not for dimension reduction, but to peel apart the layers of covariation. The result is three principal components that together account for 100\% of the variance.

Our analysis in Fig. 3 and Fig. 4 shows that the largest source of variation is \textbf{PC1: Intensity} (54.1\% of variance). Here, three actions load positively: Mutual Interaction (0.543), Enforcement (0.585), Bot oversight (0.602). This axis captures the r/news pattern: communities where the topic itself generates regulatory demand across all actions. Contentious subjects produce more user comments, which in turn require more supervision and more bot oversight. Intensity is the shared regulatory load that some communities carry beyond what their size predicts. This is not surprising: we would expect that some communities are simply ``hotter'' than others, and that this heat spills across all regulatory actions. In contrast, r/LoadingIcons, with its quiet GIF-sharing culture, sits lower on this intensity axis.

Now removing intensity from the picture; imagine all communities have the same overall regulatory load. What remains? The next axis, \textbf{PC2: One-way vs.\ Two-way Communication} (24.8\% of variance), answers this. Mutual Interaction loads $-0.826$; enforcement loads $+0.500$; bot oversight loads $+0.260$. Among communities with similar intensity, some achieve regulation through top-down control (moderators and bots direct behavior), while others achieve it through peer dialogue (users regulate each other and enforce norms). Returning to the example of r/LoadingIcons; its low mutual interaction and high bot oversight places it towards the one-way end of this axis. Users receive bot reminders about formatting rules but rarely discuss content with each other. A community like r/CasualConversation (in Fig 4) sits at the opposite end where users regulate through conversation, not through moderator intervention. This too is intuitive as governance can flow downward from authorities or emerge sideways from peers.

Now, if we remove both Intensity and Coordination directions, among communities with similar regulatory load and similar balance of top-down/one-way versus peer regulation/two-way, what still varies? The final axis, \textbf{PC3: Impersonal vs.\ Personal Moderation} (21.1\% of variance), captures this. Bot oversight loads $+0.755$; enforcement loads $-0.639$. Some communities rely on bots that apply uniform rules automatically; others rely on humans who judge each case individually. r/LoadingIcons again illustrates the pattern: its high bot oversight and minimal human removal places it toward the impersonal end. The bots enforce formatting rules that require no interpretation. A community moderating harassment or misinformation, by contrast, requires human judgment about context and intent. This parallels what organizational theorists call thick versus thin rules: thick rules are codified and leave no room for discretion; thin rules require interpretation \cite{daston2022rules}. Once more, this is a distinction we would expect to find---the choice between automation and human judgment is a fundamental design decision in any governance system.

Together, Intensity (54.1\%), One-way vs.\ Two-way Communication (24.8\%), and Impersonal vs.\ Personal Moderation (21.1\%) sum to 100\%. What the PCA reveals is not absurd. These three axes map onto principles that governance literature has long spoken of: the overall burden of regulation, the direction of information flow, and the role of discretion in enforcement. The contribution here is not discovering new concepts, but showing that these familiar dimensions emerge empirically from the covariance structure of 44,471 communities, and that they are orthogonal: each captures variation the others do not. r/news and r/LoadingIcons, introduced as contrasting cases, now have precise coordinates in this three-dimensional governance space.

Are these patterns real structure or artifacts of the method? To confirm that the three components reflect substantive governance features, we conduct permutation tests that selectively destroy specific correlations in the data. The tests show that the 54/25/21 variance split is far from random (a null model yields 33/33/33), and that each axis re-emerges only when we preserve the specific feature it encodes. Section~\ref{sec:pca_validation} reports the full validation results.

\paragraph{Building on earlier validation.} This analysis rests on foundations established in earlier sections. Section~\ref{sec:S1} documented that our 2015--2020 observation window captures a stable regulatory regime where removal data offers
a more interpretable measure of realized enforcement, no longer bottlenecked by the capacity constraints of the earlier, transitive phase. Sections~\ref{sec:S3} and \ref{sec:S4} demonstrated that scaling exponents are temporally stable (consistent year-over-year from 2015--2020) and robust to sample selection (nearly identical between our 2,828 labeled subreddits and the full universe of 107,006 communities). Section~\ref{sec:robustness} validated our comment removal measure against independent moderator logs, confirming that our Pushshift-based proxy tracks genuine moderation activity. With these foundations in place, we can confidently analyze the \emph{residual} variation---the governance differences that remain after accounting for community size knowing that this variation reflects meaningful community characteristics rather than measurement artifacts.

\subsection{Methods: Finding Governance Patterns in Residuals}
\label{sec:pca_methods}

\subsubsection{Extracting Size-Adjusted Residuals}

After establishing baseline scaling relationships, we extract residuals that capture each community's deviation from size-based expectations:
\[
\xi_{i,j} \equiv \log_{10} Y_{i,j} - \Big(\widehat{\log_{10} Y_0} + \widehat{\beta}_j \log_{10} N_i\Big), \quad j \in \{\text{mutual}, \text{enforcement}, \text{bot}\}
\]

\subsubsection{Standardization: Creating Comparable Scales}

\begin{figure}[!ht]
\centering
\includegraphics[width=\linewidth]{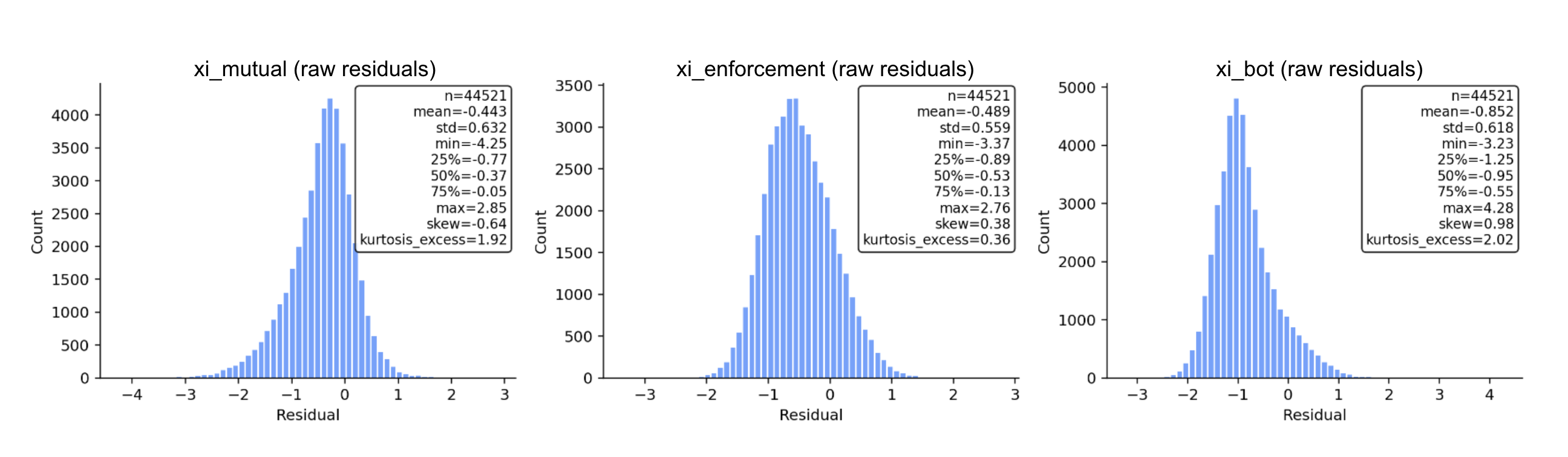}
\caption{Distributions of standardized residuals for mutual interaction, enforcement, and bot oversight after removing size effects. Each panel displays the empirical distribution with descriptive statistics. The heavy right tails, particularly visible in enforcement and bot residuals, are characteristic of social media data where extreme values occur more frequently than in normal distributions. Shapiro-Wilk tests confirm significant departures from normality (all $p < 0.001$), justifying the use of nonparametric bootstrap methods in subsequent analyses.}
\label{fig:residual_hists}
\end{figure}

The three governance measures have vastly different distributions even after size adjustment. To enable meaningful comparison, we standardize each dimension:
\[
Z_{i,j} = \frac{\xi_{i,j} - \mu_j}{\sigma_j}
\]
where $\mu_j$ and $\sigma_j$ are the mean and standard deviation of residuals $j$ across all communities. Figure~\ref{fig:residual_hists} shows these standardized distributions, revealing heavy right tails characteristic of social media data.

\subsubsection{Principal Component Analysis}

We perform PCA on the $3 \times 3$ correlation matrix of the standardized variables to identify orthogonal dimensions of variation. The correlation matrix $\mathbf{R}$ equals the covariance matrix of the z-scored data:
\[
\mathbf{R} = \frac{1}{n-1}\mathbf{Z}^T\mathbf{Z}
\]

\newpage
Using the correlation matrix rather than raw covariance ensures:
\begin{itemize}
\item \textbf{Scale invariance}: All variables contribute equally regardless of measurement scale
\item \textbf{Interpretable loadings}: Loadings are correlations bounded [-1, 1]
\item \textbf{Normalized variance}: Total variance = 3, so eigenvalues directly give percentages
\end{itemize}

PCA solves the eigenvalue problem $\mathbf{R}\mathbf{V} = \mathbf{V}\boldsymbol{\Lambda}$ where $\mathbf{V}$ contains loadings and $\boldsymbol{\Lambda}$ contains eigenvalues.

\subsubsection{Resampling for Visualization}

While our PCA analysis was performed on all 44,471 subreddits, visualizing this many points would create an overcrowded, illegible scatter plot where patterns become obscured by point density. To create the clear, interpretable visualizations shown in \textbf{Figure 3} of the main text, we developed a resampling strategy that preserves the full distribution's structure while reducing visual clutter.

\begin{figure}[!ht]
\centering
\includegraphics[width=0.5\linewidth]{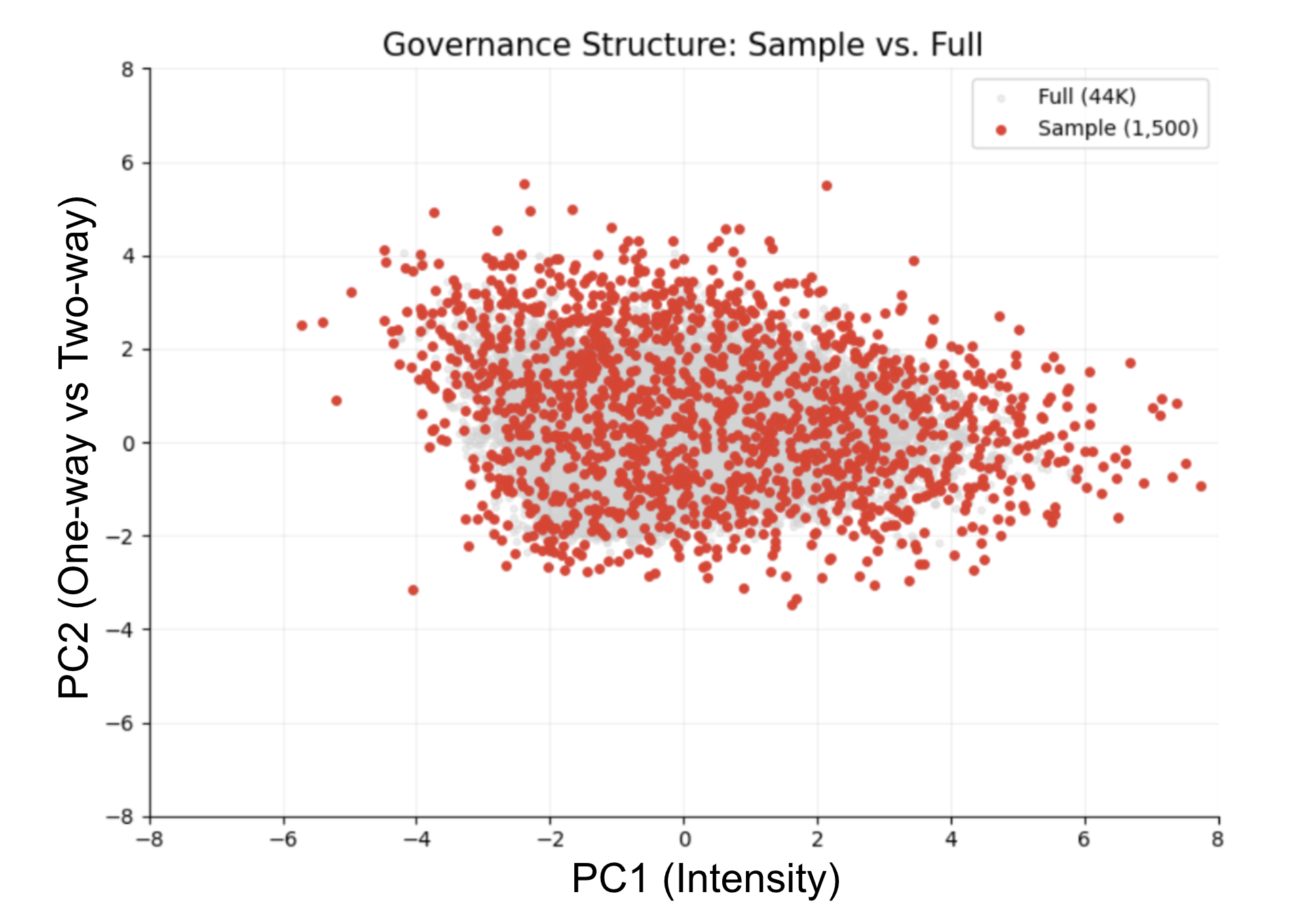}
\caption{Validation of resampling strategy for visualization clarity. The 1,500 sampled points (red) used in main text figures overlay the full dataset of 44,471 subreddits (gray background), demonstrating that inverse-probability weighting based on kernel density estimation successfully preserves the global governance structure while creating readable visualizations. Rare configurations at the extremes remain visible, the central tendency is maintained, and the full range of variation along both PC1 (intensity) and PC2 (one-way vs. two-way) is represented. This sampling is used only for creating scatter plots in the main text; all statistical analyses use the complete dataset.}
\label{fig:resampling}
\end{figure}

After computing principal component scores for all 44,471 communities, we selected a representative subset of 1,500 points for visualization. We estimated point density in the PC1--PC2 plane using a Gaussian kernel density estimator, then assigned each observation a sampling probability inversely proportional to this density. Drawing 1,500 observations without replacement according to these weights preferentially selects points from low-density (extreme) regions while down-weighting highly redundant points in dense clusters. This ensures rare governance configurations remain visible in the scatter plots while preventing the dense central cluster from creating an opaque mass. Figure~\ref{fig:resampling} demonstrates that our sampled points (red) faithfully preserve the structure of the full dataset (gray background) while creating a readable visualization. We verified this preservation by comparing means, covariance matrices, and PCA explained variance ratios between the full dataset and the sampled points.

\subsection{Results}
\label{sec:pca_results}

\subsubsection{Principal Component Structure: Three Independent Dimensions of Governance}

Analyzing 44,471 subreddits reveals three orthogonal governance dimensions that explain how communities deviate from size predictions.

Why divide eigenvalues by 3? When we standardized our three regulatory actions (mutual interaction, enforcement, bot oversight), we gave each variable variance = 1. Therefore:
\begin{itemize}
\item Total variance in the system = 1 + 1 + 1 = 3
\item The correlation matrix has trace (sum of diagonal) = 3
\item The eigenvalues must also sum to 3 (mathematical property of PCA)
\end{itemize}

To convert eigenvalues to percentages of total variance:
\begin{itemize}
\item PC1: eigenvalue 1.623 $\div$ total variance 3 = 0.541 = 54.1\%
\item PC2: eigenvalue 0.744 $\div$ total variance 3 = 0.248 = 24.8\%
\item PC3: eigenvalue 0.633 $\div$ total variance 3 = 0.211 = 21.1\%
\item Check: 54.1\% + 24.8\% + 21.1\% = 100\%
\end{itemize}

This tells us PC1 captures over half the variation in how communities govern beyond size--it's the dominant pattern. The remaining 46\% is split between PC2 and PC3. Figure~\ref{fig:governance-space}a visually confirms the wide spread of subreddit behavior in 3D standardized residual space across mutual interaction, enforcement, and bot oversight. The dominant direction of variation becomes clearer after rotation to the principal component axes, shown in Figure~\ref{fig:governance-space}b.

\subsubsection{Loading Matrix and Interpretation}

The loading matrix tells us how our original regulatory actions combine to form each principal component:

\begin{table}[h!]
\begin{center}
\begin{tabular}{lccc}
\hline
& \textbf{PC1 (54.1\%)} & \textbf{PC2 (24.8\%)} & \textbf{PC3 (21.1\%)} \\
\hline
Mutual interaction & 0.543 & $-0.826$ & $-0.149$ \\
Enforcement & 0.585 & 0.500 & $-0.639$ \\
Bot oversight & 0.602 & 0.260 & 0.755 \\
\hline
\label{tab:loadingmatrix}
\end{tabular}
\caption{Loading Matrix}
\end{center}
\end{table}

\begin{figure}[!ht]
\centering
\includegraphics[width=0.86\linewidth]{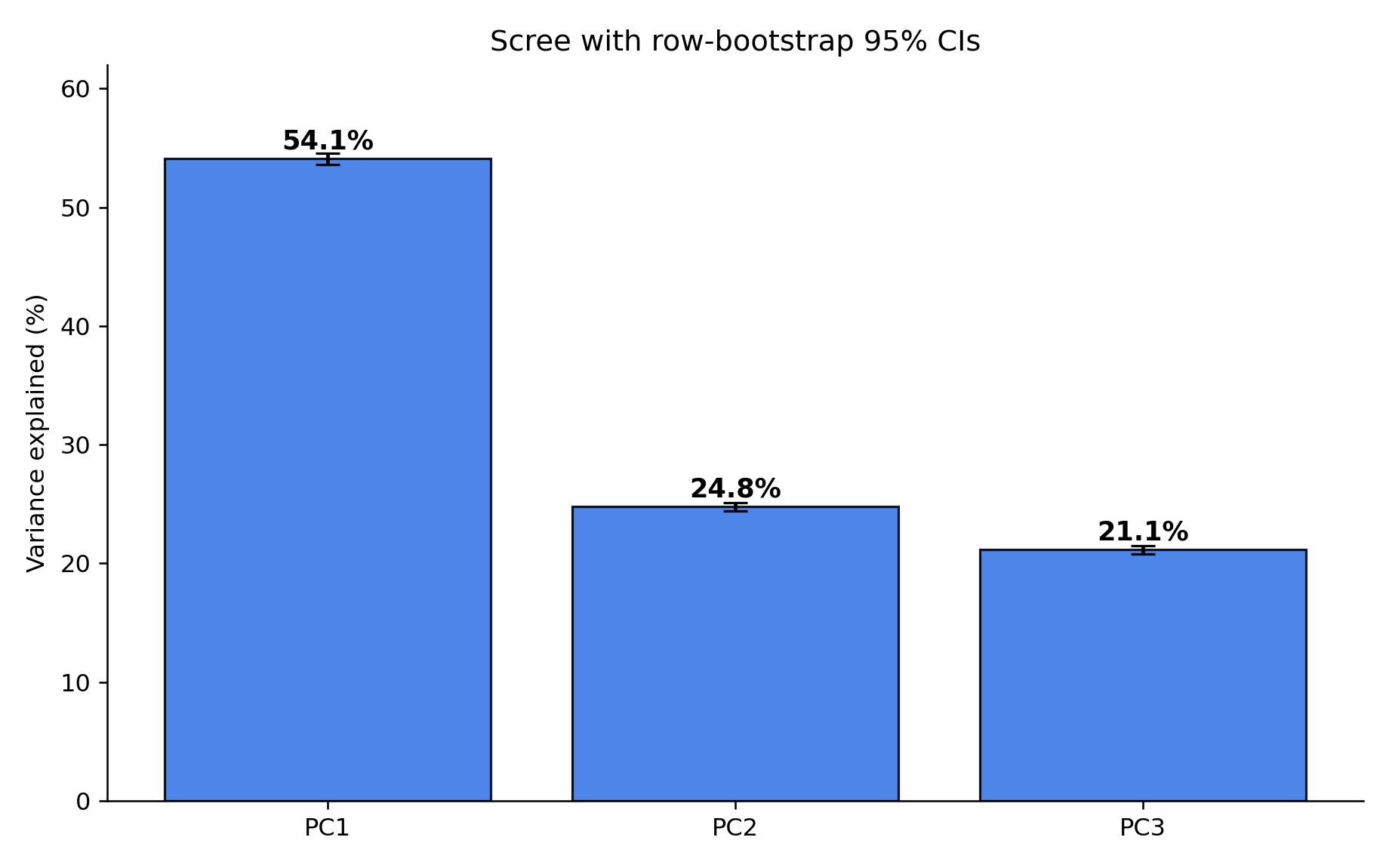}
\caption{Scree plot with bootstrap confidence intervals demonstrating component stability. The three principal components explain 54.1\%, 24.8\%, and 21.1\% of variance respectively, totaling 100\% for this three-variable system. Error bars represent 95\% confidence intervals from 400 bootstrap resamples, with remarkably tight bounds (PC1: [53.6, 54.6]\%, PC2: [24.5, 25.1]\%, PC3: [20.8, 21.4]\%) indicating robust and stable component extraction. The steep drop from PC1 to PC2 indicates the dominance of the intensity factor.}
\label{fig:scree}
\end{figure}

So PC1 is a weighted average of all three modes (all positive), PC2 contrasts mutual interaction two-way communication against the control one-way coordination (negative vs. positive), and PC3 contrasts bot oversight against enforcement (positive vs. negative). This rotation and loading structure, especially the prominent first component, are also visible in the visual layout of Figure~\ref{fig:governance-space}b and the annotated loading arrows in Figure~\ref{fig:governance-space}d. These patterns are also stable---400 bootstrap resamples yield tight confidence intervals (Figure~\ref{fig:scree}), confirming our three-component structure is robust, not a statistical artifact.

Coming back to these equations and what the PC axes mean, we look at the loading bars and try to understand the patterns. Each column reveals a distinct governance strategy:

\begin{figure}[!ht]
\centering
\includegraphics[width=\linewidth]{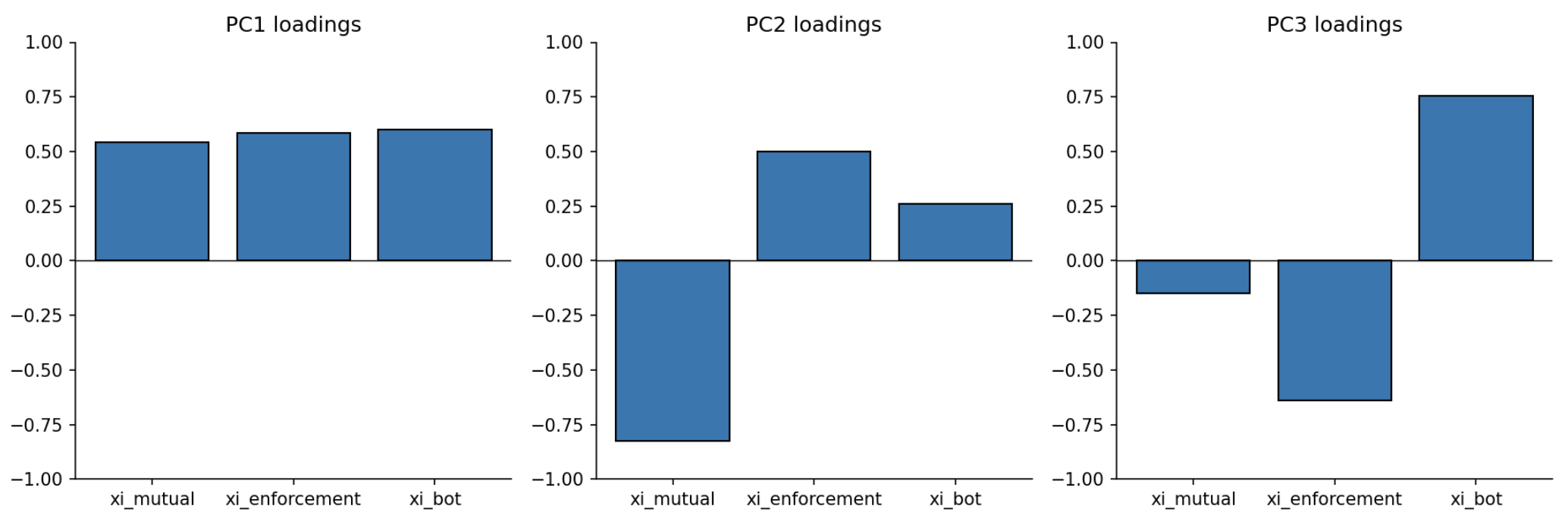}
\caption{Loading structure revealing the interpretation of each principal component. \textbf{PC1 (left):} All variables load positively (0.543--0.602), confirming this axis captures overall governance intensity. \textbf{PC2 (center):} Mutual interaction loads strongly negative ($-0.826$) while enforcement (0.500) and bots (0.260) load positive, indicating a one-way vs two-way control choice. \textbf{PC3 (right):} bot oversight load strongly positive (0.755) while enforcement loads negative ($-0.639$), with mutual nearly orthogonal ($-0.149$), cleanly separating impersonal from per moderation tactics.}
\label{fig:loadings}
\end{figure}

\textbf{PC1 Intensity hypothesis:} All three modes load positively (0.54--0.60), nearly equal. This suggests communities vary in overall governance ``temperature''---some run hot on everything, others cold on everything, beyond what size predicts.

\textbf{PC2 One-way vs. Two-way Communication hypothesis:} Mutual interaction loads strongly negative ($-0.826$) while enforcement (0.500) and bot oversight (0.260) load positive. This suggests a choice of coordination: at fixed intensity, emphasizing if a subreddit wants to control through top-down one-way regulation or bottom-up two-way norm enforcement through conversation.

\textbf{PC3 Impersonal vs. Personal Moderation hypothesis:} Bot oversight loads positive (0.755) while enforcement loads negative ($-0.639$), with mutual interaction nearly zero ($-0.149$). This suggests communities with similar control levels differ in the orientation of whether they want to do personal moderation through removed comments or for impersonal routine line bot oversight through comments.

To visualize how these abstract loadings manifest in real subreddit behavior, we now examine the governance space through different projections.

\subsubsection{Governance Space Visualization}

How do 44,471 subreddits actually distribute across our governance dimensions? Let's start with the big picture and zoom in on each pattern.

\begin{figure}[!ht]
\centering
\includegraphics[width=\linewidth]{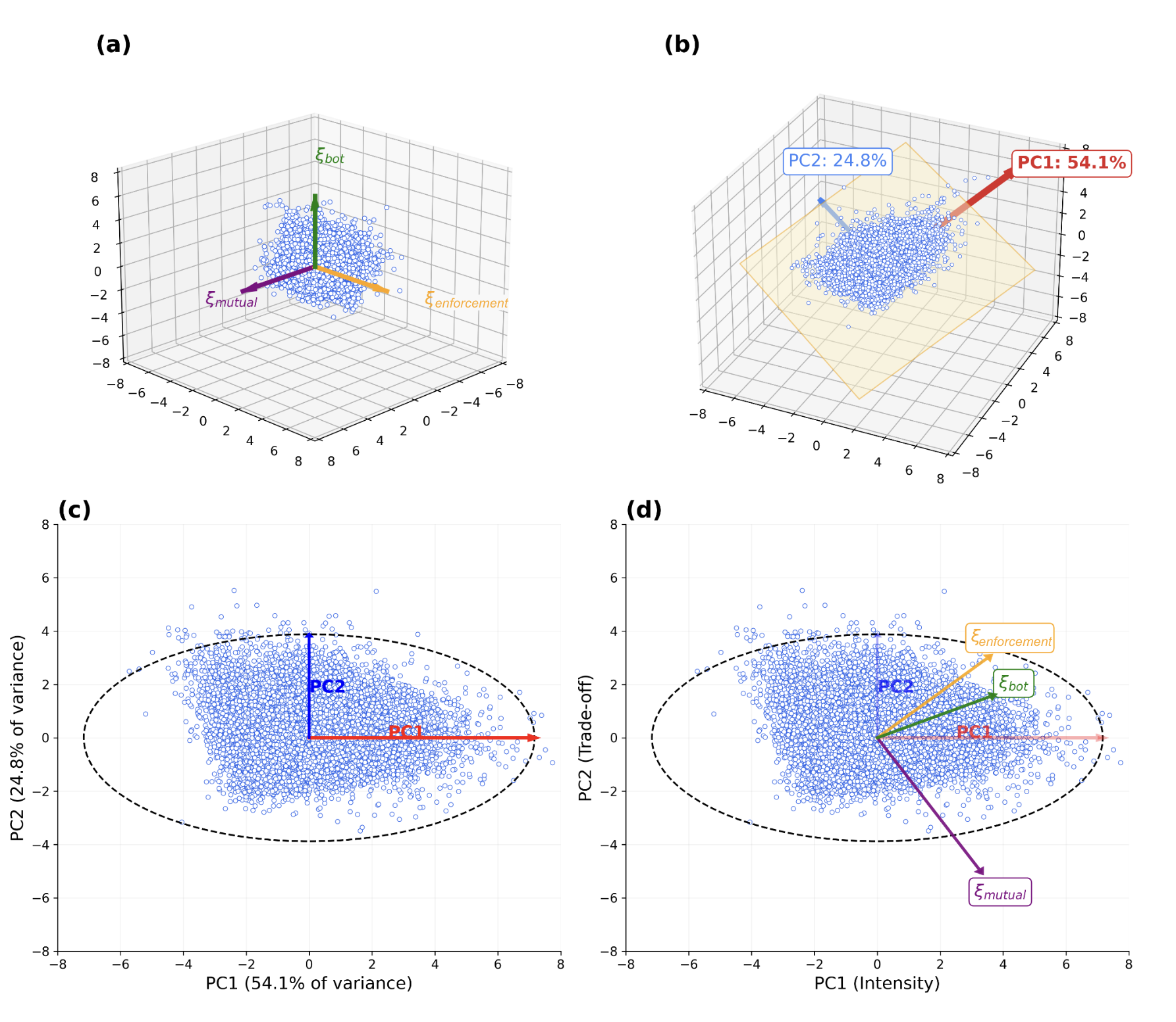}
\caption{
Principal component view of subreddit governance space.
\textbf{(a)} The raw 3D space of standardized residuals after removing size effects. Each axis corresponds to one governance mode--mutual interaction (purple), enforcement (orange), and bot oversight (green). Notice that the cloud of 44,471 subreddits is not perfectly spherical: the spread is longer in some directions than others, hinting at underlying structure.
\textbf{(b)} The same data rotated to align with the top two principal components. This reveals a tilted, slanted disc-shaped cloud--most variation lies in a 2D plane. The red and blue arrows show PC1 and PC2, which together explain 79\% of all variance. PC1 is clearly longer and steeper: its vector reaches twice as far as PC2, because it captures 2.2 times more variance (54.1\% vs. 24.8\%). The yellow plane marks this dominant subspace, where most behavioral diversity unfolds.
\textbf{(c)} A full 2D projection of all subreddits onto the PC1--PC2 plane. This compresses the original 3D cloud into its most meaningful dimensions, revealing a clear gradient of governance behavior across communities.
\textbf{(d)} The same projection, now annotated with colored arrows showing how the original axes contribute. All three governance modes point in a similar direction along PC1, confirming that this axis reflects overall governance intensity. PC2, by contrast, reflects a trade-off: mutual interaction points downward while enforcement and bot oversight point upward. Together, these panels reveal that subreddit governance is not random--it follows some consistent, interpretable structure that we try to understand.
}
\label{fig:governance-space}
\end{figure}

\paragraph{PC1: Governance Intensity (54.1\%).}

Figure~\ref{fig:governance-space}d shows how all three arrows--mutual interaction, enforcement, and bot oversight--point in roughly the same direction (towards the right)? That's the intensity pattern. Communities toward the right do \textit{more} of everything; those toward the left do \textit{less} of everything.

What does this mean specifically in the context of governance? We've already removed size effects. Take r/politics on the far right--with 8 million subscribers, we'd expect about 100 comments per post based on our scaling laws. Instead, it generates 600 comments per post, has four times the expected removals, and three times the expected bot oversight. It behaves like a community several times its actual size--conversations explode, moderators work overtime, bots fire constantly. Move toward the center and we find r/CasualConversation, which hits exactly the activity levels our equations predict--no surprises, perfectly average for its size. Then on the far left sits small hobby subreddits that run quiet even accounting for their modest size, with minimal discussion, rare removals, and dormant bots. This horizontal spread captures pure intensity after size is factored out (Fig 3 in the main text has the labelled subreddits to see this pattern).

\paragraph{PC2: One-way vs. Two-way Communication (24.8\%).}

In Figure~\ref{fig:governance-space}d, we now notice the vertical axis where Mutual interaction points downward while enforcement and bots point upward. This creates a fundamental fork in the road. Communities that move upward, like r/science, choose on-way control over two-way conversations. Heavy moderation creates pristine, curated discussions where every comment meets strict standards, but at a cost--fewer comments survive the filter, discussions feel formal, and spontaneity dies. Communities that move downward, like r/WritingPrompts, choose the opposite path. Comments flow freely with minimal interference, creating vibrant discussion where tangents flourish and quality varies wildly. We can't have both the pristine curation of r/science and the freewheeling conversation of r/WritingPrompts. That's the choice encoded in PC2. (Fig 3 in the main text has the labelled subreddits to see this pattern)

\paragraph{PC3: Impersonal vs. Personal Moderation (21.1\%).}

The first two dimensions tell most of the story (79\%), but there's a third choice communities make: \textit{how} to intervene. We need new angles to see this:

\begin{figure}[!ht]
\centering
\includegraphics[width=\linewidth]{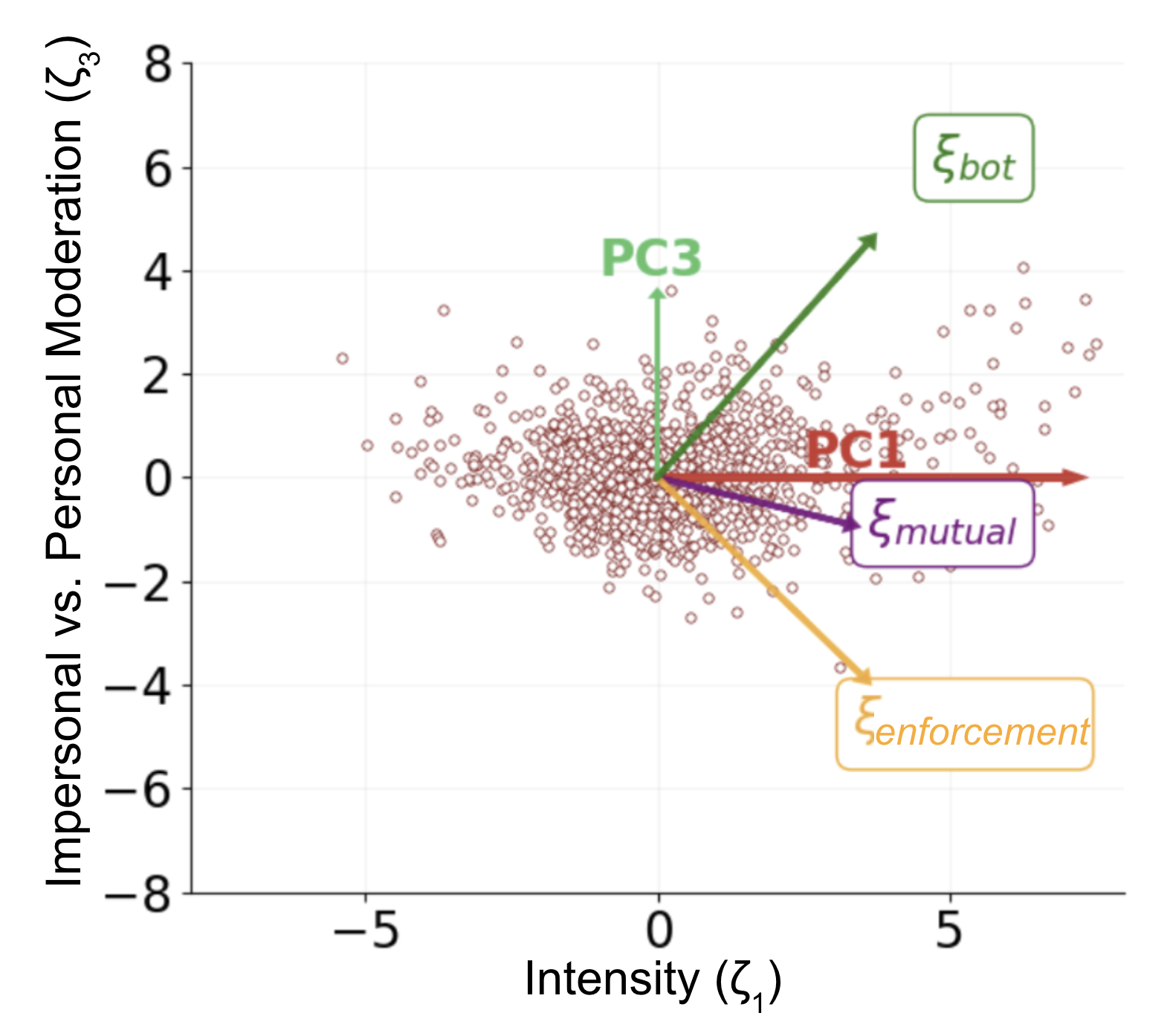}
\caption{The impersonal vs personal moderation split. bot oversight points up and enforcement points down. Same intensity (horizontal position) but different orientations of how to handle moderation (vertical position).}
\label{fig:biplot13}
\end{figure}

Let us look at the PC3-PC1 project in Figure \ref{fig:biplot13}. Two communities can sit at identical horizontal positions having the same intensity but can occupy opposite vertical positions. Communities at the top deploy AutoModerator preemptively, with bots warning ``Be civil!'' before problems start, automatic reminders about rules appearing in every thread, and preventive measures dominating. Communities at the bottom take the opposite approach, letting conversations unfold naturally then having human moderators sweep through removing violations after they appear in a more personal nuanced way.

\begin{figure}[!ht]
\centering
\includegraphics[width=\linewidth]{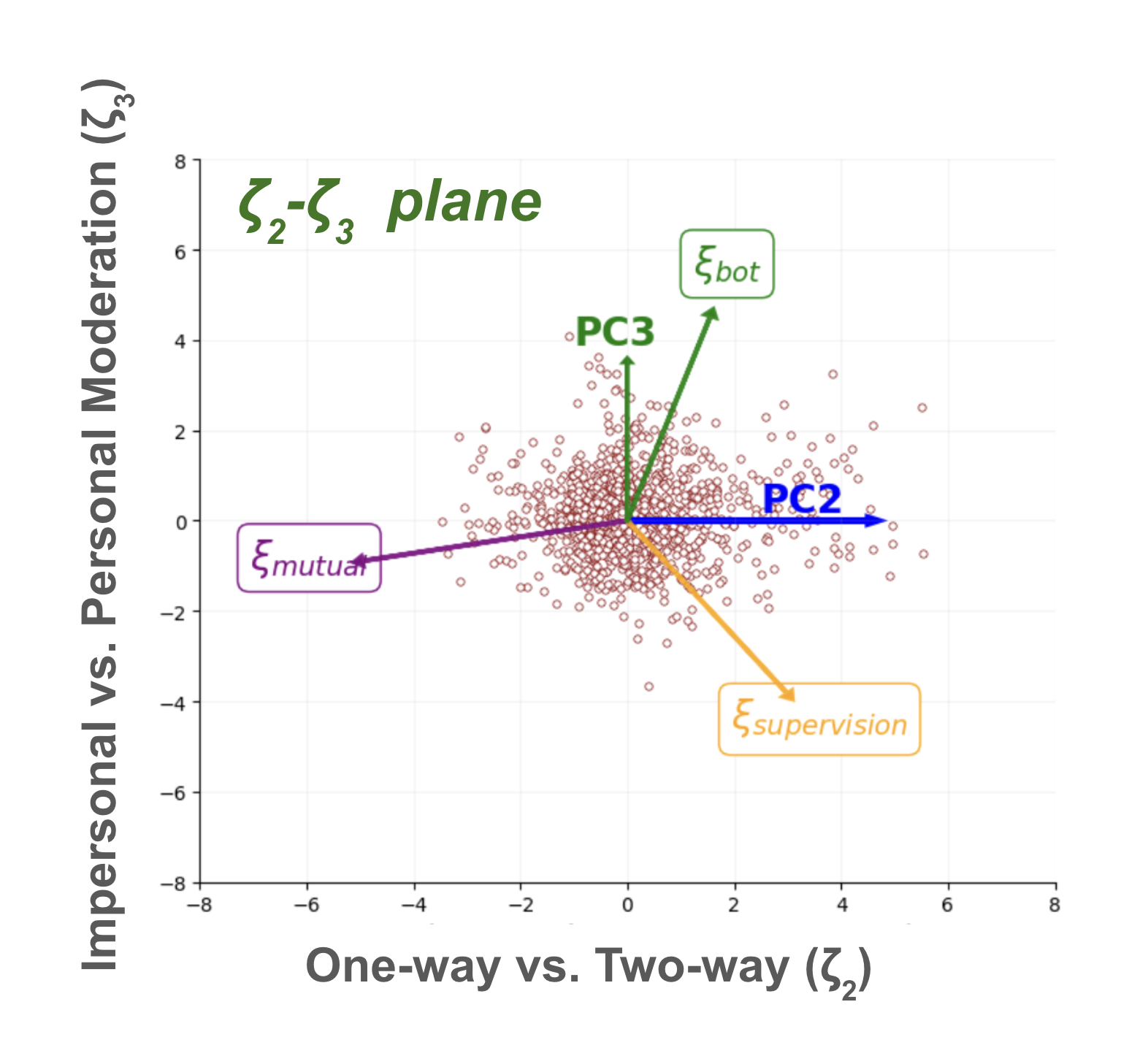}
\caption{Independence of choices. A community's control level (horizontal) doesn't determine its tactics (vertical). High-control communities exist at both top (preventive) and bottom (reactive).}
\label{fig:biplot23}
\end{figure}

The key insight from Figure~\ref{fig:biplot23} is that these are independent dials. A high-control community (right side) might prevent problems with extensive AutoModerator rules (top-right) or maintain control through vigilant post-hoc removal (bottom-right). Similarly, low-control communities (left side) might use gentle bot reminders (top-left) or simply let things run with minimal intervention of any kind (bottom-left). Every combination exists in the data because communities have three separate governance choices: how hot to run (PC1), whether to choose control or conversation (PC2), and how to act, personal or impersonal (PC3).

But how do we know these interpretations aren't just stories we're telling ourselves? The loadings suggest intensity, coordination choices, and moderation tactics--but are these patterns mathematically real or just convenient narratives? To find out, we need to test whether our principal components align with what we'd theoretically expect if our interpretations were correct.

\subsection{Validation: Geometric and Statistical Analyses}
\label{sec:pca_validation}

We validate our PCA interpretations using three complementary approaches:

\begin{itemize}
\item \textbf{Stability validation}: Resampling 44,471 communities with replacement (400 iterations) to obtain 95\% confidence intervals for variance shares and loadings (Figure~\ref{fig:scree}), confirming our three-component structure is statistically robust

\item \textbf{Geometric validation}: Testing whether PCs align with theoretically-motivated reference directions (Figure~\ref{fig:pc_angles_validation}) for example, does PC1 align with ``all modes together'' $[1,1,1]/\sqrt{3}$? Does PC2 capture ``one-way vs. two-way'' $[1,-1,-1]/\sqrt{3}$?

\item \textbf{Bootstrap permutation tests}: Using targeted randomization null models to test specific structural hypotheses (Figure~\ref{fig:perm_progression}) for example: within-row permutation preserves community intensity while breaking mode-specific patterns; swapping enforcement$\leftrightarrow$bot oversight tests if PC2 captures total cooridnation choice (one-way vs two-way).
\end{itemize}

\subsubsection{Geometric Validation}

If PC1 really captures ``intensity,'' it should point in the direction where all three regulatory actions increase together. If PC2 really represents ``one-way versus two-way,'' it should contrast mutual interaction against the enforcement+oversight modes. If PC3 captures ``impersonal versus personal,'' it should contrast bot oversight against enforcement. Let's test these hypotheses geometrically.

\begin{figure}[!ht]
\centering
\includegraphics[width=\textwidth]{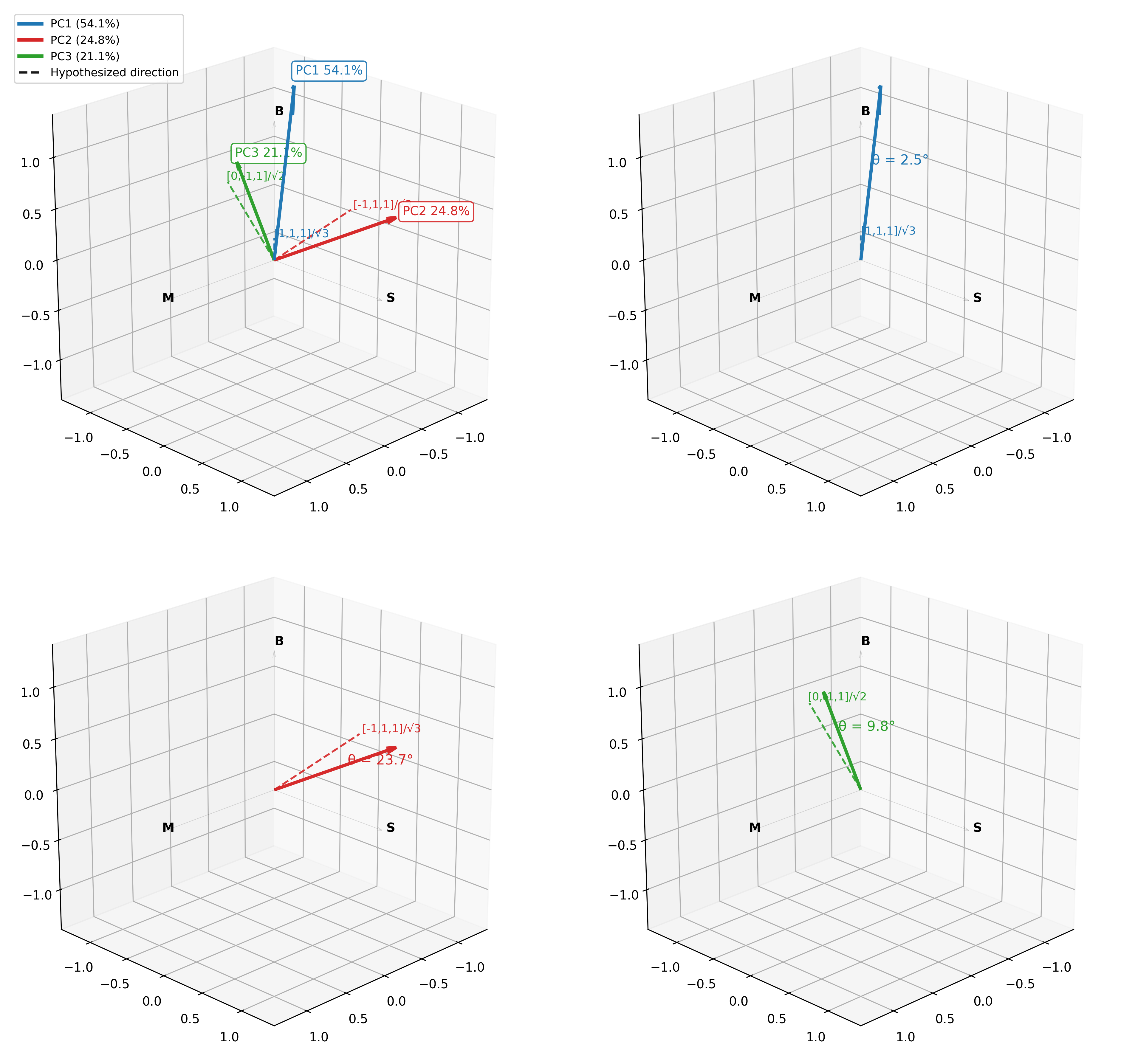}
\caption{Geometric validation of PC interpretations. \textbf{(A)} Overview: PCs (solid arrows, lengths show variance) and theoretical references (dashed). \textbf{(B)} PC1 vs. ``all modes together'' [1,1,1]: angle 2.5°. \textbf{(C)} PC2 vs. ``one-way minus two-way'': angle 23.7°. \textbf{(D)} PC3 vs. ``bots(impersonal) minus enforcement(personal)'' [0,$-1$,1]: angle 9.8°.}
\label{fig:pc_angles_validation}
\end{figure}

\begin{figure}[!ht]
    \includegraphics[width=\linewidth]{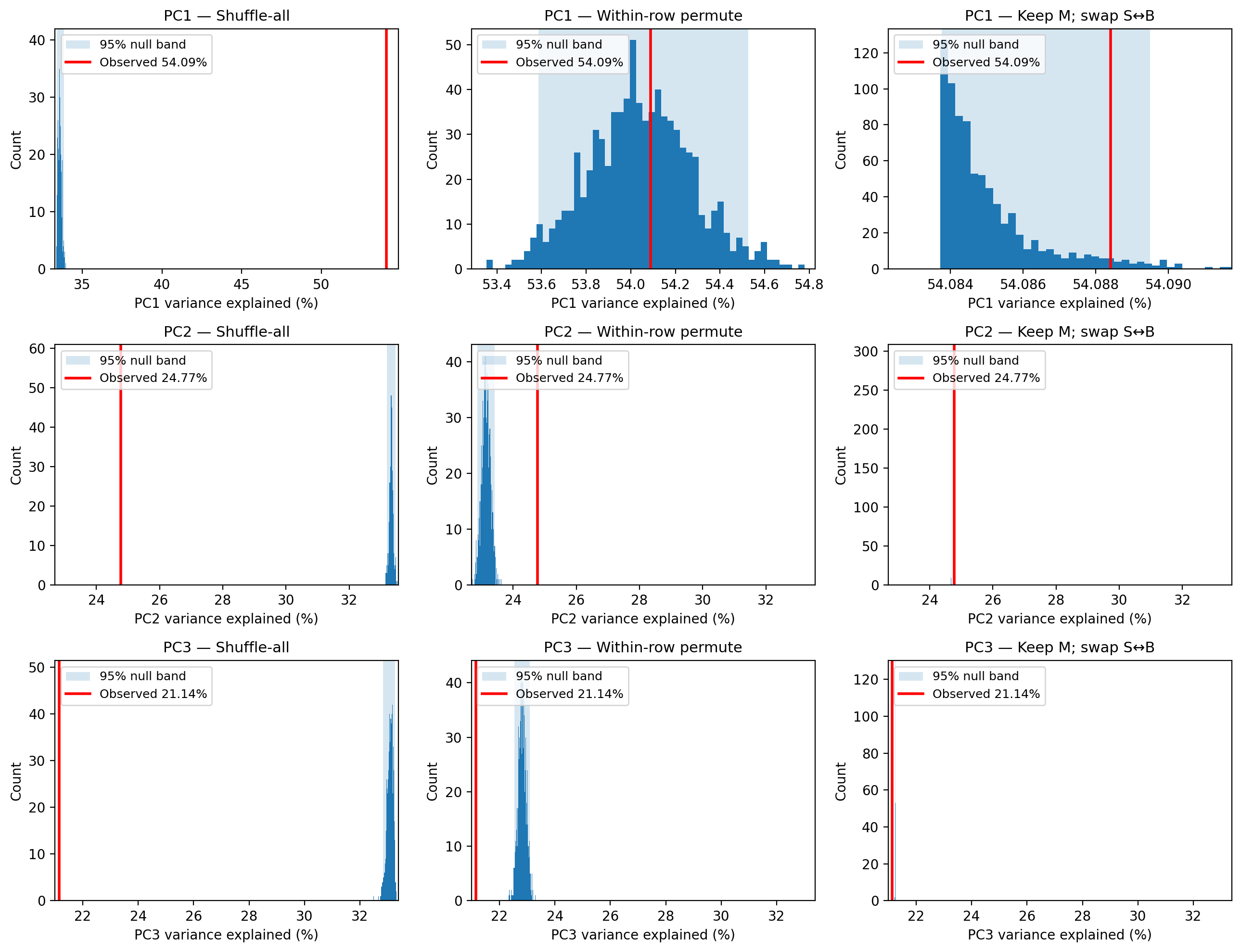}
    \caption{\textbf{Bootstrap checks for PC variance shares.}
    Each row corresponds to one principal component (top: PC1, middle: PC2, bottom: PC3).
    Each column adds structure to the null:
    \emph{(left)} ``shuffle--all'': independently permute each column across subreddits, destroying both row structure and cross--channel alignment;
    \emph{(middle)} ``within--row permute'': keep each subreddit's overall three--channel magnitude but randomize which entry is M, E, or B;
    \emph{(right)} ``keep $M$, swap $E\leftrightarrow B$'': preserve conversation and total control $E{+}B$ in every row, but erase which part of control comes from removals vs.\ bots.
    Red vertical lines show the observed variance shares (PC1 $\approx 54.1\%$, PC2 $\approx 24.8\%$, PC3 $\approx 21.1\%$). Shaded regions are 95\% null bands.}
    \label{fig:perm_progression}
\end{figure}

Figure~\ref{fig:pc_angles_validation} presents our mathematical test in four panels. Panel A shows the big picture--our three PCs as solid arrows (sized by variance explained) alongside dashed reference vectors representing theoretical governance patterns. Panels B--D zoom in on each PC individually.

\textbf{Testing PC1--Does it capture intensity?} Panel B shows PC1 (blue solid arrow) against the theoretical ``all modes together'' vector $[1,1,1]/\sqrt{3}$ (blue dashed). If PC1 represents balanced intensity, these should align. The angle between them? Just 2.5°--essentially perfect (cosine = 0.999). This indicates PC1 genuinely captures communities doing more of everything: more mutual interaction, more enforcement, more bot oversight, all in proportion.

\textbf{Testing PC2--Does it capture the trade-off of One-way vs. Two-way Communication?} Panel C tests whether PC2 represents one-way versus two-way. The red dashed vector $[1,-1,-1]/\sqrt{3}$ represents ``two-way minus one-way.'' PC2 points nearly opposite (156.3°), which means it captures the reverse: one-way minus two-way. The 23.7° deviation from perfect opposition reveals something interesting--enforcement dominates the coordination signal. The loading matrix indicates this, with enforcement at 0.500 versus bots at 0.260, essentially double the weight. Real communities like r/science achieve control primarily through removals, with bots playing a supporting role.

\textbf{Testing PC3--Does it capture Impersonal vs. Personal Moderation tactics?} Panel D shows PC3 against $[0,-1,1]/\sqrt{2}$, which contrasts bot oversight (positive) against enforcement (negative) while ignoring mutual interaction. The 9.8° angle indicates close alignment (cosine = 0.985). PC3 truly separates impersonal moderation (AutoModerator warnings) from personal nuanced (post-hoc removals). Communities like r/AskScience lean proactive with extensive bot rules, while others rely on human moderators sweeping through afterwards.

These aren't just suggestive patterns--the mathematical alignments support that our interpretations are encoded in the data's actual structure. Communities really do vary along three independent dimensions: how intensely they govern (PC1), whether they emphasize one-way over two-way coordination (PC2), and whether they orient towards more impersonal or personal moderation(PC3).

But geometric alignment is just one test. Do these structures hold up under statistical scrutiny? What happens if we systematically break different aspects of the data? If our interpretations are correct, PC1 should survive when we preserve community-level intensity, PC2 should persist when we maintain total coordination/control levels, and PC3 should remain when we keep the orientation split. Let's test these predictions with targeted permutations.

\subsubsection{Statistical Analysis Using Bootstrap Permutations}

The geometric angles in Fig.~\ref{fig:pc_angles_validation} tell us \emph{what} the three PCs point to. The permutations in Fig.~\ref{fig:perm_progression} ask the harder question: \emph{do those directions still explain the data when we deliberately break parts of the structure?} We build three nulls that add constraints step by step.

\paragraph{Null 1: shuffle all (no structure).}
For each governance channel $(M,E,B)$ we permute its values across subreddits. This wipes out correlations across channels and across communities, so with three exchangeable directions the PCs should all sit near $33\%$. That is exactly what we see in the left column in Fig.~\ref{fig:perm_progression}: all three observed shares (54.1, 24.8, 21.1) lie far to the right of the nulls, so the pattern in the real data is not an accident of sampling.

\paragraph{Null 2: within--row permutation (keep subreddit ``heat'').}
Next we keep each subreddit's \emph{total} three--channel level but shuffle which channel gets it. This is the right test for PC1: if PC1 is really ``do more of everything,'' then as long as we keep per-subreddit heat, PC1 should come back. In the top--middle panel of Fig.~\ref{fig:perm_progression} PC1's null is now centered right on the observed 54\%, and the 95\% band covers it, so PC1 is fully explained by community-level intensity. PC2 and PC3, however, are still off to the right in this middle column, which means there is structure left over after we account for heat.

\paragraph{Null 3: keep $M$, swap $S\leftrightarrow B$ (preserve one-way vs.\ two-way).}
Finally we hold conversation $M$ fixed and, for each subreddit, flip enforcement and bot entries with probability $1/2$. This preserves each row's ``how much one-way control overall'' $(E{+}B)$ but erases whether that control is delivered through removals or through bots. In the middle row, the PC2 null under this constraint moves right onto the observed 24.8\%, which is exactly what our interpretation says: PC2 cares about \emph{how much one-way control relative to two-way conversation}. The bottom row shows that PC3 does not collapse under this swap--there is still a smaller, tactic-level signal (bots vs.\ removals), but it is visibly more modest than PC1 and PC2.

\paragraph{Takeaway.} On reading the plots from left to right in Fig.~\ref{fig:perm_progression} here are our main findings: with no structure (shuffle-all), every PC is $\sim33\%$; once we keep subreddit heat (within--row permute), PC1 is fully reproduced; once we also keep ``one-way vs.\ two-way,'' (keep M, swap $E\leftrightarrow B$) PC2 is reproduced; what remains is a smaller, coupled tactic axis (PC3). This progression shows that the three governance dimensions we described are not artifacts of PCA but are the ones the data itself insists on once we hold fixed the pieces that moderators plausibly choose or inherit.

\subsection{Discussion: What This Reveals and What It Doesn't}
\label{sec:pca_discussion}

Our analysis maps how 44{,}471 subreddits organize their governance \emph{after} removing size effects. We showed that once scale is factored out, communities do not scatter randomly in a three--channel space; instead, their behaviors fall along a small set of systematic axes that we call modes. The geometric checks and the three-step permutation progression together tell the same story: if we preserve what a community plausibly ``is'' (its overall heat, and then its one-way vs.\ two-way coorindation choice), the PCs we found reappear. That means the structure is in the data, not an artifact of PCA settings.

\paragraph{What we've established.}
\begin{itemize}
    \item \textbf{A low-dimensional structure.} Three mathematically orthogonal dimensions (we call them modes) capture the bulk of size--adjusted variation: overall \emph{intensity} (all actions high together), \emph{one-way vs.\ two-way} (total $E{+}B$ set against $M$), and \emph{impersonal vs.\ personal} (bots vs.\ removals).
    \item \textbf{Robustness to targeted disruptions.} When we destroy \emph{all} structure (shuffle-all), the PCs collapse toward $33\%$ each; when we keep per--subreddit heat, PC1 comes back; when we additionally keep $M$ and the total budget $E{+}B$, PC2 comes back; only the small tactic dial remains. This progressive recovery means the interpretation is tied to substantive features of communities, not to fragile correlations.
    \item \textbf{Governance choices beyond size.} Size explains a lot (bigger subreddits simply do more), but even at the same scale communities pick different balances among conversation, enforcement choice, and moderation orientation.
\end{itemize}

\paragraph{What we cannot claim.}
Our analysis does not establish causality: although communities and topics that appear ``hot’’ align with higher values along the primary governance mode (PC1), we cannot determine whether these patterns are driven by topic characteristics, user composition, platform policy, or moderator labor. Nor do we make claims about optimality. Positions in the residual governance space—--defined by the relative balance of mutual interaction, human enforcement, and automated oversight are not inherently ``good’’ or ``bad,’’ but instead reflect different governance modes suited to distinct community goals or government styles. A highly controlled science forum and a lightly regulated casual forum may both be functioning effectively relative to their own objectives, and our analysis does not model success, retention, or content quality. We also observe only the measurable outputs of governance----comment density as mutual interaction, removals as human enforcement, and bot activity as automated oversight rather than the internal deliberations or negotiations that produced them. We can therefore characterize which governance mode a community exhibits at scale, but not the decision processes through which moderators arrived at that configuration.

\subsubsection{First Steps Toward Understanding Complex Governance}
The three dimensions we extracted should be read as a \emph{coordinate system}, not a complete theory of online moderation. What the permutation results show is that if we hold fixed the elements moderators plausibly decide or inherit—how active a community is and how much control it seeks relative to conversation—the same axes re-emerge. This indicates that we have identified core dimensions along which governance actions systematically vary.

At the same time, real governance is more granular than three summary dimensions. Communities differ in rule content, moderator capacity, sensitivity to platform-wide events, and the types of conflicts they face. Our method does not observe these processes directly; it observes only the configuration of regulatory actions they produce across mutual interaction ($M$), enforcement ($E$), and automation ($B$).

This opens the next set of questions rather than closing them: Why do some communities settle into high-control modes while others sustain high-conversation modes? Why do some adopt proactive automated tools while others rely more heavily on reactive human enforcement, even at similar overall intensity? How do changes in topic, membership turnover, or platform policy shift a community within this governance space? And, once linked to outcome measures such as participation health, newcomer survival, or content quality, can these dimensions help predict which governance bundles are most sustainable?

Till now in this SI, we show that governance on Reddit is \emph{structured} once size is removed; we stress-test that structure through targeted null models; and we provide a compact language—intensity, one-way vs.\ two-way coordination, and impersonal vs.\ personal moderation—for describing governance modes beyond scale.

\paragraph{Connection to the broader analysis.} The PCA results complete our empirical picture of Reddit governance. The main text established that regulatory actions scale systematically with community size. This Supplementary Information has shown that these scaling relationships are grounded in valid data (Section~\ref{sec:S1}), temporally stable (Section~\ref{sec:S3}), sample-robust (Section~\ref{sec:S4}), and externally validated (Section~\ref{sec:robustness}). The PCA analysis in this section reveals that the residual variation-what remains after accounting for size-is not noise but structured governance choice. A natural next question, however, is whether these same dimensions also structure change over time. If the extracted modes reflect substantive governance organization, communities should adjust along these axes rather than move randomly. We therefore turn to temporal analysis, examining whether year-to-year changes in regulatory intensity ($\zeta_1$) and coordination structure ($\zeta_2$) are coupled, persistent, and locally bounded within the same governance space (see Fig.~5 in the main text). This allows us to test whether the cross-sectional modes identified earlier also organize the dynamics of governance over time.

\section{Temporal dynamics in governance space}
\label{sec:temporal_dynamics}

\subsection{Overview}

Figure 4 in the main text characterizes how communities are positioned within the governance mode space after removing size effects for a cross-section. Here we examine how communities \emph{move} within that same space over time. If governance modes capture meaningful structure, temporal adjustments should occur along these dimensions rather than arbitrarily across the quadrants of our modes space.

Specifically, we analyze whether changes in regulatory intensity ($\zeta_1$) and coordination structure ($\zeta_2$) occur independently or in a coupled manner, whether governance positions persist across years, and whether communities undergo large structural transitions or primarily local adjustments. Figure~5 in the main text visualizes these dynamics.

\subsection{Year-to-year changes in governance modes}

We compute annual changes
\[
\Delta \zeta_{k,i,t} = \zeta_{k,i,t+1} - \zeta_{k,i,t}
\]
for each subreddit $i$ and mode $k\in\{1,2\}$, allowing us to quantify how governance configurations evolve from one year to the next.

Figure~5B (main text) plots $\Delta\zeta_1$ versus $\Delta\zeta_2$ across all subreddits and years.

Two primary patterns emerge: firstly, the joint distribution of changes is tightly concentrated around zero for both modes. Most communities exhibit modest year-to-year adjustments rather than dramatic reconfigurations. This suggests that governance modes are not highly volatile but instead fluctuate within relatively constrained ranges. Secondly, changes in the two modes are positively associated. Increases in regulatory intensity tends to coincide with shifts toward more one-way coordination (greater reliance on enforcement and automation relative to conversation), while decreases in intensity coincide with shifts toward more conversational coordination. This coupling indicates that changes in how much regulation communities deploy are systematically linked to changes in how regulation is structured. Communities do not typically increase enforcement without also shifting their overall coordination orientation. So, even though these communities occupy different positions, their fluctuations are local and their movements follow a similar mechanism of increase in intensity leads to increase in one-way.

\subsection{Bounded movement and local adjustment}
We next examine the magnitude of displacement within governance space. Rather than frequently crossing into qualitatively different regimes, most communities exhibit bounded fluctuations around a characteristic region (Fig.~5A). Large quadrant-crossing transitions are rare. At the aggregate level, topic-level centroids remain well separated across years, suggesting that domain-specific governance styles persist. At the individual subreddit level, communities primarily move within local neighborhoods associated with their broader category (Fig.~5A). This pattern indicates that governance evolution is typically gradual and path-dependent rather than characterized by abrupt structural resets.

\subsection{Quadrant-specific adjustment dynamics}
To assess whether the observed coupling between $\Delta\zeta_1$ and $\Delta\zeta_2$ is driven by a particular governance style, we repeat the temporal analysis separately within each quadrant of the governance space. Quadrants are defined by the signs of $(\zeta_1,\zeta_2)$, distinguishing communities that are relatively high vs.\ low intensity and more one-way vs.\ more conversational in coordination structure. Within each quadrant, year-to-year changes in regulatory intensity and coordination structure remain positively associated. That is, communities that are already relatively high-control, low-control, conversational, or enforcement-oriented all exhibit similar directional coupling: increases in intensity are accompanied by shifts toward more one-way coordination, and decreases in intensity coincide with shifts toward more conversational coordination. Importantly, this association need not have occurred by construction. The axes are orthogonal cross-sectionally, and thus uncorrelated at a fixed time point. The persistence of positive coupling within quadrants therefore reflects a feature of governance adjustment rather than a statistical artifact of axis construction. The results indicate that the mechanism of adjustment is broadly shared across governance styles. Communities occupying different regions of governance space do not follow fundamentally different temporal rules; rather, they adjust along the same structured dimensions, even if their baseline positions differ.

\subsection{Interpretation}

Taken together, these results indicate that the governance modes identified from cross-sectional covariance also structure temporal adjustment. Communities move through governance space primarily via incremental, locally constrained adjustments rather than abrupt transitions between fundamentally different governance configurations.

\section{Labelled Subreddit List}
\label{sec:S7}

We used a curated list of 2,828 subreddits categorized into 88 distinct groups (e.g., Science, Politics, Images) obtained from the r/ListOfSubreddits community. Table~\ref{tab:subreddit_sample} presents a representative sample; the complete list is available in the online repository.

\begin{longtable}{ll}
\caption{Representative sample of labeled subreddits used in our analyses.}
\label{tab:subreddit_sample} \\
\toprule
\textbf{Subreddit Category} & \textbf{Subreddit Name} \\
\midrule
\endfirsthead
\caption[]{(continued)} \\
\toprule
\textbf{Subreddit Category} & \textbf{Subreddit Name} \\
\midrule
\endhead
\midrule
\multicolumn{2}{r}{\textit{Continued on next page}} \\
\endfoot
\bottomrule
\endlastfoot
Gifs & /r/gifs \\
Gifs & /r/behindthegifs \\
Gifs & /r/gif \\
Gifs & /r/Cinemagraphs \\
Gifs & /r/WastedGifs \\
Gifs & /r/educationalgifs \\
Gifs & /r/perfectloops \\
Gifs & /r/highqualitygifs \\
Gifs & /r/gifsound \\
Gifs & /r/combinedgifs \\
Gifs & /r/retiredgif \\
Gifs & /r/michaelbaygifs \\
Gifs & /r/gifrecipes \\
Gifs & /r/mechanical\_gifs \\
Gifs & /r/bettereveryloop \\
Gifs & /r/gifextra \\
Gifs & /r/slygifs \\
Gifs & /r/gifsthatkeepongiving \\
Gifs & /r/wholesomegifs \\
Gifs & /r/noisygifs \\
Gifs & /r/brokengifs \\
Gifs & /r/loadingicons \\
Gifs & /r/splitdepthgifs \\
Gifs & /r/blackpeoplegifs \\
Gifs & /r/whitepeoplegifs \\
Gifs & /r/scriptedasiangifs \\
Gifs & /r/reactiongifs \\
Gifs & /r/shittyreactiongifs \\
Gifs & /r/chemicalreactiongifs \\
Gifs & /r/physicsgifs \\
Gifs & /r/babyelephantgifs \\
Gifs & /r/weathergifs \\
Images & /r/pics \\
Images & /r/PhotoshopBattles \\
Images & /r/perfecttiming \\
Images & /r/itookapicture \\
Images & /r/Pareidolia \\
Images & /r/ExpectationVSReality \\
Images & /r/dogpictures \\
Images & /r/misleadingthumbnails \\
Images & /r/FifthWorldPics \\
Images & /r/TheWayWeWere \\
Images & /r/pic \\
Images & /r/nocontextpics \\
Images & /r/miniworlds \\
Images & /r/foundpaper \\
Images & /r/images \\
Images & /r/screenshots \\
Images & /r/mildlyinteresting \\
Images & /r/interestingasfuck \\
Images & /r/damnthatsinteresting \\
Images & /r/beamazed \\
Images & /r/reallifeshinies \\
Images & /r/thatsinsane \\
Videos & /r/videos \\
Videos & /r/youtubehaiku \\
Videos & /r/artisanvideos \\
Videos & /r/DeepIntoYouTube \\
Videos & /r/nottimanderic \\
Videos & /r/praisethecameraman \\
Videos & /r/killthecameraman \\
Videos & /r/perfectlycutscreams \\
Videos & /r/donthelpjustfilm \\
Videos & /r/abruptchaos \\
General Discussion & /r/ShowerThoughts \\
General Discussion & /r/DoesAnybodyElse \\
General Discussion & /r/changemyview \\
General Discussion & /r/crazyideas \\
General Discussion & /r/howtonotgiveafuck \\
General Discussion & /r/tipofmytongue \\
General Discussion & /r/quotes \\
General Discussion & /r/casualconversation \\
General Discussion & /r/makenewfriendshere \\
Advice & /r/relationship\_advice \\
Advice & /r/raisedbynarcissists \\
Advice & /r/legaladvice \\
Advice & /r/bestoflegaladvice \\
Advice & /r/advice \\
Advice & /r/amitheasshole \\
Advice & /r/mechanicadvice \\
Advice & /r/toastme \\
Advice & /r/needadvice \\
AMA & /r/IAmA \\
AMA & /r/ExplainlikeIAmA \\
AMA & /r/AMA \\
AMA & /r/casualiama \\
Question/Answer & /r/whatisthisthing \\
Question/Answer & /r/answers \\
Question/Answer & /r/NoStupidQuestions \\
Question/Answer & /r/AskReddit \\
Question/Answer & /r/AskScience \\
Question/Answer & /r/askhistorians \\
Question/Answer & /r/askwomen \\
Question/Answer & /r/askmen \\
Stories & /r/tifu \\
Stories & /r/self \\
Stories & /r/confession \\
Stories & /r/talesfromtechsupport \\
Stories & /r/talesfromretail \\
Stories & /r/nosleep \\
Stories & /r/LetsNotMeet \\
Stories & /r/pettyrevenge \\
Stories & /r/prorevenge \\
Support & /r/depression \\
Support & /r/SuicideWatch \\
Support & /r/Anxiety \\
Support & /r/foreveralone \\
Support & /r/offmychest \\
Support & /r/socialanxiety \\
General Educational & /r/YouShouldKnow \\
General Educational & /r/todayilearned \\
General Educational & /r/explainlikeimfive \\
General Educational & /r/IWantToLearn \\
Art & /r/Art \\
Art & /r/redditgetsdrawn \\
Art & /r/drawing \\
Art & /r/pixelart \\
Art & /r/learnart \\
Science & /r/Science \\
Science & /r/AskScience \\
Science & /r/Space \\
Science & /r/astronomy \\
Science & /r/biology \\
Science & /r/chemistry \\
Science & /r/physics \\
History & /r/history \\
History & /r/AskHistorians \\
History & /r/ColorizedHistory \\
History & /r/HistoryPorn \\
History & /r/historymemes \\
Computer Science/Engineering & /r/gamedev \\
Computer Science/Engineering & /r/engineering \\
Computer Science/Engineering & /r/cscareerquestions \\
Computer Science/Engineering & /r/learnprogramming \\
Computer Science/Engineering & /r/python \\
Computer Science/Engineering & /r/machinelearning \\
Economics & /r/Economics \\
Economics & /r/wallstreetbets \\
Economics & /r/stocks \\
Anime/Manga & /r/anime \\
Anime/Manga & /r/manga \\
Anime/Manga & /r/pokemon \\
Anime/Manga & /r/onepiece \\
Anime/Manga & /r/naruto \\
Books/Writing & /r/Books \\
Books/Writing & /r/WritingPrompts \\
Books/Writing & /r/writing \\
Books/Writing & /r/harrypotter \\
Books/Writing & /r/gameofthrones \\
Books/Writing & /r/lotr \\
Video games & /r/gaming \\
Video games & /r/Games \\
Video games & /r/pcmasterrace \\
Video games & /r/nintendo \\
Video games & /r/PS4 \\
Video games & /r/xboxone \\
Video games & /r/steam \\
Video games & /r/minecraft \\
Video games & /r/leagueoflegends \\
Video games & /r/overwatch \\
Video games & /r/fortnitebr \\
Games (Tabletop) & /r/DnD \\
Games (Tabletop) & /r/boardgames \\
Games (Tabletop) & /r/magicTCG \\
\end{longtable}

\textit{Note: This table shows a representative sample of 164 subreddits across 19 categories. The full list of 2,828 subreddits across 88 categories is available in the online repository.}

\bibliographystyle{unsrtnat}
\bibliography{references}